\documentclass[pra,floatfix,twocolumn,showpacs,superscriptaddress,notitlepage,amsmath,english]{revtex4-2} 

\usepackage{pifont}
\usepackage[colorlinks=true,citecolor=blue,linkcolor=blue,urlcolor=blue]{hyperref}
\usepackage{graphicx}
\usepackage{nicefrac}
\usepackage{blindtext}
\usepackage{float}
\usepackage{multirow} 
\usepackage{xcolor}
\usepackage{colortbl}
\usepackage{booktabs}

\usepackage{babel}

\def\ZZ{\mathbb Z}

\usepackage{amsmath,amssymb,bm,color}
\usepackage{dsfont}
\usepackage{framed}
\usepackage{braket}
\definecolor{shadecolor}{gray}{0.95}
\usepackage{amsthm}
\usepackage[normalem]{ulem}
\usepackage{siunitx}
\usepackage{wasysym}
\usepackage{tikz}
\usetikzlibrary{quantikz2}
\usepackage[compat=0.6]{yquant}
\usepackage{physics}
\usetikzlibrary{arrows.meta}

\def\ds3{$\mathcal{D}(S_3)$}

\def\mubar{\mkern 1.5mu\overline{\mkern-1.5mu \mu \mkern-1.5mu}\mkern 1.5mu}
\def\mus{ \mu \sigma}
\def\mubars{ \mkern 1.5mu\overline{\mkern-1.5mu \mu \mkern-1.5mu}\mkern 1.5mu \sigma}
\def\ZZ{\mathbb Z}

\newcommand{\cbox}[2]{\vcenter{\hbox{\includegraphics[width=#1em]{#2}}}}

\begin{document}

\begin{abstract}
Abstract
\end{abstract}







\title{Universal Gates from Braiding and Fusing Anyons on Quantum Hardware}


\author{Chiu Fan Bowen Lo}
\affiliation{Department of Physics, Harvard University, Cambridge, MA 02138, USA}

\author{Anasuya Lyons}
\affiliation{Department of Physics, Harvard University, Cambridge, MA 02138, USA}

\author{Dan Gresh}
\author{Michael Mills}
\author{Peter E. Siegfried}
\author{Maxwell D. Urmey}
\affiliation{Quantinuum, 303 S Technology Ct, Broomfield, CO 80021, USA}

\author{Nathanan Tantivasadakarn}
\affiliation{C. N. Yang Institute for Theoretical Physics, Stony Brook University, Stony Brook, NY 11794, USA}

\author{Henrik Dreyer}
\affiliation{Quantinuum, Leopoldstrasse 180, 80804 Munich, Germany}

\author{Ashvin Vishwanath}
\affiliation{Department of Physics, Harvard University, Cambridge, MA 02138, USA}

\author{Ruben Verresen}
\affiliation{Pritzker School of Molecular Engineering, University of Chicago, Chicago, IL 60637, USA}

\author{Mohsin Iqbal}
\affiliation{Quantinuum, Leopoldstrasse 180, 80804 Munich, Germany}

\date{\today}
\begin{abstract}
Topological quantum computation encodes quantum information in the internal fusion space of non-Abelian anyonic quasiparticles, whose braiding implements logical gates. This goes beyond Abelian topological order (TO) such as the toric code, as its anyons lack internal structure. However, the simplest non-Abelian generalizations of the toric code do not support universality via braiding alone. Here we demonstrate that such minimally non-Abelian TOs can be made universal by treating anyon fusion as a computational primitive.
We prepare a 54-qubit TO wavefunction associated with the smallest non-Abelian group, $S_3$, on Quantinuum’s H2 quantum processor. This phase of matter exhibits cyclic anyon fusion rules, known to underpin universality, which we evidence by trapping a single non-Abelian anyon on the torus. We encode logical qutrits in the nonlocal fusion space of non-Abelian fluxes and, by combining an entangling braiding operation with anyon charge measurements, realize a universal topological gate set and read-out, which we further demonstrate by topologically preparing a magic state.
This work establishes $S_3$ TO as simple enough to be prepared efficiently, yet rich enough to enable universal topological quantum computation.
\end{abstract}
\maketitle

\section{Introduction}
The appeal of topologically ordered (TO) phases of matter lies in their ability to encode information in global properties insensitive to local perturbations~\cite{kitaev2003fault,wen_topological_1990,wen_book_2004,sachdev_book_2023,simon_book_2023}. The toric code~\cite{kitaev2003fault} exemplifies this idea: an emergent $\mathbb Z_2$ gauge theory~\cite{Wegner71,Kogut79,Fradkin79} whose anyonic quasiparticles exhibit nontrivial exchange statistics~\cite{leinaas_theory_1977,wilczek_quantum_1982, Frohlich1990, wen1991}, forming the foundation of the paradigmatic surface code~\cite{bravyi_quantum_1998,freedman_projective_2001,fowler_surface_2012}, where quantum information is stored nonlocally in the degenerate ground-state subspace~\cite{dennis_topological_2002}. A qualitatively different approach is to encode information directly in the internal degrees of freedom of anyons~\cite{kitaev2003fault,freedman_modular_2002,nayak_non-abelian_2008}, which requires non-Abelian anyons~\cite{Goldin85} and is minimally realized in quantum double models based on finite non-Abelian groups~\cite{kitaev2003fault,kitaev_finite_group_2007,preskill_topo_notes}.

Until recently, this latter approach was out of experimental reach. However, the discovery that, for solvable groups, constant-depth adaptive circuits can prepare non-Abelian quantum double models~\cite{tanti2023hierarchy,sptmeasure,verresen2021efficiently,tantivasadakarn_shortest_2023,bravyi2022adaptiveconstantdepthcircuitsmanipulating} as well as their anyons~\cite{bravyi2022adaptiveconstantdepthcircuitsmanipulating,lyons-2025,ren_efficient_2025}, has prompted renewed theoretical interest in their utility~\cite{huang_generating_2025, davydova_universal_2025,sajith_non-clifford_2025,huang2025hybrid,kobayashi2025clifford,warman2025transversal}. These measurement-based approaches substantially lower the experimental barriers to realizing such phases on quantum hardware, with a recent experimental realization of $D_4$ TO~\cite{iqbal2024non}.

Despite increased experimental accessibility to non-Abelian defects and TOs~\cite{andersen_non-abelian_2023, xu2023digital, iqbal2024non, xu_non-abelian_2024, iqbal_qutrit_2025, minev2025realizing,aghaee_distinct_2025,aghaee_interferometric_2025}, a fundamental limitation remains. While TOs based on finite groups are desirable due to being minimal non-Abelian generalizations of the toric code, their braiding operations do not provide a universal set of quantum gates~\cite{mochon_anyons_2003,mochon_smaller_groups_2004,etingof2008braid}. This stands in contrast to more exotic topological phases such as Fibonacci TO where braid-only universality can be achieved~\cite{freedman_modular_2002,nayak_non-abelian_2008}, albeit at the cost of significantly greater complexity in their realization \cite{minev2025realizing,xu_non-abelian_2024}, conjectured no-go's of scalable preparation protocols \cite{tanti2023hierarchy} and challenges in finding leakage-free gate sets \cite{ainsworth_topological_2011,cui_search_2019,burke_topological_2024}.

Even in idealized settings for topological quantum computation, however, anyon fusion and topological charge measurement are required to extract information encoded in non-Abelian anyons~\cite{overbosch_inequivalent_2001,kitaev2003fault,nayak_non-abelian_2008}. This observation raises a natural question: if measurement has already proven instrumental for preparing and manipulating non-Abelian topological phases, can it also be harnessed to enhance their computational power?
Indeed, theoretical work has suggested that TOs associated with \textit{cyclic} anyon fusion rules~\cite{Dauphinais2017}, like $a \times \bar a = a + \cdots$,
become universal when anyon fusion measurement is incorporated into the gate set~\cite{galindo_acyclic_2018,mochon_anyons_2003,mochon_smaller_groups_2004,ogburn_preskill_topological_1999,kitaev_finite_group_2007,SU24,Cui2015_metaplectic,measurementonly}. While this cyclic criterion excludes the previously realized $D_4$ TO, it applies to the quantum double of the smallest non-Abelian group, $S_3$~\cite{kitaev_finite_group_2007,cui_universal_2015,chen_universal_2025,byles_demonstration_2025,lo_universal_2025}.

\begin{figure*}[t]
	\centering \includegraphics[width=1.00\linewidth]{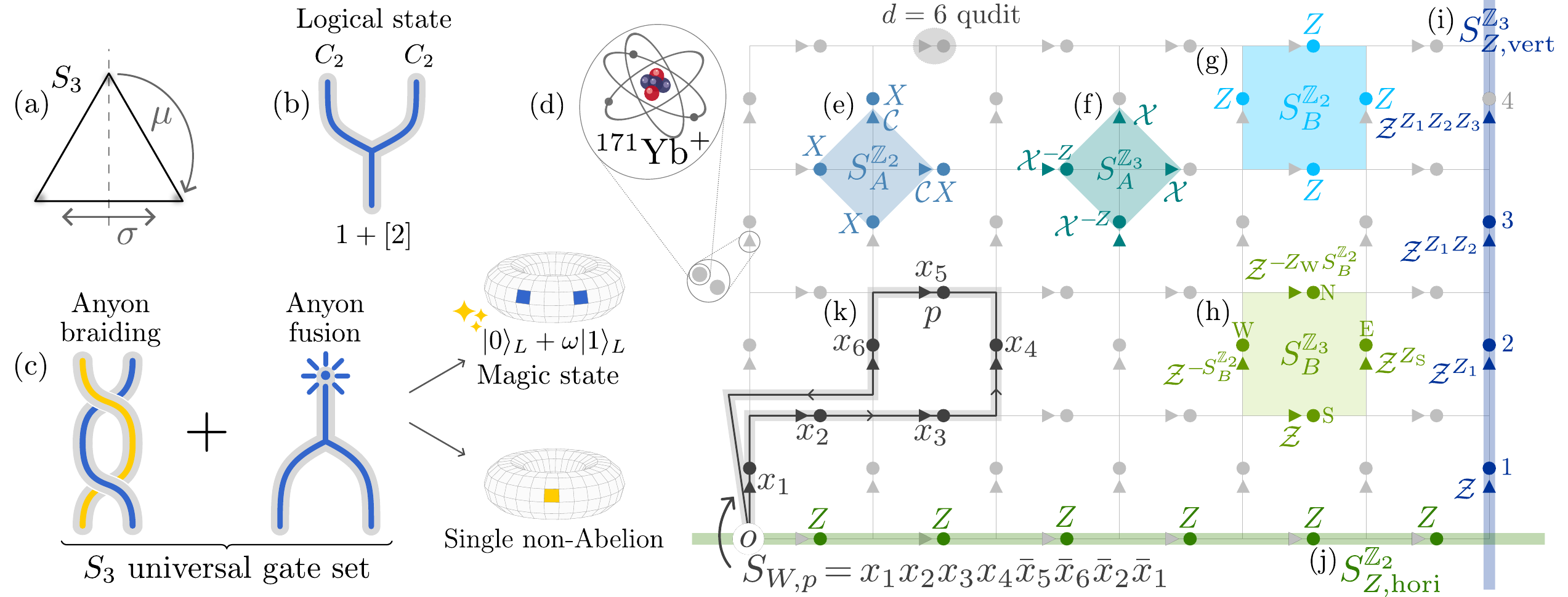}
	\caption{\textbf{Topological order based on $S_3$.} (a) Symmetry group $S_3 = \mathbb Z_3 \rtimes \mathbb Z_2$ of an equilateral triangle, generated by rotation $\mu$ and reflection $\sigma$. (b) Logical state encoded in the flux-neutral fusion space of a pair of $C_2$ fluxes. (c) Anyon braiding together with fusion realizes a universal gate set for quantum computation. (d) The code is defined on a square lattice on a torus with oriented edges hosting $d\!=\!6$ qudits, composed of a qubit (solid circle) and a qutrit (solid triangle). The triangle direction specifies the edge orientation. Physically, each qubit is encoded in the hyperfine states of $^{171}\mathrm{Yb}^+$ ions. (e--k) The stabilizers decompose into Pauli operators $X,Z$ on qubits and generalized Pauli operators $\mathcal{X},\mathcal{Z}$ together with charge conjugation $\mathcal{C}$ on qutrits (see Section~\ref{app:qubit-qutrit-encoding}). There are (e) qubit vertex stabilizers $S^{\ZZ_2}_A$, (f) qutrit vertex stabilizers $S^{\ZZ_3}_A$ (these act on qutrits, conditioned on qubits), (g) qubit plaquette stabilizers ($S_B^{\ZZ_2}$) and (h) qutrit plaquette stabilizers ($S_B^{\ZZ_3}$).
    There are also non-contractible versions of the (i) $S^{\ZZ_3}_Z$ and (j) $S^{\ZZ_2}_Z$ stabilizers.
	The stabilizer $S_{W,p}$ associated with the global flux $W$ is defined as the product of $S_3$ group elements $x_i$ along the directed loop in~(k), beginning at the origin $o$ and encircling exactly one plaquette~$p$ (see Section~\ref{app:z2z3-decomposition}).
	}
	\label{fig:notation}
\end{figure*}
This work presents the first experimental realization of a universal topological gate set based on the braiding and fusion of non-Abelian anyons in the scalable non-Abelian TO, in particular the $S_3$ quantum double model on a trapped-ion quantum computing architecture~\cite{moses_race-track_2023}. This is achieved via a two-step process which can be interpreted as coupling a $\mathbb Z_2$ and $\mathbb Z_3$ toric code~\cite{verresen2021efficiently,sptmeasure,bravyi2022adaptiveconstantdepthcircuitsmanipulating,tanti2023hierarchy}. This enables the $S_3$ TO ground state as well as the realization of a \textit{single} non-Abelian anyon on a torus~\cite{bombin_family_2008}, which we link to the cyclicity of the anyon fusion rules. We nonlocally encode logical information in the internal states of non-Abelian $C_2$-fluxes and demonstrate a universal gate set on these topological qutrits, consisting of three primitive operations: (i)~an entangling gate implemented by coherent anyon braiding, (ii)~measurement of the topological qutrit in the $\mathcal{X}$-basis, and (iii)~measurement of the topological qutrit in the $\mathcal{Z}$-basis. Any quantum operation on pairs of qutrits (and therefore also qubits) can be synthesized from these three primitives~\cite{kitaev_finite_group_2007, lo_universal_2025}. We further demonstrate how to prepare reference states (the so-called ``bureau of standards"~\cite{preskill_topo_notes}) and how our primitives can be used to prepare a magic state on a topological qutrit (Fig.~\ref{fig:notation}(a-c)).

\section{The $S_3$ quantum double}\label{subsec:S3-basics}

The internal Hilbert space consists of 6-state qudits labeled by the elements of the group $S_3=\langle\mu,\sigma|\mu^3= \sigma^2 = (\mu \sigma)^2=1\rangle$~(Fig.~\ref{fig:notation}a). We encode this as a qubit-qutrit pair $g=\mu^a\sigma^b$, where $a$ is the qutrit and $b$ the qubit value. On the Quantinuum hardware, the qutrit is itself composed of two qubits, with each qubit being encoded in an ion (Fig.~\ref{fig:notation}d).

The $S_3$ quantum double~\cite{kitaev_finite_group_2007} ground states are stabilized by the following qubit and qutrit projectors
\begin{align} 
    A^{\mathbb{Z}_2}_v &= \frac{1}{2}(1+S_A^{\mathbb{Z}_2})\;,\; \quad A^{\mathbb{Z}_3}_v=\frac{1}{3}(1+(S_A^{\mathbb{Z}_3} +\textrm{h.c.}))\;,\; \label{eq:qutrit-qubit-vertex-projectors}\\
    B^{\mathbb{Z}_2}_p &= \frac{1}{2}(1+S_B^{\mathbb{Z}_2})\;,\; \quad B^{\mathbb{Z}_3}_p = \frac{1}{3}(1+(S_B^{\mathbb{Z}_3} +\textrm{h.c.}))\;,\;\label{eq:qutrit-qubit-plaquette-projectors}
\end{align}
where the qubit and qutrit vertex and plaquette stabilizers, $S_A^{\mathbb{Z}_2}$,  $S_A^{\mathbb{Z}_3}$, $S_B^{\mathbb{Z}_2}$, and  $S_B^{\mathbb{Z}_3}$ are defined in Fig.~\ref{fig:notation}(e-h). The  $A^{\mathbb{Z}_2}_v$ and $A^{\mathbb{Z}_3}_v$ project into the zero-charge sector at vertex $v$, while $B^{\mathbb{Z}_2}_p$ and $B^{\mathbb{Z}_3}_p$ project into the trivial-flux sector at plaquette $p$. The resulting state can be interpreted as a deconfined $S_3$ gauge theory, whose quasiparticles exhibit anyonic exchange statistics due to the Aharonov-Bohm effect.

There are $8$ anyon types in the $S_3$ quantum double model (see Section~\ref{append-sub:anyon-type}). We will work with three types of non-Abelian anyons: the charge transforming under the two-dimensional $[2]$ representation of $S_3$, detected by a violation of $A^{\mathbb{Z}_3}$ (`$[2]$ charge'); the two-dimensional flux which violates a $B_p^{\mathbb Z_3}$ plaquette, with flux state labeled by the conjugacy class $C_3$ represented by $\mu$ (`$C_3$ flux'); and the three-dimensional flux, which violates both $B_p^{\mathbb Z_2,\mathbb Z_3}$ plaquettes, with flux state labeled by the conjugacy class $C_2$ represented by $\sigma$ (`$C_2$ flux').

\section{State Preparation}\label{sec:low-energy-state-prep}

\subsection{Ground state preparation}\label{subsec:ground-state-prep}

We use the all-to-all connectivity of the trapped ion platform to realize the $S_3$ quantum double ground states on a torus, where there are $8$ ground states in the $S_3$ quantum double, in one-to-one correspondence with the $8$ anyon types. We label the ground states by non-contractible line operators (Fig.~\ref{fig:notation}(i-j)). 

\begin{figure}[t]
    \centering
    \begin{tikzpicture}
    \node at (0,0){
    \includegraphics[width=1\linewidth]{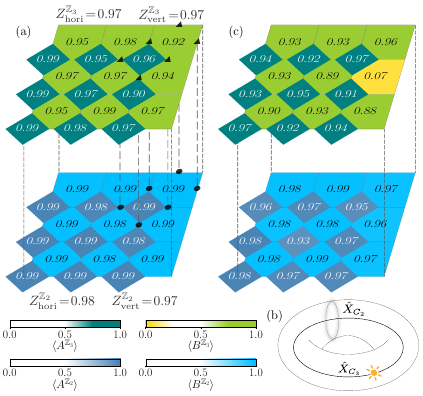}
    };
    \node[orange!80!yellow]at (3.4,-3.3){$C_3$};
     \end{tikzpicture}
    \caption{\textbf{Ground and single anyon states on 54 qubits.} (a) $\ZZ_2$ (bottom) and $\ZZ_3$ (top) vertex and plaquette projector expectation values on a $3\times3$ periodic square lattice (see Eqs.~\ref{eq:qutrit-qubit-vertex-projectors}, \ref{eq:qutrit-qubit-plaquette-projectors}). Lighter (darker) squares (rhombi) denote plaquette (vertex) projectors.
    The mean $\mathbb Z_2$ and $\mathbb Z_3$ projector values are 0.987(2) and 0.962(4), respectively and the average (maximum) standard error is 0.014 (0.029). For the non-contractible projectors, the average (maximum) standard error is 0.017 (0.022). 
    (b)~Single $C_3$ anyon creation by applying a horizontal $\hat X_{C_3}$ to a ground state with vertical $C_2$ flux loop.
    (c) The anyon is identified by a single excited $B^{\ZZ_3}$ plaquette projector; average (maximum) standard error is 0.020 (0.034).
	}
    \label{fig:gs}
\end{figure}

The state preparation protocol mirrors the decomposition of $S_3$ group elements into a qutrit and a qubit, using the fact that $S_3$ can be expressed as a semidirect product $\mathbb{Z}_3 \rtimes \mathbb{Z}_2$. We use a gauging procedure to prepare a $\mathbb{Z}_3$ qutrit toric code \cite{iqbal_qutrit_2025} and subsequently apply another gauging operation (that by isolation would produce a $\mathbb{Z}_2$ qubit TC) to obtain the ground states of $S_3$ TO \cite{verresen2021efficiently}. We implement this construction using a fully unitary circuit (see Section~\ref{app:gs_prep}). The measured projector expectation values are shown in Fig.~\ref{fig:gs}a (see Methods Section~\ref{method:gs-prep} for details of the expectation value estimation). Because the target state $\ket{\mathrm{gs}}$ is fully specified by its projectors, we bound the per-qudit fidelity of the prepared state $\rho$ on the $3\times3$ lattice with 18 sites as (see Section~\ref{app:fidelity_s3})
\begin{align}
0.970(5) \leq \bra{\mathrm{gs}} \rho \ket{\mathrm{gs}}^{1/18} \leq 0.988(3).
\end{align}

Since $S_3$ is a solvable group, it also admits a constant-depth adaptive preparation \cite{verresen2021efficiently,bravyi2022adaptiveconstantdepthcircuitsmanipulating}. We find a comparable fidelity between 0.930(8) and 0.978(2) (see Fig.~\ref{fig:appendix_gs_measurement_based} and Section~\ref{app:adaptive-prep}). We use the ground states prepared unitarily for the remaining demonstration due to the slightly better fidelity for the current system size; however, the measurement-based preparation protocol guarantees scalability of future experiments.

%
%
\subsection{A single non-Abelian anyon on the torus}

A hallmark of non-Abelian TO is the existence of low-lying excited states hosting a single anyon on the torus. A recent realization of $D_4$ TO demonstrated an excited state with a single unpaired \emph{Abelian} anyon~\cite{iqbal2024non}. While such an unpaired anyon certifies the non-Abelianness of that state, here we target a more complex and computationally powerful quantum phase: a \emph{cyclic} TO. 
Cyclicity is expected to enable universal computation via braiding and fusion~\cite{galindo_acyclic_2018,mochon_smaller_groups_2004}. A diagnostic of cyclic fusion rules is the ability to generate a single non-Abelian anyon on the torus by braiding an \textit{identical} anyon pair around a noncontractible loop.

Starting from a nontrivial ground state with a $C_2$ flux along the vertical noncontractible cycle, we create a single $C_3$ anyon by braiding a $C_3$ flux around the horizontal noncontractible cycle of the torus (Fig.~\ref{fig:gs}b). The nontrivial braiding between the two flux types toggles the fusion channel of the $C_3$ pair from the vacuum to $C_3$, leaving a single $C_3$ flux anyon (Fig.~\ref{fig:gs}c); see Section~\ref{sec:single-anyon} for more details on the preparation protocol. This constitutes the first experimental realization of a single \emph{non-Abelian} flux anyon in a lattice model with TO. 



We note that we will not use this single non-Abelion to store quantum information. Rather, the above demonstrates the cyclicity of the fusion rules, which is a signature of the computational power of the $S_3$ quantum double. We next demonstrate explicit building blocks of a universal gate set.

\begin{figure*}[t]
	\centering    \includegraphics[width=0.9\linewidth]{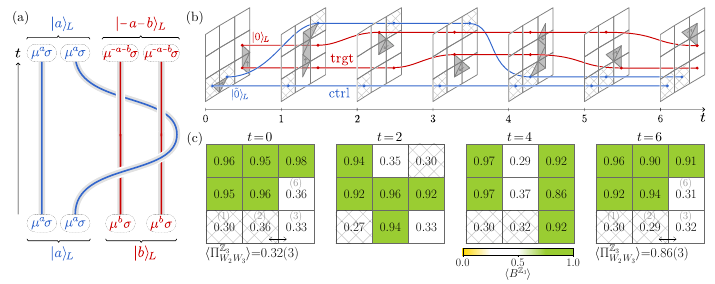}
	\caption{\textbf{Pull-through gate entangles topological qutrits.}
    (a) Braiding between the logical pairs (Eq.~\eqref{eq:absolute-logical}) gives an entangling gate (Eq.~\eqref{eq:pullthrough}) via the conjugation $(\mu^a \sigma) \mu^b \sigma (\mu^a \sigma)^{-1}$.
    (b) Spacetime diagram of the protocol on the $3\times3$ torus. Two pairs of $C_2$ flux anyons, control (blue worldlines) and target (red worldlines), are initialized in the logical state $\ket{\tilde{0}}_L\ket{0}_L$. One endpoint of the control qutrit is braided (pulled through) around both endpoints of the target and returned to its initial position at $t=6$. Oriented gray triangles indicate the ribbon operators implementing anyon motion. (c) Expectation values of the local projectors $B^{\mathbb{Z}_3}$ remain close to $1$ except on plaquettes hosting logical qutrit endpoints, where they are approximately $1/3$; the average (maximum) standard error is $0.028$ ($0.044$). The nonlocal correlator $\langle \Pi^{\mathbb{Z}_3}_{W_2 W_3}\rangle$ certifies $\mathcal{Z}$ correlations between the control and target qutrits. Its measured value is consistent with the predicted value of $1/3$ at $t=0$ and increases to $0.86(3)$ at $t=6$, in good agreement with the theoretical value of $1$ for the prepared Bell state.
	}\label{fig:pull-through}
\end{figure*}

\section{Universal topological gate set}
\subsection{Logical qutrit encoding}\label{subsec:logical-encoding}

To robustly store quantum information in nonlocal degrees of freedom, we encode it in the fusion space of spatially separated anyons. In particular, we use a pair of $C_2$ fluxes with trivial total flux (Fig.~\ref{fig:notation}b), which ensures that the encoded information is unaffected by transporting pairs across the system. Despite its flux neutrality, the pair retains delocalized charge degrees of freedom due to the fusion rule $C_2 \times C_2 = 1 + [2] + \cdots$. The associated fusion space already yields a two-dimensional logical subspace. Moreover, the internal structure of the [2] charge sector is itself protected: the two charge states cannot be distinguished or toggled by any local operation that does not enclose both fluxes. We thus see that the separation distance between anyons gives us a code distance. Provided the fluxes remain well separated, we obtain a protected three-dimensional fusion space, realizing a topological qutrit.

We can label this logical qutrit in terms of the fluxes of the constituent $C_2$ fluxes:
\begin{equation}\label{eq:absolute-logical}
\ket{j}_L = \ket{\mu^j \sigma, \mu^j \sigma}, \quad j = 0,1,2,
\end{equation}
defining our $\mathcal Z$-basis. Moreover, the nonlocal charges are diagonal in the logical $\mathcal X$-basis:
\begin{equation}
    \ket{\tilde j}_L= \frac{1}{\sqrt{3}}(\ket0_L+ \omega^j \ket1_L+ \bar\omega^{j}\ket2_L), \quad j=0,1,2
\end{equation}
where $\omega=e^{2\pi i/3}$. Indeed, it can be checked that the $\ket{\tilde0}_L$ flux pair is in the trivial anyon sector, whereas the $\ket{\tilde1}_L$ and $\ket{\tilde2}_L$ flux pairs behave as a (delocalized) $[2]$ charge.

In order for the flux labels in Eq.~\eqref{eq:absolute-logical} to be well-defined for spatially-separated $C_2$ anyons, we can consistently measure the global flux $W$ associated with a loop based at a fixed-but-arbitrary choice of origin (see Fig.~\ref{fig:notation}k for an example of a global flux). We call this the ``absolute'' encoding. While this is sufficient to define topological flux and demonstrate universal gate sets (Section~\ref{subsec:pull-through}-\ref{subsec:z-measurement}), the origin defines an Achilles' heel where quantum information can leak out via local charge measurement. This is avoided by a ``relative'' encoding, where instead of using a fixed origin, the flux values are relative to other pairs. This requires building a ``bureau of standards'' \cite{preskill_topo_notes}, which we demonstrate in Section~\ref{sec:bureau}. We note that this approach is fully topological, and if so desired it can be rephrased in terms of fusion trees (see Section~\ref{app:bureau-theory}).

In Ref.~\onlinecite{lo_universal_2025}, it is explained that three primitives (pull-through gate, $\mathcal{Z}$-basis measurement, and $\mathcal{X}$-basis measurement) constitute a universal gate set.
Here we demonstrate the experimental realization of these primitives.

\subsection{Pull-through gate}\label{subsec:pull-through}


We begin with the entangling primitive, implemented by braiding a constituent $C_2$ flux of one logical qutrit around the flux pair of another logical qutrit (see Fig.~\ref{fig:pull-through}a). This operation, known as the pull-through gate, exploits the non-Abelian braiding relations of the $S_3$ quantum double. In general, when fluxes braid around each other, they are each conjugated by their total flux. In this particular case, this leads to the logical two-qutrit gate
\begin{equation}
U\ket{a,b}_L  = \ket{a,-a-b}_L. \label{eq:pullthrough}
\end{equation}

We will demonstrate the action of this pull-through gate on $\ket{\tilde0}_L\ket{0}_L$, which creates a qutrit Bell state, 
\begin{equation}
\begin{split}
U\ket{\tilde{0}}_L\ket{0}_L &=\frac{1}{\sqrt{3}}\left(\ket{0,0}_L +\ket{1,2}_L+\ket{2,1}_L\right),
\end{split}
\end{equation}
where the logical information is entangled across the flux pairs (Fig.~\ref{fig:pull-through}b). 
Fig.~\ref{fig:pull-through}c shows the evolution of the $B^{\mathbb{Z}_3}$ projectors expectation values during the protocol (see Methods Section~\ref{method:pull-through} for protocol details and Fig.~\ref{fig:appendix_pull_through} for further data). The $B^{\mathbb{Z}_3}$ projectors identify the locations of logical qutrit endpoints, revealing local information but do not reveal information about the encoded logical state.

To verify the resulting entanglement between the two flux pairs, we certify correlations in the $\mathcal{Z}$-basis, quantified by $\langle W_{p_1}^{\mathbb{Z}_3} W_{p_2}^{\mathbb{Z}_3} \rangle$, where $W_{p_i}$ denotes the topological $W$-flux operator encircling plaquette $p_i$ (see Fig.~\ref{fig:w-paths} for the precise path). The observed increase of this correlator from its initial value at $t=0$ to near unity at $t=6$, as both qutrits return to their original lattice positions, provides strong evidence that the prepared state closely approximates the ideal target state. This braiding of non-Abelian fluxes is enabled by technical advancements in protocols to coherently move non-Abelian anyons while preserving logical information~\cite{lo_coherent} (see Section~\ref{app:coherent-moving} for theory overview).


\subsection{$\mathcal{X}$-basis measurement}\label{subsec:x-measurement}

\begin{figure}
	\centering\includegraphics[width=1\linewidth]{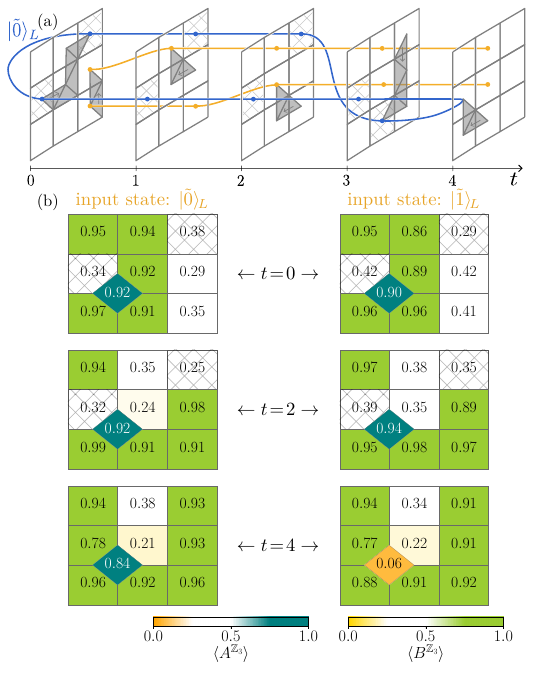}
	\caption{\textbf{$\mathcal{X}$-basis measurement.} (a) A symmetric ``measurement" flux pair (blue worldlines), created from the vacuum in the state $\ket{\tilde0}_{\! L}$, is braided around an endpoint of the input qutrit's flux pair (yellow worldlines) and then fused. The fusion outcome diagnoses the input qutrit state: annihilation to the vacuum corresponds to $\ket{\tilde 0}_{\! L}$, while a remnant charge anyon indicates $\ket{\tilde1}_{\! L}$. (b) Experimental verification at $t=0,2,4$. Left (right) column: input state $\ket{\tilde 0}_{\! L}$ ($\ket{\tilde 1}_{\! L}$). Local stabilizers remain insensitive to the logical content of the qutrits, with $B^{\mathbb Z_3}$ values near $1/3$ at all flux endpoints (average/maximum standard error $0.034/0.059$). At fusion $(t=5)$, the vertex projector $A^{\mathbb Z_3}$ yields $0.84(4)$ for $\ket{\tilde 0}_{\! L}$, consistent with vacuum annihilation, and $0.06(3)$ for $\ket{\tilde 1}_{\! L}$, indicating a remnant charge anyon. }
	\label{fig:x-basis-measure}
\end{figure}

Besides braiding logical anyons, fusion of anyons also implements logical operations, in addition to providing a read-out mechanism. The partial $\mathcal{X}$-basis measurement, which distinguishes between the charge-free $\ket{\tilde0}_L$ state and the charged $\ket{\tilde1}_L,\ket{\tilde2}_L$ subspace, 
is implemented by braiding and fusion of $C_2$ fluxes (Fig.~\ref{fig:x-basis-measure}a). This coherent detection of $\ket{\tilde 0}_L$ is sufficient for a universal gate set on the $\{ \ket{0}_L, \ket{1}_L$\} qubit \cite{lo_universal_2025}.

We demonstrate the $\mathcal{X}$-basis measurements on two input states: $\ket{\tilde 0}_L$ and $\ket{\tilde 1}_L$. The $\ket{\tilde 0}_L$ state is charge neutral, whereas the $\ket{\tilde 1}_L$ state stores a $[2]$ charge nonlocally (i.e. the flux pair fuses to a $[2]$ charge). Therefore, the way to measure in the $\mathcal{X}$-basis is by braiding with an ancilla $C_2$ flux pair and checking the fusion outcome: if the pair fuses to vacuum, then we have measured in the $\ket{\tilde 0}_L$ state, whereas if the pair fuses to a $[2]$ charge, then it is orthogonal to the $\ket{\tilde 0}_L$ state. In the latter case, one can repeat the braiding to remove the $[2]$ charge.

The projection violations for the input states $\ket{\tilde{0}}_L$ (left column) and $\ket{\tilde{1}}_L$ (right column) are shown in Fig.~\ref{fig:x-basis-measure}b (see Section~\ref{app:xmeasure-imple} for implementation details and Fig.~\ref{fig:appendix_x_basis_measurement} for further data). For the $\ket{\tilde{0}}_L$ input state, the vertex projector $A^{\mathbb{Z}_3}$ is not violated after fusion (maintaining a value close to $1$), demonstrating that we have measured the state to be in the $\ket{\tilde 0}_L$ state. For the $\ket{\tilde 1}_L$ input state, $A^{\mathbb{Z}_3}$ is violated after the fusion (with the stabilizer value close to 0), indicating the presence of a $[2]$ charge at the vertex. This demonstrates that the state is in the subspace spanned by the $\ket{\tilde 1}_L,\ket{\tilde 2}_L$ state.

\subsection{$\mathcal{Z}$-basis measurement}\label{subsec:z-measurement}

\begin{figure}[t]
	\centering \includegraphics[width=1\linewidth]{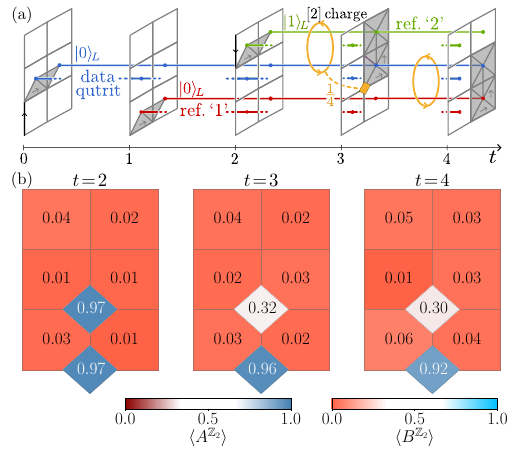}
	\caption{\textbf{$\mathcal{Z}$-basis measurement.} (a) Worldlines on the \(3\times2\) lattice for a data qutrit (blue) and two reference qutrits: reference 1 (red) in \(\ket{0}_L\) and reference 2 (green) in \(\ket{1}_L\). 	A \([2]\) charge pair (yellow) encircles the data and reference-2 qutrits between \(t=2\) and \(t=3\); because the logical states differ, the pair annihilates to the vacuum with probability \(1/4\) or leaves a remnant charge with probability \(3/4\). Between \(t=3\) and \(t=4\), a second \([2]\) charge pair encircles the data and reference-1 qutrits, both in \(\ket{0}_L\), resulting in deterministic vacuum annihilation. (b) Experimental demonstration. Local plaquette projectors \(B^{\mathbb Z_2}\) remain near zero at \(t=2,3,4\), confirming persistent qutrit endpoints (avg/max error \(0.014/0.021\)). At \(t=3\), the vertex projector \(A^{\mathbb Z_2}\) at the first fusion site yields \(0.32(4)\), consistent with the expected value \(1/4\); at \(t=4\), \(A^{\mathbb Z_2}\) at the second fusion site is \(0.92(2)\), indicating deterministic annihilation to the vacuum.
    }\label{fig:z-basis-measure}
\end{figure}

As the final primitive that completes the universal gate set, the $\mathcal{Z}$-basis measurement is a comparison measurement: the measurement outcome indicates whether the data state is the same as the reference state or orthogonal to it. At the physical level, the $\mathcal{Z}$-basis measurement determines the flux content of the logical flux pair by braiding and fusion of a $[2]$ charge pair around one flux each from the data and the reference pair (Fig.~\ref{fig:z-basis-measure}a). If the data and reference flux pairs are in the same internal state, then the $[2]$ charge pair winds around a trivial flux and fuses back to vacuum. If the data and reference flux pairs are in different internal states, then the total flux is a $C_3$ group element, which results in non-trivial braiding: there is a $3/4$ probability of obtaining a remnant $[-]$ charge upon fusion \footnote{We can wind fresh $[2]$ charge pairs $n$ times to decrease the probability of a measurement error to $(1/4)^n$.}.

We first instantiate the data logical state (here $\ket{0}_L$) in the middle row and the reference logical states $\ket{0}_L$ and $\ket{1}_L$ in the bottom and top row (Fig.~\ref{fig:z-basis-measure}a). 
We then braid $[2]$ charge pairs to do comparison measurement of the data flux pair with the reference pairs; the projection violations are shown in Fig.~\ref{fig:z-basis-measure}b (see Section~\ref{app:zmeasure-imple} for implementation details and Fig.~\ref{fig:appendix_z_basis_measurement} for further data). Comparing the data state with the $\ket{1}_L$ state, there is a violation in the vertex projector $A^{\mathbb{Z}_2}$ where the $[2]$ charge pair is fused at $t=3$ (Fig.~\ref{fig:z-basis-measure}b), indicating that the data qutrit is not in the $\ket{1}_L$ state. Finally, comparing the data state and the $\ket{0}_L$ state, there is no additional violation in the $A^{\mathbb{Z}_2}$ vertex projector (at $t=4$ in Fig.~\ref{fig:z-basis-measure}b), supporting that the data flux pair is indeed in the $\ket{0}_L$ state.  
This comparison measurement between logical states also allows us to transition from the absolute encoding (Eq.~\ref{eq:absolute-logical}) to the relative encoding.

\subsection{Bureau of standards}\label{sec:bureau}

\begin{figure}[t]
		\centering    \includegraphics[width=1\linewidth]{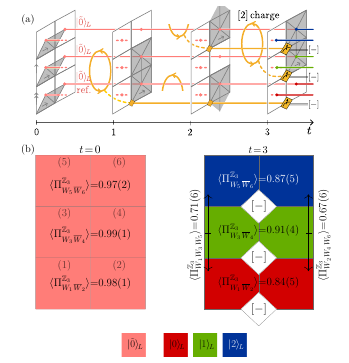}
		\caption{\textbf{Bureau of standards.} (a) Schematic of the protocol used to define a logical qutrit in the $\mathcal{Z}$-basis via relative flux comparisons. Three $C_2$ flux pairs are initialized in the $\ket{\tilde{0}}_{\! L}$ state, with the lowest pair chosen as the reference $\ket{0}_{\! L}$. The logical states of the remaining pairs are determined through three comparison measurements implemented by braiding a $[2]$ charge pair. (b) Experimental verification of the prepared states. At $t=1$, the $\Pi^{\ZZ_3}_{W_{p1} W_{p2}}$ projectors take values close to unity, demonstrating that the prepared states closely approximate the ideally prepared (i.e.\ noiseless) $\ket{\tilde{0}}_L$ pairs. At $t=3$, strong intra-pair and inter-pair $\mathcal{Z}$ correlations are confirmed by measurements of $\Pi^{\ZZ_3}_{W_{p1} W_{p2}}$ and $\Pi^{\ZZ_3}_{W_{p1} W_{p2} W_{p3}}$, respectively.
		}\label{fig:bureau}
\end{figure}

To define the relative encoding, one requires a bureau of standards, as mentioned in Section~\ref{subsec:logical-encoding}. Here we demonstrate the preparation of states in the $\mathcal{Z}$-basis bureau of standards (the procedure is shown schematically in Fig.~\ref{fig:bureau}a). Three charge-neutral $\ket{\tilde0}_L$ states are instantiated by being pair-created from the vacuum. Without loss of generality, we designate the lowest $C_2$ flux pair as the reference logical $\ket{0}_L$ state, and define all other logical states relative to this reference. Then, $\mathcal{Z}$-basis comparison measurements are performed among the three possible pairings. By post-selecting on the $[-]$ fusion channel via mid-circuit measurements, indicating each of the three pairs occupies distinct $\mathcal{Z}$-basis states, the reference $\ket{1}_L$ and $\ket{2}_L$ state are designated. This completes the preparation of the $\mathcal{Z}$-basis bureau of standards, certified by the inter-pair 3-body correlation $\left<\Pi^{\mathbb{Z}_3}_{W_{p_1} W_{p_2}W_{p_3}}\right>$ (see Fig.~\ref{fig:bureau}b).

In the experiment, for simplicity, we have post-selected for the $[-]$ charge outcome for all three $\mathcal{Z}$-basis measurements to obtain all three logical states. But post-selection is not necessary. If the measurement outcome is not $[-]$ but the vacuum $1$, that implies we have partially prepared the bureau of standards (we documented the other possible outcomes in Fig.~\ref{fig:appendix_z_bureau_of_standards}). We can repeat until success: create fresh $C_2$ flux pair in the $\ket{\tilde0}_L$ state and compare until we obtain the $[-]$ fusion outcome to prepare the full $\mathcal{Z}$-basis reference states. 

\section{Demonstrating magic}\label{sec:magic}
\begin{figure}[t]
	\centering\includegraphics[width=1\linewidth]{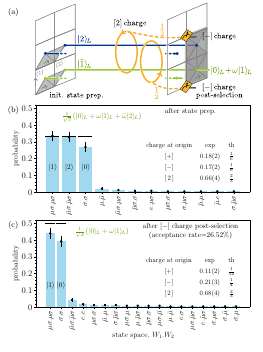}
	\caption{\textbf{Illustrating universality by creating a topological magic state.} (a) Spacetime diagram illustrating the procedure for preparing a magic state. The protocol begins by initializing the lowest two plaquettes, labeled (1) and (2), with the input state $\ket{\tilde 1}_L = \frac{1}{\sqrt{3}}(\ket{0}_L + \omega\ket{1}_L + \bar{\omega}\ket{2}_L)$ and the two middle plaquettes with the reference state $\ket{2}_L$. A $\mathcal{Z}$-basis comparison measurement with respect to $\ket{2}_L$ is performed using $[2]$ charges. Post-selection on the $[-]$ fusion outcome (occurring with probability 0.5) projects onto the subspace orthogonal to $\ket{2}_L$, yielding the magic state $\frac{1}{\sqrt{2}}(\ket{0}_L + \omega \ket{1}_L)$. Two successive $[2]$-charge comparison measurements are required to ensure coherence of the projection (see Methods Section~\ref{method:magic-state-coherence}). 
    (b) The bar plots display the joint probability of measuring the $W$-fluxes around plaquettes (1) and (2). Following initial state preparation, the three possible flux configurations, $\sigma.\sigma$, $\mus.\mus$, and $\mubars.\mubars$, which correspond to the logical basis states, occur with approximately equal probability of $1/3$. Successful preparation of the magic state is validated by a significant increase in the probability of the $\sigma.\sigma$ and $\mus.\mus$ flux configurations, accompanied by a suppression of the $\mubars.\mubars$ configuration.
	}\label{fig:magic}
\end{figure}

To demonstrate the power of the $S_3$ universal gate set we have constructed, we exhibit a simple procedure to create a magic state using topological operations. Starting with the basis state $\ket{\tilde 1}_L \propto \ket{0}_L + \omega \ket{1}_L + \bar \omega \ket{2}_L$, we use the $\mathcal{Z}$-basis measurement primitive to find this in `not $\ket{2}_L$', collapsing the state into $\ket{0}_L+\omega\ket{1}_L$ (see Fig.~\ref{fig:magic} for the protocol). This gives a minimal demonstration of how measurement on qutrits enables the creation of non-Clifford states---in this case using only topological operations.

To certify the magic state, we measure the state in both $\mathcal{Z}$- and $\mathcal{X}$-basis. First, in the $\mathcal{Z}$-basis, the resulting state indeed has equal weights only on the $\ket0_L,\ket1_L$ basis states as expected (Fig.~\ref{fig:magic}c). For the $\mathcal{X}$-basis, we use the Achilles' heel of the absolute encoding to our advantage: by interrogating the logical charge violations at the origin, we can infer the $\mathcal{X}$-basis measurement outcome (see Methods Section~\ref{method:charge-origin} and Table~\ref{table:magic-state} for theory prediction). The charge measurement at the origin provides a measure of the coherence of the magic state we create and is sensitive to the relative phase $\omega$. In the inset of Fig.~\ref{fig:magic}c, the experimental value of the charge violation weight at the origin matches with the theory prediction for when the state is $\ket{0}_L+\omega\ket{1}_L$, thus confirming the coherence of the magic state created.

\section{Conclusion}

In this work, we demonstrate a purely topological approach to universal quantum computation on quantum hardware. Rather than encoding information in a ground-state subspace, as in conventional topological codes, we encode logical states directly in the nonlocal fusion space of non-Abelian anyons, and we use their braiding and fusion as computational primitives. These capabilities are enabled by the realization of a deconfined $S_3$ gauge theory, the non-Abelian topological order based on the smallest gauge group. Exploiting its cyclic anyon fusion rules, we realize a universal logical gate set, demonstrate the topological preparation of a magic state, and trap a single non-Abelian anyon on the torus. Together, these results establish non-Abelian anyons as programmable and computationally powerful degrees of freedom, while the solvability of $S_3$ supports scalable preparation and control.

The realization of these capabilities reflects a convergence of advances in quantum hardware and theoretical understanding of non-Abelian topological phases. Improvements in qubit number and coherence now enable the preparation and manipulation of complex topological states, while recent theoretical progress in lattice-scale anyon creation~\cite{bravyi2022adaptiveconstantdepthcircuitsmanipulating, ren_efficient_2025,lyons-2025}, encoding~\cite{lo_universal_2025}, and coherent moving~\cite{lo_coherent} establish a setting in which non-Abelian anyons can be controlled rather than merely detected.

Looking ahead, encoding logical information directly in non-Abelian anyonic excitations opens new directions for fault-tolerant quantum computation. Due to their close similarities with the surface code, non-Abelian quantum double models offer a promising route to extending surface code techniques while enabling new capabilities including universal topological gate sets. Scaling system sizes will allow for studies of logical error mechanisms and decoding strategies \cite{wootton2014,Wootton_improved_HDRG, BurtonIsing, BurtonFibonacci, Verstraete, Wootton2016proof,Dauphinais2017, HarringtonFibonacci,Chirame, davydova_universal_2025,jing2025intrinsicheraldingoptimaldecoders}, as well as systematic comparisons between ground-state-based approaches and fusion-space encodings \cite{mochon_smaller_groups_2004, SU24, kitaev_finite_group_2007, measurementonly,cui_universal_2015,Cui2015_metaplectic,lo_universal_2025,chen_universal_2025,byles_demonstration_2025}. 

More generally, our work highlights how finite-group non-Abelian topological orders offer a compelling balance between computational power and experimental accessibility. While more exotic phases may admit braid-only universality, solvable gauge theories such as $S_3$ occupy a ``sweet spot'': they are sufficiently rich to support universal computation, yet sufficiently structured to admit efficient preparation and coherent anyon transport. $S_3$ quantum double provides a versatile platform for studying both scalable quantum computation beyond the stabilizer paradigm and new regimes of real-time dynamics of lattice gauge theories.

\section*{Data availability}
The data generated in this study have been deposited in the Zenodo repository 10.5281/zenodo.18054264~\cite{supporting_s3_2025} database under open access.

\section*{Code availability}
The code used for numerical simulations is available from Zenodo repository 10.5281/zenodo.18054264~\cite{supporting_s3_2025}.

\section*{Author Contributions}
C.F.B.L. and M.I. wrote the code generating the circuits for all experiments. The experiment was built and carried out by D.G., M.M., P.E.S., and M.D.U. The data analysis and interpretation was done by C.F.B.L., M.I., A.L., N.T., H.D., A.V., and R.V.. C.F.B.L., A.L., N.T., A.V. and R.V. contributed to the ideation, theory and experiment design, including the definition of the operators for creation, coherent moving, and fusion of anyons; in particular, C.F.B.L., A.L., and R.V. contributed to the theory for bureau of standards for the relative logical encoding. C.F.B.L. drafted the initial paper, which was refined by contributions from all authors, especially M.I., A.L., N.T., A.V., and R.V.. C.F.B.L., M.I. and A.L. contributed to the Supplementary Materials; in particular, C.F.B.L. contributed to the protocol implementations, M.I. contributed to state preparation protocols and circuit compilation techniques, and A.L. contributed to the bureau of standards theory.

\section*{Acknowledgments}
We thank the broader team at Quantinuum for comments. R.V. thanks Jason Alicea for discussions on the manuscript. C.F.B.L. and A.L. acknowledge support from the National Science Foundation Graduate Research Fellowship Program (NSF GRFP). C.F.B.L., A.L., and A.V. are supported by the Simons Collaboration on Ultra Quantum Matter, which is a grant from the Simons Foundation (618615). This work is in part supported by the DARPA MeasQuIT program (R.V. and A.V.). All quantum data in this work were produced on the Quantinuum System Model H2 (specifically H2-1) quantum computer, powered by Honeywell, between December 2024 and December 2025.

\section*{Competing interests} H.D. is a shareholder of Quantinuum. All other authors declare no competing interests. 

\section*{Additional information}
Correspondence and requests for materials should be addressed to C.F.B.L. at chiufanbowenlo@g.harvard.edu, R.V. at verresen@uchicago.edu, or M.I. at mohsin.iqbal@quantinuum.com.
\clearpage

\section{Methods}

\subsection{Ground state preparation}\label{method:gs-prep}
We prepare the ground state with trivial flux along both noncontractible loops of the torus. At the end of the preparation circuit, we non-destructively measure an implicitly prepared $S_A^{\ZZ_2}$ stabilizer together with an adjacent $S_B^{\ZZ_2}$ stabilizer. Empirically, we find that, in the event of a state-preparation error, these stabilizers are more likely to be violated~\cite{iqbal_qutrit_2025}. We therefore discard any experimental shot in which either of the two non-destructive measurement outcomes is $-1$. Furthermore, we exploit the qutrit encoding---implemented using two physical qubits per qutrit---to detect errors that take qutrits outside the computational subspace; this serves as an independent heralding condition (see Section~\ref{app:s3_encoding_and_circs}).
\\

Before destructive readout of the edge qudits, a circuit barrier is inserted to ensure complete creation of the ground state. Expectation values of stabilizer and logical projectors are obtained from destructive measurements of each qubit-qutrit pair in the following bases:
\begin{itemize}
	\item $Z$ and $\mathcal{Z}$, yielding $B^{\ZZ_2}$, $B^{\ZZ_3}$, and horizontal and vertical logical $Z$ projectors;
	\item $Z$ and $\mathcal{X}$, yielding $A^{\ZZ_3}$ and $B^{\ZZ_2}$ projectors;
	\item $X$ and $\mathcal{C}$, yielding $A^{\ZZ_2}$ projectors.
\end{itemize}

Mean values of logical $Z$ projectors of the $S_3$ quantum double are obtained by averaging across rows for the horizontal logical operators $Z_{\mathrm{hori}}^{\ZZ_2}$ and $Z_{\mathrm{hori}}^{\ZZ_3}$, and across columns for the vertical logical operators $Z_{\mathrm{vert}}^{\ZZ_2}$ and $Z_{\mathrm{vert}}^{\ZZ_3}$. After applying all heralding criteria, approximately $24\%$ of shots are discarded. For the smaller $3\times2$ lattice, shown in Fig.~\ref{fig:appendix_gs_3x2}, the corresponding discard rate is $\sim13\%$.
\\

For comparison, we also prepare the ground state using an implementation with mid-circuit measurements and feedforward (see Section~\ref{app:adaptive-prep}), which yields slightly lower fidelities (Fig.~\ref{fig:appendix_gs_measurement_based}).




\subsection{Pull-through gate}\label{method:pull-through}

First, we show that the action of flux braiding (conjugation by the total flux) indeed reproduces the qutrit action of the pull-through gate as specified in the main text. We start with the state $\ket{\tilde0}_L\ket{0}_L=(\ket{\sigma,\sigma} + \ket{\mu \sigma,\mu \sigma}  +\ket{\bar{\mu}\sigma,\bar{\mu}\sigma} )\ket{\sigma, \sigma}$. At the level of the flux, the fluxes are conjugated by the total flux participating in the braiding:

\begin{equation}
	\begin{split}
		&U(\ket{\sigma,\sigma} + \ket{\mu \sigma,\mu \sigma}  +\ket{\bar{\mu}\sigma,\bar{\mu}\sigma} ) \ket{\sigma,\sigma} \\
		&= \ket{\sigma,\sigma}\ket{\sigma ,\sigma}  + \ket{\mu \sigma,\mu \sigma} \ket{\bar{\mu}\sigma,\bar{\mu}\sigma}  + \ket{\bar{\mu}\sigma,\bar{\mu}\sigma} \ket{\mu \sigma,\mu \sigma} \\
		&= \ket{0}_L\ket{0}_L +\ket{1}_L\ket{2}_L+\ket{2}_L\ket{1}_L.
	\end{split}
\end{equation}

In implementing the pull-through gate, as well as all other experiments presented in this work, logical qutrit states are prepared in the $\mathcal{X}$-basis (i.e., $\ket{\tilde{\bullet}}_L$) using an ancilla qutrit (see Section~\ref{app:pullthrough-imple}). Given the constrained budget of available qutrits (56 in total), ancilla qutrits are obtained by employing a qubit-reuse technique, in which physical qubits that are no longer acted on in subsequent stages of the circuit are measured early and reset for use as ancillas.

After successful preparation, the ancilla qutrit is disentangled and ideally returns to the qutrit zero state. Measurement outcomes corresponding to a nonzero ancilla state---predominantly arising from memory errors in deeper circuits (cf. Table~\ref{table_gate_counts})---are therefore used to herald unsuccessful logical qutrit preparation, and the corresponding shots are discarded.

The correlator $\langle W_{p_1}^{\mathbb{Z}_3} W_{p_2}^{\mathbb{Z}_3} \rangle$, which we use to quantify correlations, ensures that
\begin{itemize}
	\item when the control global flux is $\sigma$, the target is $\sigma$;
	\item when the control global flux is $\mu\sigma$, the target is $\bar{\mu}\sigma$;
	\item when the control global flux is $\bar{\mu}\sigma$, the target is $\mu\sigma$.
\end{itemize}
See Fig.~\ref{fig:appendix_pull_through} for further data on nontrivial $\mathcal{Z}$ correlators.

\subsection{Magic state creation}

\subsubsection{Coherent projection for magic state creation}\label{method:magic-state-coherence}

We elaborate on why two pairs of $[2]$ charges is needed for coherent projection in the procedure for creating the magic state $\ket{0}_L+\omega \ket{1}_L$.

After braiding the first $[2]$ charge pair around the end of both flux pairs, we post-select on the $[-]$ fusion outcome. This implies that the measured state is orthogonal to the $\ket2_L$ state, so the state is projected to the orthogonal subspace. To ensure the correct coherence of the state in the subspace spanned by $\ket{0}_L$ and $\ket{1}_L$, we need to make sure that there is no net charge extracted from the flux pair constituting the logical state. With only one round of $[2]$ charge braiding, there is one local $[-]$ charge in the system, and by neutrality, the flux pair is storing another $[-]$ charge nonlocally. Therefore, we repeat the $\mathcal{Z}$ comparison measurement and wind another $[2]$ charge pair and again post-select on the $[-]$ fusion outcome, such that there are two local $[-]$ charges in the system; this ensures that the charge content of the resulting state is correct such that we obtain the state $\ket0_L+\omega\ket1_L$ (bottom panel of Fig.~\ref{fig:magic}). 

\subsubsection{Charge measurement at origin}\label{method:charge-origin}

Recall from Section~\ref{subsec:x-measurement}, the $\mathcal{X}$-basis measurement determines whether the state is $\ket{\tilde0}_L$ or in the orthogonal subspace by measuring the $[2]$ charge of the flux pair. In the absolute logical state, the origin contains the charge information of the logical state. This is due to the neutrality condition, the charges at the origin are exactly the partners of the charges shared by the flux pairs, such that fusing all anyons will return to vacuum. Therefore, by interrogating the logical charge violations at the origin, we can infer the $\mathcal{X}$-basis measurement outcome.

The prediction for the expected charge violation weights for the $\ket{0}_L+\omega\ket{1}_L$ magic state and other possible states are shown in Table~\ref{table:magic-state}.

\begin{table}[h]
\centering
\setlength{\tabcolsep}{8pt}
\begin{tabular}{c c c c c}
\toprule
\textbf{State} & $1$ & $[-]$ & $[2]$\\
\midrule
$\ket{\psi_\omega}=\frac{1}{\sqrt2}(\ket{0}_L+\omega\ket{1}_L)$  & $\frac{1}{12}$ & $\frac{1}{4}$ & $\frac{2}{3}$\\
\midrule
$\ket{\psi_+}=\frac{1}{\sqrt2}(\ket{0}_L+\ket{1}_L)$ &$\frac{1}{3}$ & $0$ & $\frac{2}{3}$\\
\midrule
$\rho=p\ketbra{0}{0}_L+(1-p)\ketbra{1}{1}_L$ & $\frac{1}{6}$ & $\frac{1}{6}$ & $\frac{2}{3}$\\
\bottomrule
\end{tabular}
\caption{Theory prediction for the weights of charge violations at the origin for the magic state $\ket{0}_L+\omega\ket{1}_L$ and other possible states. The state without a relative phase ($\ket{\psi_+}$) and the completely decohered state ($\rho$) have distinct weights of charge violations. \label{table:magic-state}}
\end{table}

\bibliography{references}
\clearpage
\onecolumngrid

\begin{center}
	\textbf{\large Supplemental Material}
\end{center}
\setcounter{section}{0}
\renewcommand{\thesection}{S\arabic{section}} 
\renewcommand{\theequation}{S\arabic{equation}} 
\setcounter{equation}{0}
\renewcommand{\thefigure}{S\arabic{figure}} 
\renewcommand{\theHfigure}{S\arabic{figure}}
\setcounter{figure}{0}
\renewcommand{\thetable}{S\arabic{table}} 

\section{Basics of $S_3$ quantum double model}

For more details, we refer the reader to the comprehensive pedagogical notes in Refs.~\onlinecite{cui_lecture_notes,lo_universal_2025}.

\subsection{Review of $S_3$ group theory and representation theory}\label{append-sub:S3-review}

We use the group presentation for $S_3=\left<\mu,\sigma \vert \mu^3=\sigma^2=e, \mu\sigma = \sigma \mu^{-1} \right>$, where $\mu$ and $\sigma$ corresponds to the $120^\circ$ rotation and reflection of an equilateral triangle. Another common notation is to recognize that $S_3$ is a symmetric group on three elements, so group elements are permutations that can be represented in terms of cycles. The identification is $\mu=(123)$, $\sigma = (23)$ (the remaining group elements can be determined from these two generators). 

There are three conjugacy classes that partitions $S_3$: $C_1=\set{e}$, $C_2=\set{\sigma,\mu\sigma,\bar\mu\sigma}$, $C_3=\set{\mu,\bar\mu}$.

There are three irreducible representations (irreps) of $S_3$: $[+],[-],[2]$. The trivial representation $[+]$ sends every group element to $1$. The sign representation $[-]$ sends group element $g \mapsto \text{sgn}(g)$, where $\text{sgn}$ gives the sign of the permutation; so even permutations (e.g. $C_3$ group elements) are mapped to $+1$, whereas odd permutations (e.g. $C_2$ group elements) are mapped to $-1$. The standard representation $[2]$ is a two-dimensional representation, and the representation matrices are:
\begin{align*}
	\Gamma_{[2]}(e)=\begin{pmatrix}
		1 & 0\\
		0 & 1\\
	\end{pmatrix}\quad\quad 
	\Gamma_{[2]}(\mu)&=\begin{pmatrix}
		\omega & 0\\
		0 & \bar\omega\\
	\end{pmatrix}\quad\quad
	\Gamma_{[2]}(\bar\mu)=\begin{pmatrix}
		\bar\omega & 0\\
		0 & \omega\\
	\end{pmatrix}\quad\quad \\
	\Gamma_{[2]}(\mu \sigma)=\begin{pmatrix}
		0 & \omega\\
		\bar\omega & 0\\
	\end{pmatrix}\quad\quad
	\Gamma_{[2]}(\bar\mu \sigma)&=\begin{pmatrix}
		0 & \bar\omega\\
		\omega & 0\\
	\end{pmatrix}\quad\quad
	\Gamma_{[2]}(\sigma)=\begin{pmatrix}
		0 & 1\\
		1 & 0\\
	\end{pmatrix}\quad\quad
\end{align*}
where the indexing order is $\{\ket{2+},\ket{2-}\}$ and $\omega = \exp(2 \pi i/3)$.

The representation theory of the $\mathbb{Z}_3$ subgroup of $S_3$ also warrants a mention, as it will be important for defining some of the anyon types in $S_3$ quantum double. Since $\mathbb{Z}_3$ is an Abelian group, all irreps are one-dimensional. The three possible irreps are labeled by the third root of unity. The trivial representation $[1]$ maps all group elements to $1$. The $[\omega]$ representation is given by the map $a \mapsto \omega^a$. The $[\bar{\omega}]$ representation is given by the map $a \mapsto (\bar{\omega})^{a}$. 

\subsection{Qutrit-qubit encoding}\label{app:qubit-qutrit-encoding}

We use the qutrit-qubit encoding for the $S_3$ group elements, such that for a given $S_3$ group element basis $\ket{g} = \ket{\mu^a\sigma^b}$, it is encoded as a qutrit $\ket{a \text{ (mod 3)}}$ and a qubit $\ket{b \text{ (mod 2)}}$, where $a,b$ are integers.

The left- and right-group multiplication operators are
\begin{equation}
    L^g\ket{h} = \ket{gh}, \quad \quad \quad R^g\ket{h} = \ket{h\bar{g}},
\end{equation}
where $\bar g$ is the inverse of $g$.
In terms of qutrit and qubit gates, the left- and right- group multiplication of $S_3$ generators are
\begin{align}
    L^\mu &= \mathcal{X}\otimes I\;,\; \quad  L^\sigma = \mathcal{C}\otimes X \;,\;\label{eq:q2q3_enc_lr_multiplication_1}\\
    R^\mu &= \mathcal{X}^{-Z}\;,\; \quad  R^\sigma = I\otimes X\;,\;\label{eq:q2q3_enc_lr_multiplication_2}
\end{align}
where $\mathcal{X}$ and $\mathcal{Z}$ denote the qutrit clock matrices, and $\mathcal{C}$ is the qutrit charge-conjugation operator,
\begin{equation}\label{eq:x_and_z_and_C_def}
    \mathcal{Z} = \begin{pmatrix}
    1 & & \\
    & \omega & \\
    & & \omega^2
    \end{pmatrix}, \quad 
    \mathcal{X} = \begin{pmatrix}
    &  & 1\\
    1 & & \\
    & 1 & \end{pmatrix}, \quad
    \mathcal{C} = \begin{pmatrix}
		1 & & \\
		&  & 1\\
		& 1 & 
	\end{pmatrix}\;,
\end{equation}
and where $Z$ and $X$ are the Pauli matrices acting on the qubit degrees of freedom.
\\

Given an $S_3$ group element $g=\mu^a\sigma^b$, its inverse is $\bar{g} = \mu^{-(-1)^ba}\sigma^b$. This inverse operation is implemented by the following circuit:
\begin{equation}\label{eq:inverse-circ}
\begin{tikzpicture}
\begin{yquant}[horizontal]

[violet] qubit {$\ket{a}$} a[1];
[violet] qubit {$\ket{b}$} b[1];

box {$\mathcal{C}$} a[0] | b[0];
box {$\mathcal{C}$} a[0];

\end{yquant}
\end{tikzpicture}
\end{equation}
Note that taking an inverse of a $S_3$ group element does not change the qubit value $b$ since inverse operation does not change its conjugacy class. 

Controlled left multiplication of a group element $h=\mu^{\textcolor{teal}{\alpha}}\sigma^{\textcolor{teal}{\beta}}$ (in \textcolor{teal}{teal}) by the group element $g=\mu^{\textcolor{violet}{a}} \sigma^{\textcolor{violet}{b}}$ (in \textcolor{violet}{violet}) (left panel) and its inverse $\bar g$ (right panel) is 

\begin{equation}\label{eq:left-multi-circ}
\begin{tikzpicture}
\draw (-4.19,-0.8) node [anchor=north west][inner sep=0.75pt]  [font=\normalsize]  {$\ket{g}\ket{h}\to \ket{g}\ket{gh}$:};
\begin{yquant}[horizontal]

[violet] qubit {$\ket{a}$} a[1];
[violet] qubit {$\ket{b}$} b[1];

[teal] qubit {$\ket{\alpha}$} a[+1];
[teal] qubit {$\ket{\beta}$} b[+1];

cnot b[1] | b[0];
box {$\mathcal{C}$} a[1] | b[0];
box {$\mathcal{X}$} a[1] | a[0];

\end{yquant}
\end{tikzpicture}
\quad \quad \quad \quad \quad \quad \quad
\begin{tikzpicture}
\draw (-4.19,-0.8) node [anchor=north west][inner sep=0.75pt]  [font=\normalsize]  {$\ket{g}\ket{h}\to \ket{g}\ket{\bar{g}h}$:};
\begin{yquant}[horizontal]

[violet] qubit {$\ket{a}$} a[1];
[violet] qubit {$\ket{b}$} b[1];

[teal] qubit {$\ket{\alpha}$} a[+1];
[teal] qubit {$\ket{\beta}$} b[+1];

cnot b[1] | b[0];
box {$\mathcal{X}^\dagger$} a[1] | a[0];
box {$\mathcal{C}$} a[1] | b[0];

\end{yquant}
\end{tikzpicture}
\end{equation}

Similarly, controlled right multiplication by a group element $g$ (left panel) and its inverse $\bar g$ (right panel) is
\begin{equation}\label{eq:right-multi-circ}
\begin{tikzpicture}
\draw (-4.19,-0.8) node [anchor=north west][inner sep=0.75pt]  [font=\normalsize]  {$\ket{g}\ket{h}\to \ket{g}\ket{hg}$:};

\begin{yquant}[horizontal]

[violet] qubit {$\ket{a}$} a[1];
[violet] qubit {$\ket{b}$} b[1];

[teal] qubit {$\ket{\alpha}$} a[+1];
[teal] qubit {$\ket{\beta}$} b[+1];

box {$\mathcal{C}$} a[1] | b[1];
box {$\mathcal{X}$} a[1] | a[0];
box {$\mathcal{C}$} a[1] | b[1];

cnot b[1] | b[0];

\end{yquant}
\end{tikzpicture}
\quad\quad\quad\quad\quad\quad 
\begin{tikzpicture}
\draw (-4.19,-0.8) node [anchor=north west][inner sep=0.75pt]  [font=\normalsize]  {$\ket{g}\ket{h}\to \ket{g}\ket{h\bar{g}}$:};

\begin{yquant}[horizontal]

[violet] qubit {$\ket{a}$} a[1];
[violet] qubit {$\ket{b}$} b[1];

[teal] qubit {$\ket{\alpha}$} a[+1];
[teal] qubit {$\ket{\beta}$} b[+1];

cnot b[1] | b[0];

box {$\mathcal{C}$} a[1] | b[1];
box {$\mathcal{X}^\dagger$} a[1] | a[0];
box {$\mathcal{C}$} a[1] | b[1];

\end{yquant}
\end{tikzpicture}
\end{equation}

Finally, controlled conjugation sends $g \to \bar{z}gz$, where $z=\mu^a\sigma^b$ is the control (in \textcolor{violet}{violet}) and $g=\mu^\alpha\sigma^\beta$ is the target (in \textcolor{teal}{teal}). So the circuit is
\begin{equation}\label{eq:conjugation-circ}
\begin{tikzpicture}
\begin{yquant}[horizontal]

[violet] qubit {$\ket{a}$} a[1];
[violet] qubit {$\ket{b}$} b[1];

[teal] qubit {$\ket{\alpha}$} a[+1];
[teal] qubit {$\ket{\beta}$} b[+1];

box {$\mathcal{C}$} a[1] | b[1];
box {$\mathcal{X}$} a[1] | a[0];
box {$\mathcal{C}$} a[1] | b[1];

box {$\mathcal{X}^\dagger$} a[1] | a[0];
box {$\mathcal{C}$} a[1] | b[0];

\end{yquant}
\end{tikzpicture}
\end{equation}

\subsection{Anyon types, fusion rules, and braiding}\label{append-sub:anyon-type}

Given a finite group $G$, the quantum double of $G$ (also known as the $G$ Drinfeld double, denoted as $\mathcal{D}(G)$)~\cite{kitaev2003fault}, has  anyon type specified by $(C,\chi)$, where $C$ is a conjugacy class and $\chi$ is an irreducible representation of the centralizer $Z(r)$ for a representative $r$ in $C$. The 8 possible anyon types in $S_3$ quantum double are given below:

\begin{table}[h]
\centering
\setlength{\tabcolsep}{6pt}
\begin{tabular}{c c c c c}
\toprule
\textbf{Conjugacy class} & \textbf{Centralizer} & \textbf{Charge} & \(d\)\\
\midrule
\multirow{3}{*}{\(C_{1}\)} &
\multirow{3}{*}{\(Z(e)=S_{3}\)} &
\([+]\) & 1\\
& & \([-]\) & 1\\
& & \([2]\) & 2\\
\midrule
\multirow{2}{*}{\(C_{2}\)} &
\multirow{2}{*}{\(Z(\sigma)\cong \mathbb{Z}_{2}\)} &
\([+]\) & 3\\
& & \([-]\) & 3\\
\midrule
\multirow{3}{*}{\(C_{3}\)} &
\multirow{3}{*}{\(Z(\mu)\cong \mathbb{Z}_{3}\)} &
\([1]\) & 2\\
& & \([\omega]\) & 2\\
& & \([\bar\omega]\) & 2\\
\bottomrule
\end{tabular}
\caption{Conjugacy classes, centralizers, charges, and the quantum dimensions ($d$) of the 8 anyon types in $\mathcal{D}(S_3)$.}
\end{table}

We denote the vacuum anyon $(C_1,[+])$ by $1$ in the main text. In the following, we also simply refer to it as $[+]$.

The most relevant fusion rules are as follows:
\begin{equation}
    [-] \times [-] = [+], \quad [2] \times [2] = [+] + [-] + [2]
\end{equation}
\begin{equation}
        C_2 \times C_2 = [+] + [2] + C_3 + (C_3, [\omega]) + (C_3, [\bar{\omega}]), \quad C_2 \times [2] = C_2 + (C_2, [-])
\end{equation}
\begin{equation}
    C_3 \times C_3 = [+] + [-] + C_3, \quad C_3 \times C_2 = C_2 + (C_2, [-])
\end{equation}

The full categorical data describing the $S_3$ quantum double, including fusion rules, the $F$ symbols, and the $R$ matrices, can be found in Ref.~\onlinecite{cui_universal_2015} and in the Supplementary Categorical Data of Ref.~\onlinecite{chen_universal_2025}.

For braiding, it suffices to describe two scenarios: 1)  braiding two fluxes and 2) braiding a flux around a charge. Braiding involving dyons can be decomposed into aforementioned scenarios. For scenario 1, when two fluxes are braided around each other, the global fluxes labels are both conjugated by the total flux. For scenario 2, the action of a flux $g$ on a charge's internal Hilbert space is given by the representation matrix $\Gamma(g)$, where $\Gamma$ is the irreducible representation of the charge.

\subsection{Ribbon operators}\label{app:ribbon-operators}
\subsubsection{Basic construction}
The operators that create pairs of non-Abelian anyons on top of the quantum double ground state are known as \emph{ribbon operators.} A ribbon operator $F^{A;u,u'}_{\rho}$ is specified by four data: (i) the path along the lattice where the ribbon acts, $\rho$, (ii) the anyon type $A=(C,\chi)$, where $C$ is the conjugacy class and $\chi$ is an irreducible representation, (iii) the local internal state $u=(v,i)$ at the start, and (iv) the local internal state $u'=(v',j)$ at the end, where $u,v$ are elements in the conjugacy class $C$ and $i,j$ are the matrix indices for the irreducible representation $\chi$. In general, a ribbon operator creates vertex and plaquette violations at both ends of the ribbon.

\begin{figure}
    \centering
    \includegraphics[width=1\linewidth]{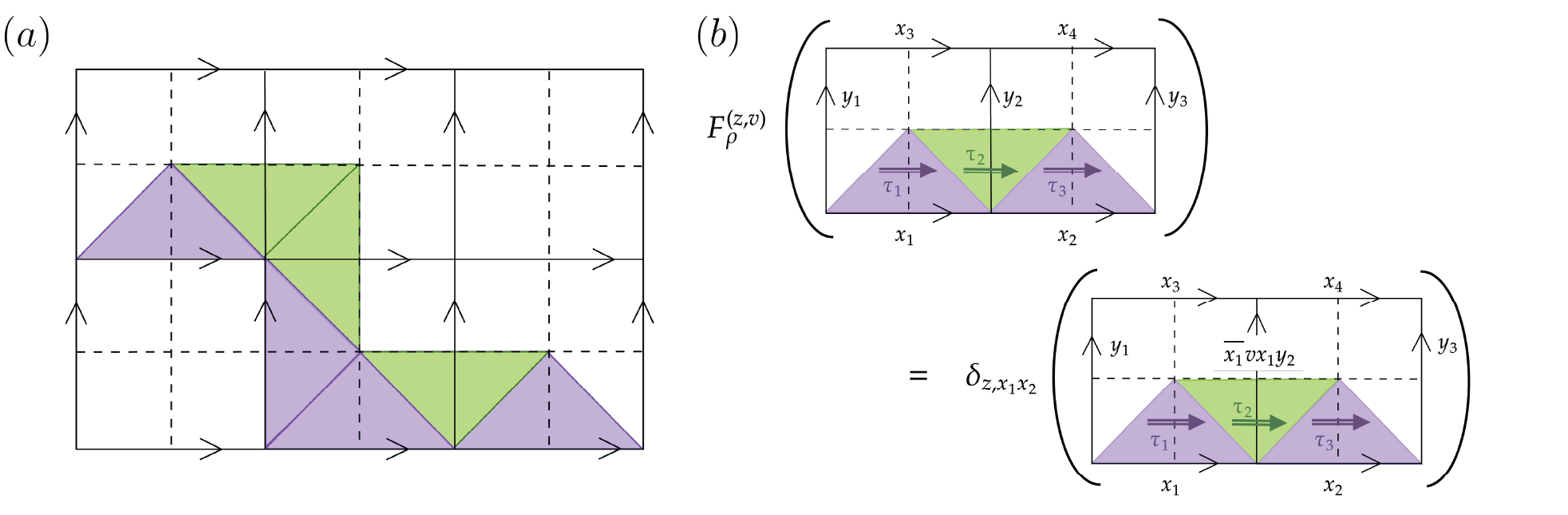}
    \caption{(a) Ribbon operators create non-Abelian anyons at their ends; they are thicker as compared to their abelian counterparts, and act both on the direct and dual edges of the lattice. (b) The action of a ribbon operator in the ``group element'' basis. This ribbon creates an anyon with local flux $v$ and nonlocal flux $z v \bar{z}$, and will generically create a superposition of charges. The anyon basis ribbons are superpositions of these ribbons.}
    \label{fig:appendix_ribbon_action}
\end{figure}

The easiest way to express the ribbon operator $F^{A;u,u'}_{\rho}$ is in terms of basis elements $F^{(z, v)}_\rho$, whose action is depicted in Fig.~\ref{fig:appendix_ribbon_action}b.
\begin{equation}\label{eq:anyon-basis-from-microscopic-basis}
    F^{A; u, u'} = \frac{|\chi|}{|\mathbf Z(r)|} \sum_{n \in \mathbf{Z}(r)} \Gamma^\chi_{j j'}(n) F^{(q_v n \overline{q}_{v'}, v)}, \quad A = (C, \chi), ~u = (v, j)
\end{equation}
Here, $r \in C$ is a representative element of the conjugacy class corresponding to the anyon in question. $\mathbf{Z}(r)$ is the centralizer of $r$, the subgroup of all elements in that commute with $r$. Furthermore, we have
\begin{equation}
    v = q_v r \bar{q}_v , \quad v' = q_{v'} r \bar{q}_{v'}
\end{equation}
The operator $F^{A;u,u'}_{\rho}$ creates two excitations at either end of $\rho$, but commutes with all stabilizers along the bulk of the ribbon. The anyons on either end will be of type $A = (C, \chi)$, and the internal state of each anyon will be labeled by $u$ and $u'$, respectively.

\begin{figure}
    \centering
    \includegraphics[width=0.6\linewidth]{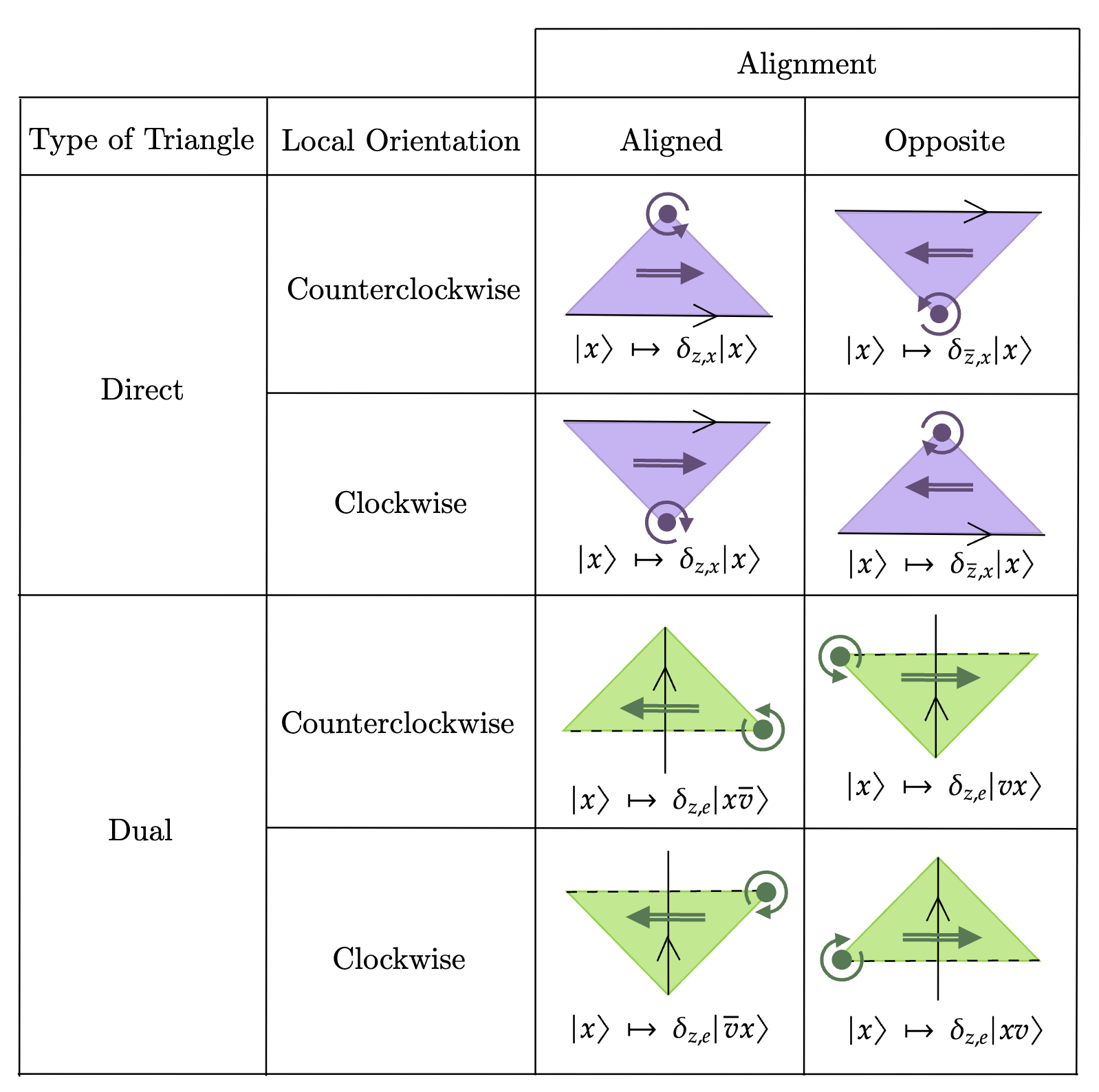}
    \caption{Definition of triangle operators, based on the type (direct or dual), alignment (aligned or opposite), and local orientation (clockwise or counterclockwise) of the triangle operator~\cite{lo_universal_2025}. Direct and dual edge is denoted by solid and dotted line respectively. The orientation of the edge is indicated by the single arrow; the direction of the ribbon is indicated by the double arrow. The local orientation of the triangle operator is determined by first fixing the hinge (dark circle) at the plaquette center touching the triangle operator, and then turn along the ribbon direction.}
    \label{fig:app-triangle-operators}
\end{figure}

Ribbon operators are constructed out of fundamental building blocks known as ``triangle'' operators: these are operators that act on a single edge of the lattice, and are strung together through a recursive formula to yield a full-length ribbon operator. Fig.~\ref{fig:app-triangle-operators} contains a table listing the eight different triangle operators, distinguished by three different characteristics: (i) whether they act on the direct or dual lattice, (ii) whether the direction of the operator is aligned with the lattice, and (iii) a property known as ``local orientation''~\cite{yan_ribbon_2022}. 

A full-length ribbon operator $F^{(z, v)}$ is built out of triangle operators recursively. Consider the three triangle ribbon pictured in Fig.~\ref{fig:appendix_ribbon_action}. Denote the triangles by $\tau_i$. We can derive the action of the full ribbon using the following formula: 
\begin{equation}\label{eq: recursive ribbon}
    F^{(z, v)}(\rho = \rho_1 \cup \rho_2) = \sum_{k\in G} F^{(k, v)}(\rho_1) F^{(\overline{k}z, \overline{k}vk)}(\rho_2) \; .
\end{equation}
Triangles $\tau_1$ and $\tau_3$ act on the direct lattice, are aligned with the lattice orientation, and have counter-clockwise local orientation, meaning the corresponding triangle operators $F^{(z, v)}(\tau_1)$ and $F^{(z, v)}(\tau_3)$ act by $\ket{x} \to \delta_{z, x}\ket{x}$. Triangle $\tau_2$ acts on the dual lattice, is opposite to the lattice orientation, and has counter-clockwise local orientation, and so the corresponding operator acts as $F^{(z, v)}(\tau_2) \ket{x} = \delta_{z, e} \ket{vx}$.

We start by breaking up the ribbon into two pieces: $\rho = \tau_1 \cup (\tau_2 \cup \tau_3)$:
\begin{equation}
    F^{(z, v)}(\rho) = \sum_{k\in G} F^{(k, v)}(\tau_1) F^{(\overline{k}z, \overline{k}vk)}(\tau_2 \cup \tau_3) = \sum_{k\in G} \delta_{x_1, k} F^{(\overline{k}z, \overline{k}vk)}(\tau_2 \cup \tau_3) = F^{(\overline{x}_1z, \overline{x}_1vx_1)}(\tau_2 \cup \tau_3).
\end{equation} Applying the recursion relation again: 
\begin{equation}
\begin{aligned}
    F^{(z, v)}(\rho) &= \sum_{k\in G} F^{(k, \overline{x}_1vx_1)}(\tau_2) F^{(\overline{k}\overline{x}_1z, \overline{k}\overline{x_1}vx_1 k)}(\tau_3)\\
    &= \sum_{k\in G} \delta_{k, e} \ket{(\overline{x}_1v x_1 )y_2}\bra{y_2} \delta_{x_2, \overline{k}\overline{x}_1z} \\
    &= \delta_{x_1 x_2, z} \ket{(\overline{x}_1v x_1 )y_2}\bra{y_2} .
\end{aligned}
\end{equation}
This is the same action quoted in Fig.~\ref{fig:appendix_ribbon_action}(b). 

\subsubsection{Unitary ribbon}\label{app:unitary-ribbon}
We now specialize to useful ribbons used in the main text. We observe that by summing over the local internal state on one end of the ribbon, the resulting ribbon operator is unitary. Such unitary ribbon can be decomposed into three steps: 1) ungauging, 2) finite-depth local unitary (FDLU) circuit in a simpler TO, and 3) regauging~\cite{lyons-2025}. For the ungauging step, it is a sequential unitary circuit that ``unzips" along the length of the ribbon to a simpler TO. Then, the FDLU circuit is either a string operator or symmetry fractionalization pattern in the simpler TO. Finally, the regauging circuit ``zips up" along the length of the ribbon back to the original TO. 

The unitary ribbon for a $[2]$ charge vacuum pair is given by:
\begin{equation}
   F^{(+, [2]); 0} = \sum_{j} F^{(+, [2]); 0,j}_{\rho} = \prod_{x \in \rho} \mathcal{Z}_x^{\prod_{g < x} Z_{g}}.
\end{equation}
Here $x, g$ are edges on the lattice along the ribbon $\rho$, and $g < x$ means that $g$ occurs before edge $x$, traveling along $\rho$ from the start of the ribbon to the end. The gate $\mathcal{Z}_x^{Z_g}$ is a controlled gate with the qutrit on edge $x$ as the target and the qubit on edge $g$ as the control:
\begin{equation}
    \mathcal{Z}_x^{Z_g} \ket{\mu^a}_x \ket{\sigma^{b}} = \omega^{(-1)^{b} a} \ket{\mu^a}_x  \ket{\sigma^b}, \quad a = 0, 1, 2, \quad b = 0, 1.
\end{equation}
There can also be multiple controls on some set of edges $\{g\}$:
\begin{equation}
    \mathcal{Z}_x^{\prod_g Z_g} \ket{\mu^a}_x \prod_g \ket{\sigma^{b_g}}_g = \omega^{(-1)^{\sum_{g} b_g} a} \ket{\mu^a}_x \prod_g \ket{\sigma^b}_g, \quad a = 0, 1, 2, \quad b_g = 0, 1.
\end{equation}
$F^{(+, [2]); 0}$ can be expressed as a circuit by using the identities in Section~\ref{app:qubit-qutrit-encoding} (where the purple arrow in the first purple triangle indicates the direction of the ribbon):

\begin{equation}\label{eq:appendix_2charge_circuit}
\raisebox{-70pt}{
    \begin{tikzpicture}
        \node at (0, 0) {
            \begin{tikzpicture}[scale=0.5]
                \draw[black, fill=purple!50!blue!30!white] (-6, 0) -- (-5, 1) -- (-4, 0) -- (-6, 0);
                \draw[black, fill=purple!50!blue!30!white] (-4, 0) -- (-3, 1) -- (-2, 0) -- (-4, 0);
                \draw[black, fill=purple!50!blue!30!white] (-2, 0) -- (-1, 1) -- (0, 0) -- (-2, 0);
                \draw[black, fill=purple!50!blue!30!white] (0, 0) -- (1, 1) -- (2, 0) -- (0, 0);
                \draw[black, fill=green!30] (-4, 0) -- (-3, 1) -- (-5, 1) -- (-4, 0);
                \draw[black, fill=green!30] (-2, 0) -- (-1, 1) -- (-3, 1) -- (-2, 0);
                \draw[black, fill=green!30] (0, 0) -- (1, 1) -- (-1, 1) -- (0, 0);
                \draw[black, fill=green!30] (2, 0) -- (3, 1) -- (1, 1) -- (2, 0);
                \draw[black] (-5 - 0.2,0.2) -- (-5+0.2,0) -- (-5 - 0.2,-0.2);
                \draw[black] (-3 - 0.2,0.2) -- (-3+0.2,0) -- (-3 - 0.2,-0.2);
                \draw[black] (-1 - 0.2,0.2) -- (-1+0.2,0) -- (-1 - 0.2,-0.2);
                \draw[black] (1 - 0.2,0.2) -- (1+0.2,0) -- (1 - 0.2,-0.2);
                \draw[black] (-4 - 0.2,1.2) -- (-4,1.6) -- (-4 + 0.2,1.2);
                \draw[black] (-2 - 0.2,1.2) -- (-2,1.6) -- (-2 + 0.2,1.2);
                \draw[black] (-0 - 0.2,1.2) -- (-0,1.6) -- (-0 + 0.2,1.2);
                \draw[black] (2 - 0.2,1.2) -- (2,1.6) -- (2 + 0.2,1.2);
                \draw  [violet][fill=violet ,fill opacity=1 ][line width=0.75]  (-5+0.2,0.3) -- (-5+0.2,0.5) -- (-5+0.4,0.4) -- cycle ;
                \draw[thick,violet] (-5-0.2,0.35) -- (-5+0.2,0.35) ;
                \draw[thick,violet] (-5-0.2,0.45) -- (-5+0.2,0.45) ;

                \draw[thick] (-6,0) -- node[below]{$\alpha_1, b_1$} (-4,0) ;
                \draw[thick] (-4, 0) -- node[right]{} (-4, 2);
                \draw[thick] (-4,0) -- node[below]{$\alpha_2, b_2$} (-2,0);
                \draw[thick] (-2, 0) -- node[right]{} (-2, 2);
                \draw[thick] (-2,0) -- node[below]{$\alpha_3, b_3$} (0,0);
                \draw[thick] (0, 0) -- node[right]{} (0, 2);
                \draw[thick] (0,0) -- node[below]{$\alpha_4, b_4$} (2,0);
                \draw[thick] (2, 0) -- node[right]{} (2, 2);
                
            \end{tikzpicture}
        };
        \node at (8, 0) {
            \begin{tikzpicture}

                \begin{yquant}[horizontal]
                qubit {$\ket{\alpha_1}$} a1;
                qubit {$\ket{b_1}$} b1;
                qubit {$\ket{\alpha_2}$} a2;
                qubit {$\ket{b_2}$} b2;
                qubit {$\ket{\alpha_3}$} a3;
                qubit {$\ket{b_3}$} b3;
                qubit {$\ket{\alpha_4}$} a4;
                qubit {$\ket{b_4}$} b4;
                
                cnot b2 | b1;
                cnot b3 | b2;
                cnot b4 | b3;

                
                align a2, a3, a4;
                box {$\mathcal{C}$} a2 | b1;
                box {$\mathcal{C}$} a3 | b2;
                box {$\mathcal{C}$} a4 | b3;
                
                barrier (a1, a2, a3, a4, b1, b2, b3, b4);
                
                align a1, a2, a3, a4;
                box {$\mathcal{Z}$} a1; 
                box {$\mathcal{Z}$} a2; 
                box {$\mathcal{Z}$} a3; 
                box {$\mathcal{Z}$} a4; 
                
                barrier (a1, a2, a3, a4, b1, b2, b3, b4);
                
                box {$\mathcal{C}$} a4 | b3;
                box {$\mathcal{C}$} a3 | b2;
                box {$\mathcal{C}$} a2 | b1;

                
                cnot b4 | b3;
                cnot b3 | b2;
                cnot b2 | b1;
                
                \end{yquant}
                \draw[decoration=brace, decorate]
            		($(0,.3)$) -- ($(2.7,.3)$)
            		node[midway, above=1pt, text width=3.0cm, align=center] {Ungauging};
                \draw[decoration=brace, decorate]
                    		($(2.8,.3)$) -- ($(3.9,.3)$)
                    		node[midway, above=1pt, text width=6.0cm, align=center] {FDLU};
                \draw[decoration=brace, decorate]
                    		($(4,.3)$) -- ($(6.7,.3)$)
                    		node[midway, above=1pt, text width=6.0cm, align=center] {Regauging};
            \end{tikzpicture}
        };
        
    \end{tikzpicture}
}
\end{equation}

The $[2]$ charge ribbon operator circuit above illustrate how a unitary ribbon operator can be decomposed into three steps (as divided by dotted line): 1) ungauging, 2) finite-depth local unitary (FDLU) circuit in a simpler TO, and 3) regauging~\cite{lyons-2025}. For the ungauging step, it is a sequential unitary circuit that "unzip" the direct edges to a simpler TO (in this case ungauging to $\mathbb{Z}_3$ toric code). Then, the FDLU circuit is just the usual string operator for the $e$ anyons in $\mathbb{Z}_3$ toric code. Finally, the regauging layer is used to "zip up" the direct edges back to $S_3$ TO. 

Unitary ribbons for a $[C_2]$ flux is more complicated that the $[2]$ charge ribbon. For example, to implement the unitary ribbon $F^{C_2;\sigma}=\sum_{c\in C_2} F^{C_2;\sigma,c}$, where the local internal state is $\ket{\sigma}$ at the start of the ribbon, the circuit for a $C_2$ ribbon (with direction indicated by the purple arrow in the first purple triangle) with counterclockwise local orientation (the issue of local orientation will be explained below) is given by:

\begin{equation}\label{eq:appendix_C2_circuit}
    \raisebox{-150pt}{
        \begin{tikzpicture}
            \node at (-8, 0) {
                \begin{tikzpicture}[scale=0.5]
                    \draw[black, fill=purple!50!blue!30!white] (-6, 0) -- (-5, 1) -- (-4, 0) -- (-6, 0);
                    \draw[black, fill=purple!50!blue!30!white] (-4, 0) -- (-3, 1) -- (-2, 0) -- (-4, 0);
                    \draw[black, fill=purple!50!blue!30!white] (-2, 0) -- (-1, 1) -- (0, 0) -- (-2, 0);
                    \draw[black, fill=purple!50!blue!30!white] (0, 0) -- (1, 1) -- (2, 0) -- (0, 0);
                    \draw[black, fill=green!30] (-4, 0) -- (-3, 1) -- (-5, 1) -- (-4, 0);
                    \draw[black, fill=green!30] (-2, 0) -- (-1, 1) -- (-3, 1) -- (-2, 0);
                    \draw[black, fill=green!30] (0, 0) -- (1, 1) -- (-1, 1) -- (0, 0);
                    \draw[black, fill=green!30] (2, 0) -- (3, 1) -- (1, 1) -- (2, 0);
                    \draw[black] (-5 - 0.2,0.2) -- (-5+0.2,0) -- (-5 - 0.2,-0.2);
                    \draw[black] (-3 - 0.2,0.2) -- (-3+0.2,0) -- (-3 - 0.2,-0.2);
                    \draw[black] (-1 - 0.2,0.2) -- (-1+0.2,0) -- (-1 - 0.2,-0.2);
                    \draw[black] (1 - 0.2,0.2) -- (1+0.2,0) -- (1 - 0.2,-0.2);
                    \draw[black] (-4 - 0.2,1.2) -- (-4,1.6) -- (-4 + 0.2,1.2);
                    \draw[black] (-2 - 0.2,1.2) -- (-2,1.6) -- (-2 + 0.2,1.2);
                    \draw[black] (-0 - 0.2,1.2) -- (-0,1.6) -- (-0 + 0.2,1.2);
                    \draw[black] (2 - 0.2,1.2) -- (2,1.6) -- (2 + 0.2,1.2);
                    \draw  [violet][fill=violet ,fill opacity=1 ][line width=0.75]  (-5+0.2,0.3) -- (-5+0.2,0.5) -- (-5+0.4,0.4) -- cycle ;
                    \draw[thick,violet] (-5-0.2,0.35) -- (-5+0.2,0.35) ;
                    \draw[thick,violet] (-5-0.2,0.45) -- (-5+0.2,0.45) ;
                
                    \draw[red, thick] (-6,0) -- node[below]{$a_1, b_1$} (-4,0) ;
                    \draw[blue, thick] (-4, 0) -- (-4, 1.5) node[right]{\footnotesize $\alpha_1, \beta_1$} -- (-4, 2);
                    \draw[red, thick] (-4,0) -- node[below]{$a_2, b_2$} (-2,0);
                    \draw[blue, thick] (-2, 0) -- (-2, 1.5) node[right]{\footnotesize $\alpha_2, \beta_2$} -- (-2, 2);
                    \draw[red, thick] (-2,0) -- node[below]{$a_3, b_3$} (0,0);
                    \draw[blue, thick] (0, 0) -- (0, 1.5) node[right]{\footnotesize $\alpha_3, \beta_3$} -- (0, 2);
                    \draw[red, thick] (0,0) -- node[below]{$a_4, b_4$} (2,0);
                    \draw[blue, thick] (2, 0) -- (2, 1.5) node[right]{\footnotesize $\alpha_4, \beta_4$} -- (2, 2);
                \end{tikzpicture}
            };
            \node at (1, 0) {
                \begin{tikzpicture}[scale=0.8]
                    \begin{yquant}[horizontal]
                    [red]
                    qubit {$\ket{a_1}$} a[1];
                    [red]
                    qubit {$\ket{b_1}$} b[1];
                    [blue]
                    qubit {$\ket{\alpha_1}$} alp[1];
                    [blue]
                    qubit {$\ket{\beta_1}$} beta[1];
                    
                    [red]
                    qubit {$\ket{a_2}$} a[+1];
                    [red]
                    qubit {$\ket{b_2}$} b[+1];
                    [blue]
                    qubit {$\ket{\alpha_2}$} alp[+1];
                    [blue]
                    qubit {$\ket{\beta_2}$} beta[+1];
                    
                    [red]
                    qubit {$\ket{a_3}$} a[+1];
                    [red]
                    qubit {$\ket{b_3}$} b[+1];
                    [blue]
                    qubit {$\ket{\alpha_3}$} alp[+1];
                    [blue]
                    qubit {$\ket{\beta_3}$} beta[+1];
                    
                    [red]
                    qubit {$\ket{a_4}$} a[+1];
                    [red]
                    qubit {$\ket{b_4}$} b[+1];
                    [blue]
                    qubit {$\ket{\alpha_4}$} alp[+1];
                    [blue]
                    qubit {$\ket{\beta_4}$} beta[+1];
                    
                    cnot b[1] | b[0];
                    cnot b[2] | b[1];
                    cnot b[3] | b[2];
                    
                    align a[1], a[2], a[3];
                    box {$\mathcal{C}$} a[1] | b[0];
                    box {$\mathcal{C}$} a[2] | b[1];
                    box {$\mathcal{C}$} a[3] | b[2];
                    
                    barrier (a, b, alp, beta);
                    
                    box {$\mathcal{X}$} a[1] | a[0];
                    box {$\mathcal{X}$} a[2] | a[1];
                    box {$\mathcal{X}$} a[3] | a[2];
                    
                    barrier (a, b, alp, beta);
                    
                    align alp;
                    box {$X$} beta; 
                    box {$\mathcal{C}$} alp;
                    
                    box {$\mathcal{C}$} a[0] | b[0];
                    box {$\mathcal{C}$} a[1] | b[1];
                    box {$\mathcal{C}$} a[2] | b[2];
                    box {$\mathcal{C}$} a[3] | b[3];
                    
                    box {$\mathcal{X}$} alp[0] | a[0];
                    box {$\mathcal{X}$} alp[1] | a[1];
                    box {$\mathcal{X}$} alp[2] | a[2];
                    box {$\mathcal{X}$} alp[3] | a[3];
                    
                    box {$\mathcal{C}$} a[0] | b[0];
                    box {$\mathcal{C}$} a[1] | b[1];
                    box {$\mathcal{C}$} a[2] | b[2];
                    box {$\mathcal{C}$} a[3] | b[3];
                    
                    
                    
                    barrier (a, b, alp, beta);
                    
                    box {$\mathcal{X}^{\dagger}$} a[3] | a[2];
                    box {$\mathcal{X}^{\dagger}$} a[2] | a[1];
                    box {$\mathcal{X}^{\dagger}$} a[1] | a[0];
                    
                    barrier (a, b, alp, beta);
                    
                    align a[1], a[2], a[3];
                    box {$\mathcal{C}$} a[3] | b[2];
                    box {$\mathcal{C}$} a[2] | b[1];
                    box {$\mathcal{C}$} a[1] | b[0];
                    
                    cnot b[3] | b[2];
                    cnot b[2] | b[1];
                    cnot b[1] | b[0];
                    
                    \end{yquant}
                    \draw[decoration=brace, decorate]
                    		($(0,.3)$) -- ($(5.1,.3)$)
                    		node[midway, above=1pt, text width=3.0cm, align=center] {Ungauging};
                    \draw[decoration=brace, decorate]
                            ($(5.2,.3)$) -- ($(7.5,.3)$)
                            node[midway, above=1pt, text width=6.0cm, align=center] {FDLU};
                    \draw[decoration=brace, decorate]
                            ($(7.6,.3)$) -- ($(13.1,.3)$)
                            node[midway, above=1pt, text width=6.0cm, align=center] {Regauging};
                \end{tikzpicture}
            };
        \end{tikzpicture}
    }
\end{equation}

Note that the ungauging and regauging steps are more complicated than that of the $[2]$ charge ribbon, because it does not stop at the level of the $\mathbb{Z}_3$ toric code but continue to the level of $\mathbb{Z}_3$ paramagnet~\cite{lyons-2025}.

As a remark, to implement a $C_2$ ribbon where the starting local internal state is $\ket{\mu\sigma}$ and $\ket{\bar\mu\sigma}$, the entire circuit needed to be conjugated with $\mathcal{X}$ and $\mathcal{X}^\dagger$ respectively on the first direct edge ($a_1,b_1$ in this case), as it modifies the action of individual dual triangle to $L^{\mu\sigma}$ and $L^{\bar\mu\sigma}$ respectively (for the counterclockwise, aligned case). 

It is important to keep track of the local orientation of the ribbon to ensure consistent action of ribbon operator~\cite{yan_ribbon_2022,lo_universal_2025}. The example ribbons displayed above are of counterclockwise local orientation. The circuit for a $C_2$ ribbon with clockwise local orientation is given by:

\begin{equation}\label{eq:appendix_C2_circuit_cw}
    \raisebox{-150pt}{
        \begin{tikzpicture}
            \node at (-8, 0) {
                \begin{tikzpicture}[scale=0.5]
                    \draw[black, fill=purple!50!blue!30!white] (-6, 2) -- (-5, 1) -- (-4, 2) -- (-6, 2);
                    \draw[black, fill=purple!50!blue!30!white] (-4, 2) -- (-3, 1) -- (-2, 2) -- (-4, 2);
                    \draw[black, fill=purple!50!blue!30!white] (-2, 2) -- (-1, 1) -- (0, 2) -- (-2, 2);
                    \draw[black, fill=purple!50!blue!30!white] (0, 2) -- (1, 1) -- (2, 2) -- (0, 2);
                    \draw[black, fill=green!30] (-4, 2) -- (-3, 1) -- (-5, 1) -- (-4, 2);
                    \draw[black, fill=green!30] (-2, 2) -- (-1, 1) -- (-3, 1) -- (-2, 2);
                    \draw[black, fill=green!30] (0, 2) -- (1, 1) -- (-1, 1) -- (0, 2);
                    \draw[black, fill=green!30] (2, 2) -- (3, 1) -- (1, 1) -- (2, 2);
                    \draw[black] (-5 - 0.2,2.2) -- (-5+0.2,2) -- (-5 - 0.2,2-0.2);
                    \draw[black] (-3 - 0.2,2.2) -- (-3+0.2,2) -- (-3 - 0.2,2-0.2);
                    \draw[black] (-1 - 0.2,2.2) -- (-1+0.2,2) -- (-1 - 0.2,2-0.2);
                    \draw[black] (1 - 0.2,2.2) -- (1+0.2,2) -- (1 - 0.2,2-0.2);
                    \draw[black] (-4 - 0.2,1.2) -- (-4,1.6) -- (-4 + 0.2,1.2);
                    \draw[black] (-2 - 0.2,1.2) -- (-2,1.6) -- (-2 + 0.2,1.2);
                    \draw[black] (-0 - 0.2,1.2) -- (-0,1.6) -- (-0 + 0.2,1.2);
                    \draw[black] (2 - 0.2,1.2) -- (2,1.6) -- (2 + 0.2,1.2);
                    \draw  [violet][fill=violet ,fill opacity=1 ][line width=0.75]  (-5+0.2,1.3+0.1) -- (-5+0.2,1.5+0.1) -- (-5+0.4,1.4+0.1) -- cycle ;
                    \draw[thick,violet] (-5-0.2,1.35+0.1) -- (-5+0.2,1.35+0.1) ;
                    \draw[thick,violet] (-5-0.2,1.45+0.1) -- (-5+0.2,1.45+0.1) ;
                
                    \draw[red, thick] (-6,2) -- node[above]{$a_1, b_1$} (-4,2) ;
                    \draw[blue, thick] (-4, 0) -- (-4, 0.5) node[right]{\footnotesize $\alpha_1, \beta_1$} -- (-4, 2);
                    \draw[red, thick] (-4,2) -- node[above]{$a_2, b_2$} (-2,2);
                    \draw[blue, thick] (-2, 0) -- (-2, 0.5) node[right]{\footnotesize $\alpha_2, \beta_2$} -- (-2, 2);
                    \draw[red, thick] (-2,2) -- node[above]{$a_3, b_3$} (0,2);
                    \draw[blue, thick] (0, 0) -- (0, 0.5) node[right]{\footnotesize $\alpha_3, \beta_3$} -- (0, 2);
                    \draw[red, thick] (0,2) -- node[above]{$a_4, b_4$} (2,2);
                    \draw[blue, thick] (2, 0) -- (2, 0.5) node[right]{\footnotesize $\alpha_4, \beta_4$} -- (2, 2);
                \end{tikzpicture}
            };
            \node at (1, 0) {
                \begin{tikzpicture}[scale=0.8]
                    \begin{yquant}[horizontal]
                    [red]
                    qubit {$\ket{a_1}$} a[1];
                    [red]
                    qubit {$\ket{b_1}$} b[1];
                    [blue]
                    qubit {$\ket{\alpha_1}$} alp[1];
                    [blue]
                    qubit {$\ket{\beta_1}$} beta[1];
                    
                    [red]
                    qubit {$\ket{a_2}$} a[+1];
                    [red]
                    qubit {$\ket{b_2}$} b[+1];
                    [blue]
                    qubit {$\ket{\alpha_2}$} alp[+1];
                    [blue]
                    qubit {$\ket{\beta_2}$} beta[+1];
                    
                    [red]
                    qubit {$\ket{a_3}$} a[+1];
                    [red]
                    qubit {$\ket{b_3}$} b[+1];
                    [blue]
                    qubit {$\ket{\alpha_3}$} alp[+1];
                    [blue]
                    qubit {$\ket{\beta_3}$} beta[+1];
                    
                    [red]
                    qubit {$\ket{a_4}$} a[+1];
                    [red]
                    qubit {$\ket{b_4}$} b[+1];
                    [blue]
                    qubit {$\ket{\alpha_4}$} alp[+1];
                    [blue]
                    qubit {$\ket{\beta_4}$} beta[+1];
                    
                    cnot b[1] | b[0];
                    cnot b[2] | b[1];
                    cnot b[3] | b[2];
                    
                    align a[1], a[2], a[3];
                    box {$\mathcal{C}$} a[1] | b[0];
                    box {$\mathcal{C}$} a[2] | b[1];
                    box {$\mathcal{C}$} a[3] | b[2];
                    
                    barrier (a, b, alp, beta);
                    
                    box {$\mathcal{X}$} a[1] | a[0];
                    box {$\mathcal{X}$} a[2] | a[1];
                    box {$\mathcal{X}$} a[3] | a[2];
                    
                    barrier (a, b, alp, beta);
                    
                    align alp;
                    
                    box {$\mathcal{C}$} a[0] | b[0];
                    box {$\mathcal{C}$} a[1] | b[1];
                    box {$\mathcal{C}$} a[2] | b[2];
                    box {$\mathcal{C}$} a[3] | b[3];

                    box {$\mathcal{C}$} alp[0] | beta[0];
                    box {$\mathcal{C}$} alp[1] | beta[1];
                    box {$\mathcal{C}$} alp[2] | beta[2];
                    box {$\mathcal{C}$} alp[3] | beta[3];
                    
                    box {$\mathcal{X}$} alp[0] | a[0];
                    box {$\mathcal{X}$} alp[1] | a[1];
                    box {$\mathcal{X}$} alp[2] | a[2];
                    box {$\mathcal{X}$} alp[3] | a[3];
                    
                    box {$\mathcal{C}$} a[0] | b[0];
                    box {$\mathcal{C}$} a[1] | b[1];
                    box {$\mathcal{C}$} a[2] | b[2];
                    box {$\mathcal{C}$} a[3] | b[3];

                    box {$\mathcal{C}$} alp[0] | beta[0];
                    box {$\mathcal{C}$} alp[1] | beta[1];
                    box {$\mathcal{C}$} alp[2] | beta[2];
                    box {$\mathcal{C}$} alp[3] | beta[3];

                    box {$X$} beta; 
                    
                    barrier (a, b, alp, beta);
                    
                    box {$\mathcal{X}^{\dagger}$} a[3] | a[2];
                    box {$\mathcal{X}^{\dagger}$} a[2] | a[1];
                    box {$\mathcal{X}^{\dagger}$} a[1] | a[0];
                    
                    barrier (a, b, alp, beta);
                    
                    align a[1], a[2], a[3];
                    box {$\mathcal{C}$} a[3] | b[2];
                    box {$\mathcal{C}$} a[2] | b[1];
                    box {$\mathcal{C}$} a[1] | b[0];
                    
                    cnot b[3] | b[2];
                    cnot b[2] | b[1];
                    cnot b[1] | b[0];
                    
                    \end{yquant}
                    \draw[decoration=brace, decorate]
                    		($(0,.3)$) -- ($(5.1,.3)$)
                    		node[midway, above=1pt, text width=3.0cm, align=center] {Ungauging};
                    \draw[decoration=brace, decorate]
                            ($(5.2,.3)$) -- ($(8.2,.3)$)
                            node[midway, above=1pt, text width=6.0cm, align=center] {FDLU};
                    \draw[decoration=brace, decorate]
                            ($(8.3,.3)$) -- ($(13.8,.3)$)
                            node[midway, above=1pt, text width=6.0cm, align=center] {Regauging};
                \end{tikzpicture}
            };
        \end{tikzpicture}
    }
\end{equation}

The circuit implementation for clockwise local orientation can be derived from that of counterclockwise local orientation by applying the inverse operation circuit (Eq.~\ref{eq:inverse-circ}) on the dual edges.

\subsubsection{Generalized ribbon operator}\label{app:generalized-ribbon}

In Ref.~\onlinecite{lo_universal_2025}, a generalized ribbon operator is introduced that relaxes the condition that a ribbon operator only create violations at two sites. It is observed the charge violations can be decoupled from the flux violations at the ends of a ribbon; i.e., the charge violations can be moved arbitrarily far away from the flux violations. This can be understood from the gauging perspective discussed in Section~\ref{app:unitary-ribbon}: the ungauging/regauging circuit can be along a length beyond the support of the FDLU part of the circuit. Then the charge violations are located at the starting and ending location of the ungauging/regauging circuit, whereas the flux violations are located at the starting and ending location of the FDLU circuit.

Generalized ribbon operators are crucial for defining the absolute logical encoding used in demonstrating the universal gate set in Section~\ref{subsec:pull-through}, \ref{subsec:x-measurement}, and \ref{subsec:z-measurement}. A generalized ribbon operator ensures global fluxes are well-defined (so that they can be compared consistently) and allows the fluxes at the two ends of a ribbon operator to be at arbitrary locations and moved for braiding. When defining a flux, we need to specify the base point where the flux loop starts and ends. Changing base point conjugate the flux by the value of the path between the old and new base point. Therefore, to compare the flux internal states between distinct fluxes, the same base point has to be used to ensure consistency. 

For the experiment, we fix the base point to be at the origin such that all global fluxes are defined with respect to the origin; an example is shown in Fig.~\ref{fig:notation}(k).

\section{Qubit–Qutrit Encoding of \ds3 Stabilizers and Projectors}\label{app:z2z3-decomposition}
In Section~\ref{app:qubit-qutrit-encoding} we detailed the encoding of $S_3$ group elements in terms of qubit--qutrit pairings and described the corresponding group multiplication rules. Here we provide a description of the \emph{standard} \ds3\ projectors, originally defined in Ref.~\onlinecite{kitaev2003fault}, expressed in terms of their action on the qubit and qutrit degrees of freedom. In particular, we derive explicit expressions for the \ds3\ vertex and plaquette projectors in terms of the operators ${A^{\ZZ_2}}$, ${A^{\ZZ_3}}$, ${B^{\ZZ_2}}$, and ${B^{\ZZ_3}}$ introduced in Eqs.~\eqref{eq:qutrit-qubit-vertex-projectors} and~\eqref{eq:qutrit-qubit-plaquette-projectors}. Similar definitions for the $W$-flux and non-contractible projectors are also given.

We begin by writing \ds3 hamiltonian
\begin{equation}
	H = -\sum_v A_v - \sum_p B_p^e,
\end{equation}
where the vertex projector $A_v = \frac{1}{6}\sum_{g \in S_3} A_v^g$ enforces the zero-charge constraint at each vertex. Pictorially, the action of $A_v^g$ on a group-element configuration is
\begin{align}
	\label{eq:vertex-stabilizer-group-basis-action}
	A_v^g\!\left(
	\cbox{6.0}{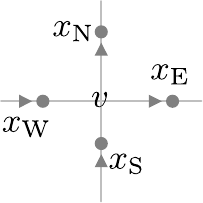}
	\right)
	\; = \;
	\cbox{6.0}{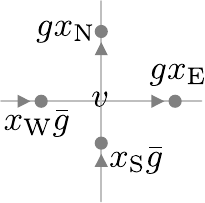}\;.
\end{align}
Here, the solid circle and triangle on each edge denote the qubit--qutrit encoding.  
Similarly, the plaquette projectors $B_p^e$ enforce trivial total flux around each plaquette. Their action on basis states is
\begin{align}
	\label{eq:plaquette-stabilizer}
	B_p^e\!\left(
	\cbox{5.5}{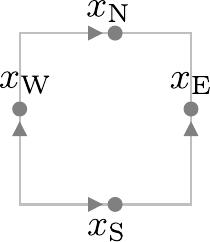}
	\right)
	\; = \;
	\delta_{e,\,x_{\textrm W}x_{\textrm N}\bar x_{\textrm E}\bar x_{\textrm S}}
	\cbox{5.5}{figures_appendix/deriv_p_1.pdf}\;,
\end{align}
where $x_{\mathrm W} x_{\mathrm N} \bar x_{\mathrm E} \bar x_{\mathrm S}$ denotes the flux around the plaquette.

\subsection{Vertex projectors}

From Eq.~\eqref{eq:vertex-stabilizer-group-basis-action}, $A_v^g$ left-multiplies the group elements on the north (N) and east (E) edges, and right-multiplies those on the south (S) and west (W) edges. The full vertex projector can be written as
\begin{align}
	\label{eq:vertex-projector1}
	A_v 
	\;=\;
	\frac{1}{6} \sum_{g \in S_3} {\cbox{5.5}{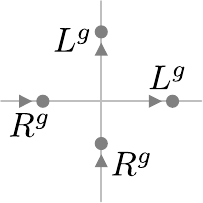}}.
\end{align}

Using Eqs.~\eqref{eq:q2q3_enc_lr_multiplication_1} and~\eqref{eq:q2q3_enc_lr_multiplication_2}, the terms $A_v^\sigma$ and $A_v^\mu$ can be expressed as
\begin{align}
	\label{eq:vertex-projector2}
	A_v^{\sigma}
	\;=\;
	\cbox{5.2}{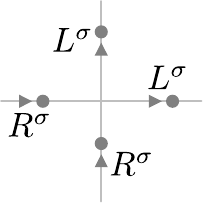}
	\;=\;
	\cbox{5.2}{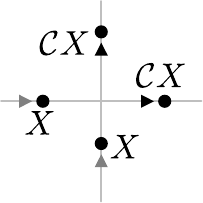}\;, \quad\quad
	A_v^{\mu}
	\;=\;
	\cbox{5.2}{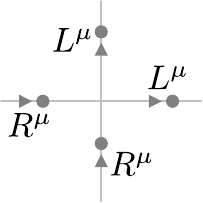}
	\;=\;
	\cbox{5.2}{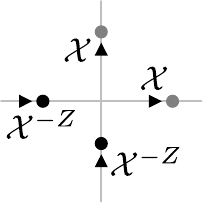}\;.
\end{align}

Expanding $A_v$, we obtain
\begin{align}
	A_v
	&= \frac{1}{6}\left(1 + A^{\mu} + A^{\mubar} + A^{\sigma} + A^{\mu\sigma} + A^{\mubars}\right) \nonumber\\
	&= \frac{1}{6}\left( 1 + A^{\sigma} \right)\left( 1 + A^{\mu} + A^{\mubar} \right) \nonumber\\
	&= \frac{1}{2}\!\left( 
	1 + 
	\cbox{5.1}{figures_appendix/deriv_v_32.pdf}
	\right)
	\cdot
	\frac{1}{3}\!\left[
	1 +
	\left(
	\cbox{5.1}{figures_appendix/deriv_v_34.pdf}
	+ \mathrm{h.c.}
	\right)
	\right] \nonumber\\
	&= A^{\ZZ_2}\, A^{\ZZ_3}.
\end{align}

\subsection{Plaquette projector}
Since $B_p^e$ enforces trivial flux around each plaquette, it can be written as
\begin{align} 
	\label{eq:B_pe_as_operator}
	B^e_p
	\; = \; 
	\sum_{g_{\mathrm W}, g_{\mathrm N}, g_{\mathrm E}, g_{\mathrm S} \in S_3}
	\delta_{e,\,g_{\mathrm S} g_{\mathrm E} \bar g_{\mathrm N}\bar g_{\mathrm W}}\
	\cbox{5.8}{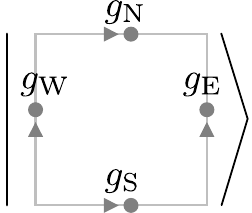}\ 
	\cbox{5.8}{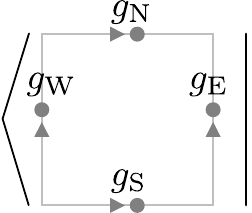}.
\end{align}

Using the qubit--qutrit encoding $g=\mu^a\sigma^b=\sigma^b\mu^{(-1)^b a}$ and $\bar g = \sigma^b\mu^{-a} = \mu^{-(-1)^{b}a}\sigma^b$, the flux  
$g_{\mathrm S} g_{\mathrm E} \bar g_{\mathrm N} \bar g_{\mathrm W}$ becomes
\begin{align}
	\textrm{flux}
	&=
	\mu^{a_{\mathrm S}}\sigma^{b_{\mathrm S}}\
	\mu^{a_{\mathrm E}}\sigma^{b_{\mathrm E}}\
	\sigma^{b_{\mathrm N}}\mu^{-a_{\mathrm N}}\
	\sigma^{b_{\mathrm W}}\mu^{-a_{\mathrm W}} \nonumber\\
	&=
	\mu^{a_{\mathrm S}}\,
	\mu^{(-1)^{b_{\mathrm S}} a_{\mathrm E}}\,
	\sigma^{b_{\mathrm S}+b_{\mathrm E}+b_{\mathrm N}}\,
	\mu^{-a_{\mathrm N}}\,
	\mu^{-(-1)^{b_{\mathrm W}} a_{\mathrm W}}\,
	\sigma^{b_{\mathrm W}} \nonumber\\
	&=
	\mu^{a_{\mathrm S}}\,
	\mu^{(-1)^{b_{\mathrm S}} a_{\mathrm E}}\,
	\mu^{-(-1)^{b_{\mathrm S}+b_{\mathrm E}+b_{\mathrm N}} a_{\mathrm N}}\,
	\mu^{-(-1)^{b_{\mathrm S}+b_{\mathrm E}+b_{\mathrm N}+b_{\mathrm W}} a_{\mathrm W}}\,
	\sigma^{b_{\mathrm S}+b_{\mathrm E}+b_{\mathrm N}+b_{\mathrm W}}.
\end{align}

Each term in the product can be written in terms of $Z$ and $\mathcal{Z}$ operators on the qubit and qutrit degrees of freedom, that is,
\begin{align}\label{eq:deriv_p_sigma}
    \sigma^{b_{\mathrm S}+b_{\mathrm E}+b_{\mathrm N}+b_{\mathrm W}}
    \;\rightarrow \; \cbox{5.2}{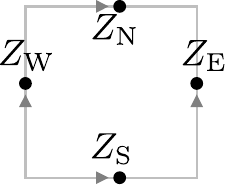}\; := \; S^{\ZZ_2},
\end{align}
\begin{align}
    \mu^{-(-1)^{b_{\mathrm S}+b_{\mathrm E}+b_{\mathrm N}+b_{\mathrm W}}a_{\mathrm W}}
    \;\rightarrow\;
    \cbox{4.3}{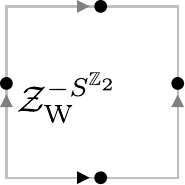}\;,\quad		\mu^{-(-1)^{b_{\mathrm S}+b_{\mathrm E}+b_{\mathrm N}}a_{\mathrm N}}
    \;\rightarrow\;
    \cbox{4.5}{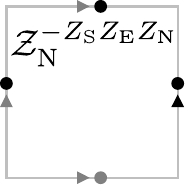}
    \;=\;
    \cbox{4.5}{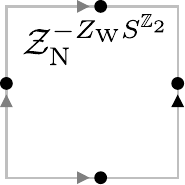}\;,\nonumber
\end{align}

\begin{align}\label{eq:deriv_p_2}
    \mu^{(-1)^{b_{\mathrm S}}a_{\mathrm E}}
    \;\rightarrow\;
    \cbox{4.5}{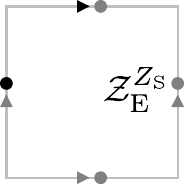}\ , \ \ \textrm{and}\ \
    \mu^{a_{\mathrm S}}	
    \; \rightarrow \;
    \cbox{4.5}{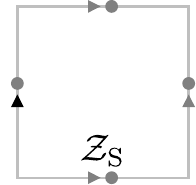}\;.
\end{align}
Collecting all contributions, the flux around the plaquette can be measured by
\begin{align}
    \label{eq:deriv_p_3}
    S^{\ZZ_2}
    \;\cdot\;
    \cbox{8.6}{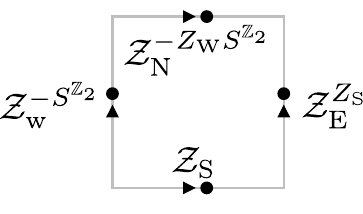} = S^{\ZZ_2} \;\cdot\; S^{\ZZ_3}.
\end{align}
Moreover, projectors onto the $+1$ eigenspaces of $S^{\ZZ_2}$ and $S^{\ZZ_3}$ enforce the trivial flux constraint, and Eq.~\eqref{eq:B_pe_as_operator} can be written as (see Ref.~\onlinecite{nonabelianfracton}):
\begin{align}
    \label{eq:deriv_p_all}
    B_p^e
    &=
    \frac{1}{2}\!\left(
    1 + 
    \cbox{5.2}{figures_appendix/deriv_p_ssigma.pdf}
    \right)
    \cdot
    \frac{1}{3}\!\left[
    1 +
    \left(
    \cbox{8.6}{figures_appendix/deriv_p_all.pdf}
    + \mathrm{h.c.}
    \right)
    \right] \nonumber\\
    &= B^{\ZZ_2}\, B^{\ZZ_3}.
\end{align}

\subsection{$W$-flux and non-contractible $Z$ projectors of \ds3}
\begin{figure}[!h]
    \centering
    \includegraphics[width=0.8\linewidth]{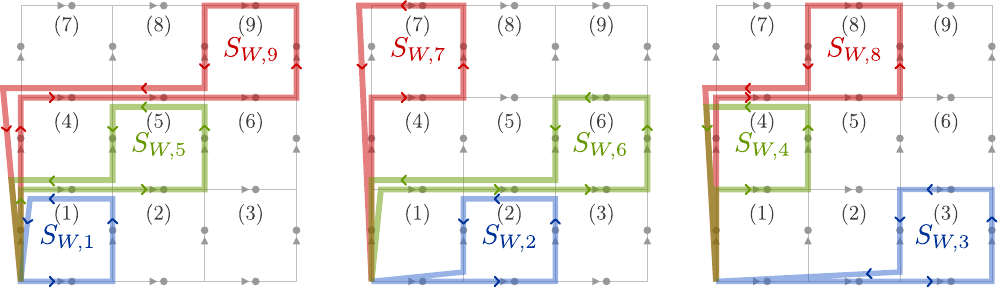}
    \caption{\textbf{$W$-flux.} Paths along which $W$-flux stabilizers are evaluated to benchmark the			experimental results presented in the main text.}
    \label{fig:w-paths}
\end{figure}

The $W$-flux is the topological flux defined by the ordered group product along a closed loop, based at a fixed origin, that encircles an excitation. Fig.~\ref{fig:w-paths} enumerates the different paths used to define the
$W$-flux stabilizer $S_{W_p}$, each of which encloses exactly one plaquette $p$. The corresponding projector $W_p$ is defined as the projector onto the $+1$ eigenspace of $S_{W_p}$. Using steps analogous to those employed in deriving the plaquette operator $B_p$, the $W$-flux projector can also be written as
\begin{align}
    W_p = W^{\mathbb{Z}_2}_p \, W^{\mathbb{Z}_3}_p .
\end{align}
Here $W^{\mathbb{Z}_2}_p = B^{\mathbb{Z}_2}_p$ (as defined in Eq.~\eqref{eq:deriv_p_sigma}) for the paths listed in Fig.~\ref{fig:w-paths}, while $W^{\mathbb{Z}_3}_p$ has a form analogous to $S^{\mathbb{Z}_3}_B$ and acts
non-trivially on all qudits along the loop.

Correlators of the $W$-flux are obtained by forming products of stabilizers, for example $S_{W_{p_1}} S_{W_{p_2}}$, and projecting onto their joint $+1$ eigenspace, yielding the projector $\Pi_{W_{p_1} W_{p_2}}$. This projector similarly factorizes into independent $\ZZ_2$ and $\ZZ_3$ components.
\begin{align}
    \Pi_{W_{p_1}W_{p_2}}	= 	\Pi^{\ZZ_2}_{W_{p_1}W_{p_2}} \Pi^{\ZZ_3}_{W_{p_1}W_{p_2}}
\end{align}

Similarly, the `$Z$-logical' projectors of the \ds3 code are defined as the ordered product of group elements along non-contractible loops. For the horizontal and vertical directions, these take the form
$Z_{\mathrm{hori}} = Z^{\mathbb{Z}_2}_{\mathrm{hori}} Z^{\mathbb{Z}_3}_{\mathrm{hori}}$
and
$Z_{\mathrm{vert}} = Z^{\mathbb{Z}_2}_{\mathrm{vert}} Z^{\mathbb{Z}_3}_{\mathrm{vert}}$,
respectively (see Fig.~\ref{fig:notation}e--f for the corresponding stabilizer definitions).

\section{Coherent moving of non-Abelian anyons: theory}\label{app:coherent-moving}

In our experiment, we only need to coherently move non-Abelian pure fluxes. We will focus on the protocol for coherently moving non-Abelian flux for the rest of the section. This is a special case of the general procedure of coherent moving~\cite{lo_coherent}.

Suppose we want to coherently move an anyon from location $A$ to $B$. Operationally, we demand a deterministic process such that there is no remnant anyon at the site $A$ after coherent moving. We implement coherent moving unitarily with the use of one ancilla qudit, where $d$ matches the quantum dimension of the non-Abelian anyon. 

Ribbon operators create pairs of excitation at their end points; to ensure that an anyon is deterministically moved from the starting site to the ending site, we need to ensure that the fusion outcome is always vacuum at the starting site. This is done by judicious choice of internal state at the ends of a ribbon operator. For a pair of non-Abelian fluxes, the state in the vacuum fusion channel is the symmetric superposition over all representatives in a conjugacy class $C$, i.e.,
\begin{equation}
    \ket{\text{vac}} =\frac{1}{|C|} \sum_{c\in \mathcal{C}}  \ket{c,\bar c}.
\end{equation}

The pairing of $c,\bar c$ ensures flux neutrality since the total flux is $c\cdot \bar c = e$; the symmetrization ensures charge neutrality as any permutation of the terms leave the state invariant, so the state is not charged under any symmetry action (which acts by conjugation). For moving non-Abelian \textit{pure} flux, the procedure is simplified because pure fluxes already satisfy the property that the local internal flux state is a symmetric superposition over all possible basis states. Therefore, we only need to ensure the correct pairing $c,\bar c$ for the internal state of the ribbon operators.

The coherent moving protocol of non-Abelian flux consists of three main steps:
\begin{enumerate}
    \item entangle the ancilla with the local internal state of the flux;
    \item conditioned on the ancilla, apply the corresponding ribbon operator;
    \item disentangle ancilla.
\end{enumerate}

For step 1, a $C_L\mathcal{X}_a$ gate is implemented to entangle the ancilla with the local internal state of the flux. 

For step 2, a $C_2$ ribbon operator is conjugated by $C_a\mathcal{X}_{dir}$, where $dir$ is the first direct edge of the ribbon operator, such that starting local internal state created by the ribbon operator matches with the ancilla state. This ensures that the ribbon operator create the same $C_2$ flux at the starting site to cancel out the flux there.

For step 3, we want to implement a $C_L\mathcal{X}_a$ gate again to disentangling the ancilla from the local internal state of the flux at the start. However, the flux at the start is annihilated to vacuum! The key is to recognize that the flux label \textit{based at the same vertex} remains the same due to the fact that ribbon operator can only create vacuum pair of anyons, i.e. ribbon operator satisfies the neutrality condition by construction~\cite{lo_universal_2025}. Therefore, the control qutrit is the global internal state (based at the starting vertex of the ribbon operator) of the new $C_2$ flux at the ending site.

See Section~\ref{app:pullthrough-imple} and Section~\ref{app:xmeasure-imple} for examples of coherent moving circuit implementations.

\section{Physical encoding and implementation of primitive gates for \ds3}\label{app:s3_encoding_and_circs}
This Section describes our implementation of qudit encoding and gates, realized with the native gate set of Quantinuum H2-series devices \cite{h2-data-sheet, noauthor_quantinuum_2024}. The native gates for the H2 architecture are as follows:
\begin{equation}
    \label{eq:1qgates}
    U_{1q}(\alpha,\beta) = e^{-\frac{1}{2}i\pi \beta Z} \ e^{-\frac{1}{2}i\pi \alpha X} \ e^{+\frac{1}{2}i\pi \beta Z}  ,\quad 
    R_Z(\alpha) = e^{-\frac{1}{2}i\pi\theta Z}\quad  \text{and}
\end{equation}
\begin{equation}
    \label{eq:ZZphase}
    \textrm{ZZPhase}(\theta) = e^{-\frac{1}{2}i\pi\theta (Z\otimes Z)}.
\end{equation}
The number of ZZPhase gates significantly affects the overall implementation cost and serves as a key metric for evaluating gate decompositions.

Each six-dimensional qudit, representing the \ds3 degrees of freedom on the edges of our square lattice, is encoded using a  qutrit and a qubit.  The qutrit, in turn, is encoded using two qubits. Therefore, each qudit requires a total of 3 qubits. The qutrit encoding is defined as:
\begin{align}
    \label{eq:encoding}
    \ket{0}_{\blacktriangle}:=\ket{00}, \quad
    \ket{1}_{\blacktriangle}:=\ket{10}, \quad
    \ket{2}_{\blacktriangle}:=\ket{11}, \quad
    \ket{\textrm{nc}} := |01\rangle.
\end{align}
The non-computational state,  $|\text{nc}\rangle$, is used for error detection; its presence indicates that the two encoding qubits have fallen outside the qutrit subspace.

Qutrits are primarily manipulated using the $\mathcal{X}$, $\mathcal{Z}$, and $\mathcal{C}$ gates, as defined in Eqs.~\eqref{eq:x_and_z_and_C_def}. $\mathcal{X}$ and $\mathcal{Z}$ satisfy the commutation relation $\mathcal{XZ} = \omega \mathcal{ZX}$. The charge-conjugation gate $\mathcal{C}$ is Hermitian, $\mathcal{C}^\dagger=\mathcal{C}$, and conjugates $\mathcal{X}$ and $\mathcal{Z}$:
\begin{align}
    \label{eq:cc_transform}
    \mathcal{C}\mathcal{X} \mathcal{C} = \mathcal{X}^\dagger\quad \mathrm{and} \quad
    \mathcal{C}\mathcal{Z} \mathcal{C} = \mathcal{Z}^\dagger.
\end{align}

The qutrit $\mathcal{Z}$ gate can be efficiently implemented using two single-qubit native $R_Z$ gates:
\begin{align}
    \label{eq:Z}
    \mathcal{Z} &:= 
    \cbox{5}{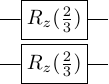} 
    = \bar \omega\ \textrm{diag}(1,\ \omega,\ \omega,\ \bar \omega).
\end{align}
The action of $\mathcal{Z}$ on $|\textrm{nc}\rangle$ is trivial (up to $\omega$ phase). The extra $\bar \omega$ phase needs to be corrected to properly implement a controlled-$\mathcal{Z}$ (C$\mathcal{Z}$) gate.
In contrast, implementing $\mathcal{X}$ requires either two CNOT gates or a single ZZPhase gate:
\begin{align}
    \label{eq:qutrit-Z}
    \mathcal{X} &:= \cbox{30}{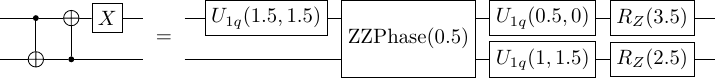} 
    = \begin{pmatrix}
        0 & 0 & 0 & 1\\
        0 & 1 & 0 & 0\\
        1 & 0 & 0 & 0\\
        0 & 0 & 1 & 0\\
    \end{pmatrix}
\end{align}
The charge-conjugation gate $\mathcal{C}$ can also be implemented using only a single ZZPhase gate. 
\begin{align}
    \label{eq:qutrit-C}
    \mathcal{C} &:=
    \cbox{28}{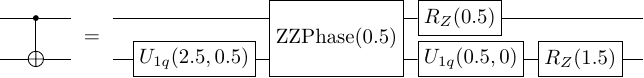} 
    =  \begin{pmatrix}
        1 & 0 & 0 & 0\\
        0 & 1 & 0 & 0\\
        0 & 0 & 0 & 1\\
        0 & 0 & 1 & 0
    \end{pmatrix}
\end{align}
$\mathcal{X}$ can be transformed into $\mathcal{Z}$ gate by performing a qutrit Fourier transform, denoted by $\mathcal{H}$
which acts on the qutrit space as follows:
\begin{align}
    \label{eq:qutrit-H}
    \mathcal{H} \ket{i}_\blacktriangle = \frac{1}{\sqrt{3}} (\ket{0}_\blacktriangle + \omega^i \ket{1}_\blacktriangle + \omega^{2i} \ket{2}_\blacktriangle),
\end{align}	
The native implementation of $\mathcal{H}$ requires three ZZPhase gates, preserving the $\ket{\textrm{nc}}$ state.
Note that we treat the action of $\mathcal{H}$ on $\ket{0}_{\blacktriangle}$ as a \emph{state preparation procedure}, which can be implemented using one ZZPhase gate:
\begin{align}
    \label{eq:state_prep}
    \mathcal{H}\ket{00} &=
    \cbox{40}{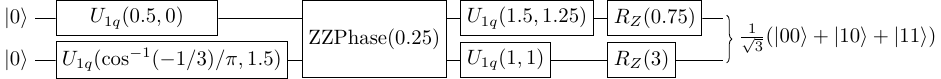}.
\end{align}
Furthermore, $\mathcal{H}$ commutes with $\mathcal{C}$, since $\mathcal{H}^2=\mathcal{C}$.
\\

Similarly, $H_{12}$ diagonalizes $\mathcal{C}$, leaving $\ket{0}_\blacktriangle$ unchanged while applying a ``Hadamard gate'' to both $\ket{1}_\blacktriangle$ and $\ket{2}_\blacktriangle$.  In terms of native gates, $H_{12}$ requires a single ZZPhase gate:
\begin{align}
    \label{eq:qutrit-hc}
    H_{12} &:=
    \cbox{28}{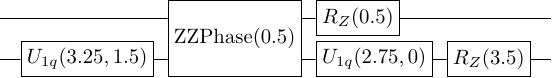} 
    =  \begin{pmatrix}
        1 & 0 & 0 & 0\\
        0 & 1 & 0 & 0\\
        0 & 0 & \frac{1}{\sqrt{2}} & \frac{1}{\sqrt{2}}\\
        0 & 0 & \frac{1}{\sqrt{2}} & -\frac{1}{\sqrt{2}}
    \end{pmatrix}.
\end{align}
Since $\mathcal{C}$ squares to the identity, the C$\mathcal{C}$ gate is defined with a qubit control and qutrit target. To avoid applying three qubit CC$Z$ gate to implement C$\mathcal{C}$, we use the fact that $H_\mathcal{C}$ disentangles the $\mathcal{C}$ gate, i.e., $H_{12}\ \mathcal{C}\ H_{12}^\dagger = IZ$. The implementation of C$\mathcal{C}$ requires three ZZPhase gates:
\begin{align}
    \label{eq:CC}
    \text{C}\mathcal{C} &:=
    \cbox{16}{circs_primitive/cc.pdf} 
\end{align}
The application of C$\mathcal{Z}$ requires four C$R_z$ gates:
\begin{align}
    \label{eq:qutrit-CZ}
    \textrm{C}\mathcal{Z} &:=
    \cbox{28}{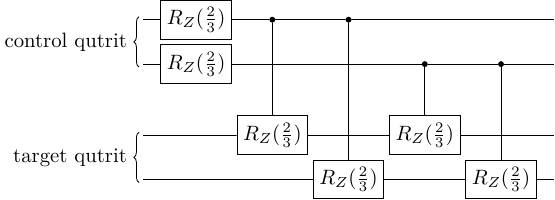}, 
\end{align}
where each of the C$R_z$ gate can be implemented using a single ZZPhase gate. The initial single qubit $R_z$ gates on control qutrit compensate for the extra phase $\bar \omega$ in Eq.~\eqref{eq:qutrit-Z}. Applying C$\mathcal{Z}$ while transforming the target qutrit by $\mathcal{H}$ yields the C$\mathcal{X}$ gate:
\begin{align}
    \label{eq:qutrit-CX}
    \textrm{C}\mathcal{X} &:=
    \cbox{9}{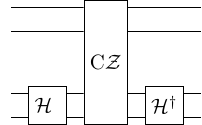} 
\end{align}
The C$\mathcal{X}$ gate requires a total of $3+4+3=10$ ZZPhase gates.

We define a $\textrm{C}_2\textrm{NOT}$ gate with a qutrit as the control and a qubit as the target.
\begin{align}
    \label{eq:qubit-CNOT}
    \textrm{C}_2\textrm{NOT} &:=
    \cbox{13}{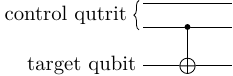} 
\end{align}
It acts trivially on the target qubit when the control qutrit is in the $\ket{0}$ or $\ket{1}$ state, and flips the target qubit when the control is in the $\ket{2}$ state. It is implemented by a single CNOT gate conditioned on the second physical qubit of the control qutrit, because in the physical encoding of the qutrit (see Eq.~\eqref{eq:encoding}), the second physical qubit is in the $\ket{1}$ state only for the $\ket{2} = \ket{11}$ qutrit state. 

Moreover, we define the C$X_{12}$ gate with a qutrit control and a qutrit target. 
\begin{align}
    \label{eq:CX12}
    \textrm{C}X_{12} &:=
    \cbox{18}{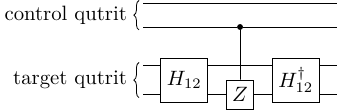} 
\end{align}
The gate acts as the identity when the control qutrit is in the $\ket{0}$ or $\ket{1}$ state. When the control is in the $\ket{2}$ state, it swaps the target states $\ket{1}$ and $\ket{2}$ while leaving $\ket{0}$ invariant.

\section{Circuit Optimization by Local Basis Transformation}\label{sec:app-local-basis}
For each experiment, before compiling the circuit unitary $U$ into native gates using TKET~\cite{Sivarajah_2021}, we locally transform every qutrit to the dual ($\mathcal{X}$) basis by applying $\mathcal{H}$, as defined in Eq.~\eqref{eq:qutrit-H}, and construct $\tilde{U}$ such that
\begin{align}
\tilde{U} = {{\mathcal{H}}^{\otimes n}}^{\dagger}\, U \,{\mathcal{H}}^{\otimes n},
\end{align}
where \(n\) is the number of qutrits. We then compile and execute $
\mathcal{H}^{\otimes n}\,\tilde{U}\,\mathcal{H}^{\otimes n}
$ instead of compiling and executing \(U\) directly. This local basis transformation reduces the number of gates after compilation and, empirically, also reduces memory errors. For comparison,	
\begin{align}
    \label{eq:dual-basis}\nonumber
    \mathcal{X}\textrm{-basis measurement:}\ 
    \renewcommand{\arraystretch}{1.2}
    \begin{tabular}{r}
        $n_{2q}$ gates = 792\\
        depth$_{2q}$ = 261
    \end{tabular}   
    & \xrightarrow[\textrm{transformation}]{\textrm{local basis}} 
    \renewcommand{\arraystretch}{1.2}
    \begin{tabular}{r}
        744\\
        232
    \end{tabular}\textrm{, and}\\
    \textrm{Pull-through gate:}\ 
    \renewcommand{\arraystretch}{1.2}
    \begin{tabular}{r}
        $n_{2q}$ gates = 911\\
        depth$_{2q}$ = 354
    \end{tabular}   
    & \xrightarrow[\textrm{transformation}]{\textrm{local basis}} 
    \renewcommand{\arraystretch}{1.2}
    \begin{tabular}{r}
        845\\
        307
    \end{tabular}.
\end{align}	
It is straightforward to compute $\tilde{U}$ because $\mathcal{H}$ commutes with the charge-conjugation gate $\mathcal{C}$ and maps a C$\mathcal{X}$ gate to another C$\mathcal{X}$ gate with the control and target qutrits swapped.

\section{Ground State Preparation}
\label{app:gs_prep}
This section details the methodology for preparing the ground state of the \ds3 model. The preparation strategy we use follows the method given in the appendix of Ref.~\onlinecite{verresen2021efficiently} and consists of a multi-step procedure. The first step is the preparation of the $\mathbb{Z}_3$ toric code ground state. Subsequently, this Abelian state is transformed into the desired non-Abelian \ds3 order by gauging its charge-conjugation symmetry. A detailed exposition of these steps follows.

\subsection{Preparation of $\mathbb{Z}_3$ toric code ground state}\label{subsec:z3_unitary_prep}
The ground state of the $\mathbb{Z}_3$ toric code with qutrits on the edges of the square lattice is stabilized by,
\begin{align}
    \label{eq:z3_stabilizer}
    S_\mathcal{A} = 
    \cbox{6.0}{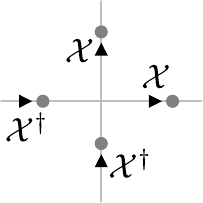}\ \;,\; \quad\quad &  
    S_\mathcal{B} = 
    \cbox{5.5}{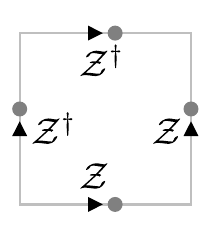}\;.
\end{align}
While these stabilizers can be prepared in constant depth using an adaptive circuit with mid-circuit measurements and ancilla qutrits, this approach incurs a significant qubit overhead. On the Quantinuum System Model~H2, which has 56 qubits, preparing the \ds3 ground state on a $3\times3$ lattice already requires 54 qubits, leaving little room
for additional ancillae. We therefore employ a purely unitary preparation protocol that requires
neither mid-circuit measurements nor ancillae for the $\mathbb{Z}_3$ toric code state preparation itself.
The protocol is described below.

\begin{figure}[!h]
    \centering
    \includegraphics[width=1\linewidth]{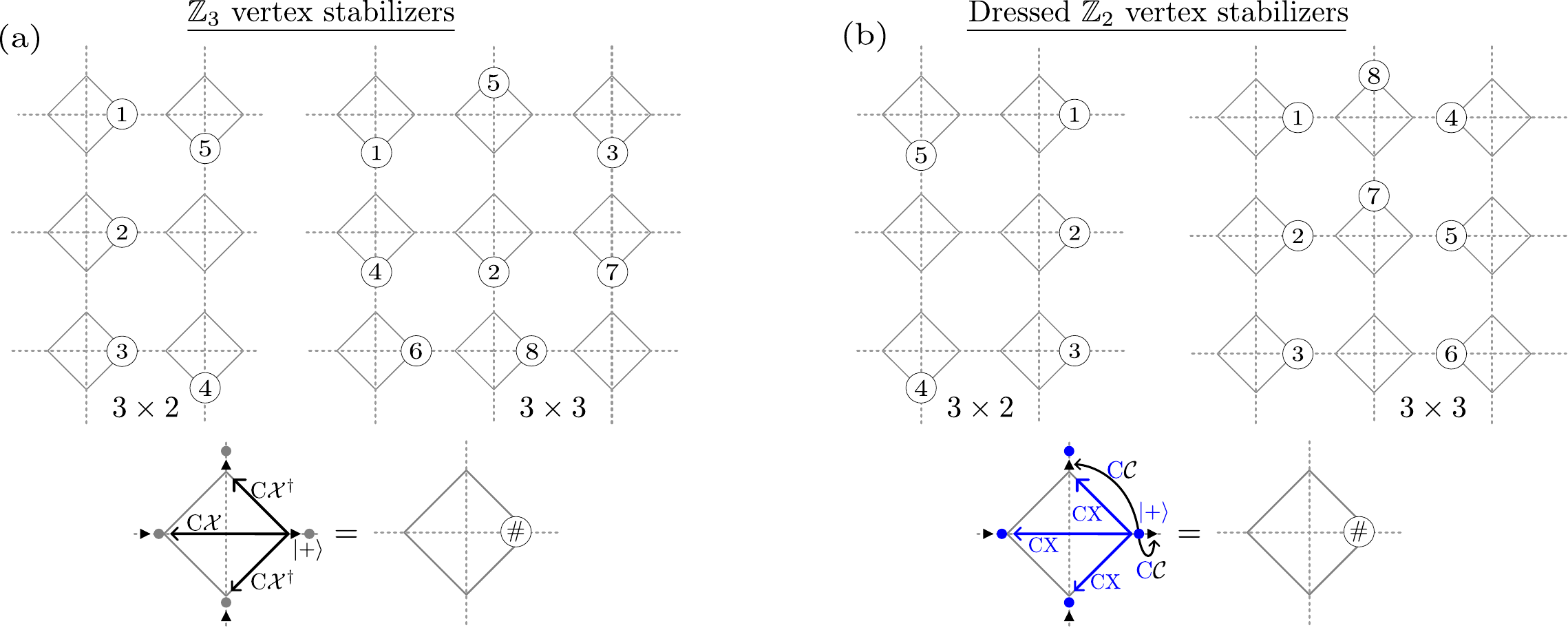}
    \caption{\textbf{Vertex Stabilizers Unitary Preparation.} 
        Ordering for preparing vertex stabilizers in (a) the $\ZZ_3$ toric code and (b) the dressed $\ZZ_2$ toric code, for the $3\times 2$ and $3\times3$ lattices. The dashed lines outline the square lattice. Rotated squares at each vertex represent the $S_\mathcal{A}$ stabilizers. Numbered edges denote the control qutrits, and the sequence of numbers indicates the order in which these vertex stabilizers are prepared.}
    \label{fig:z3z2_prep_order}
\end{figure}

\begin{enumerate}
    \item All qutrits are initialized in the state $\ket{0}$, which trivially satisfies all plaquette stabilizers $S_\mathcal{B} = +1$.
    
    \item To satisfy the vertex stabilizers $S_\mathcal{A} = +1$, we proceed in an iterative manner:
    \begin{enumerate}
        \item[(i)] Select a vertex stabilizer and designate one of its qutrits as the``representative.'' The representative qutrit is prepared in the state $\ket{+}_{\blacktriangle} := \frac{1}{\sqrt{3}}(\ket{0}_{\blacktriangle}+ \ket{1}_{\blacktriangle} + \ket{2}_{\blacktriangle}),$
        using the state-preparation circuit of Eq.~\eqref{eq:state_prep}.
        
        \item[(ii)] Apply a sequence of $\mathrm{C}\mathcal{X}$ or $\mathrm{C}\mathcal{X}^\dagger$ gates from the representative qutrit to the remaining qutrits within the same vertex stabilizer. A $\mathrm{C}\mathcal{X}$ gate is applied when both the control and target qutrits are acted on by $\mathcal{XX}$ or $\mathcal{X}^\dagger \mathcal{X}^\dagger$ in the vertex stabilizer; otherwise, a $\mathrm{C}\mathcal{X}^\dagger$ gate is applied.
    \end{enumerate}
    This procedure is repeated for all vertex stabilizers except one, ensuring that a qutrit is selected as a representative only if it is in the $\ket{0}$ state and has not been acted upon previously. The final vertex stabilizer is implicitly satisfied due to the global constraint $\prod_v S_{\mathcal{A}_v} = 1$. The vertex preparation order used to obtain the $\ZZ_3$ toric code ground state on $3\times 2$ and $3\times 3$ lattices is illustrated in Fig.~\ref{fig:z3z2_prep_order}a.
   
\end{enumerate}
The resulting state is the logical ground state $\ket{00}_L$ of the $\mathbb{Z}_3$ toric code. To prepare the logical state $\ket{++}_L$, where both logical $\mathcal{X}_L$ operators have eigenvalue $+1$, a dual procedure is employed. All qutrits are first initialized in the $\ket{+}_{\blacktriangle}$ state, satisfying all vertex stabilizers $S_\mathcal{A} = +1$. Then, plaquette stabilizers $S_\mathcal{B} = +1$ are enforced using a protocol analogous to steps (i) and (ii) above.

\subsection{Gauging the $\ZZ_2$ charge-conjugation symmetry}
The $\mathbb{Z}_3$ toric code Hamiltonian possesses a global $\mathbb{Z}_2$ charge-conjugation symmetry, denoted $\prod{\mathcal{C}}$ (see Eqs.~\eqref{eq:cc_transform} and \eqref{eq:z3_stabilizer}). Gauging this global $\mathbb{Z}_2$ symmetry means promoting it to a local gauge symmetry, where an independent $\mathbb{Z}_2$ transformation can be performed locally at each ``site''. This procedure \emph{enriches} the topological order, transforming the $\mathbb{Z}_3$ toric code into the \ds3 quantum double model \cite{verresen2021efficiently, tanti2023hierarchy}.

\begin{figure*}[t]
		\centering\includegraphics[width=0.6\linewidth]{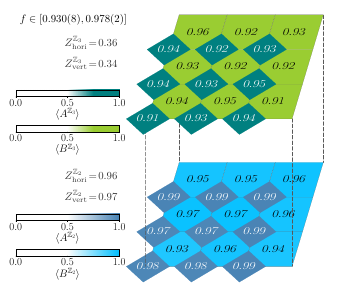}
		\caption{
        \textbf{Ground state preparation using mid-circuit measurements and feed-forward.} Expectation values of local and $Z$-logical projectors are shown. The $\ZZ_3$ toric-code ground state is prepared unitarily by enforcing plaquette stabilizers (in contrast to the vertex stabilizers used in the main-text experiments), yielding a +1 eigenstate of the $\ZZ_3$ $X$-logical operators. This is verified by $Z^{\ZZ_3}_{L,\text{hori}}$ and $Z^{\ZZ_3}_{L,\text{vert}}$ values near 1/3. The $\ZZ_2$ charge-conjugation symmetry is then gauged using an adaptive circuit (see Section~\ref{app:adaptive-prep}) with mid-circuit measurements and conditional error correction based on stabilizer outcomes. Given that $\Pi$ is the projector onto the ground-state manifold, $f = \Tr(\rho \ \Pi)^{1/18}$ denotes the per-qudit fidelity of the prepared state $\rho$. The average (maximum) standard error for local projectors is 0.012 (0.017), and for $Z$-logical projectors the average and maximum standard errors are 0.019 and 0.028, respectively.
        }\label{fig:appendix_gs_measurement_based}
\end{figure*}

The gauging process can be implemented via an adaptive quantum circuit involving measurements and feed-forward, or via a purely unitary circuit composed of a deterministic sequence of gates. For the experiments presented in the main text,
we prepare the \ds3 ground state using a unitary protocol. The data for the \ds3 ground state prepared using an adaptive circuit is shown in Fig.~\ref{fig:appendix_gs_measurement_based}.

\subsubsection{Using adaptive circuit}\label{app:adaptive-prep}
The gauging protocol augments the Hilbert space and enforces new local constraints:
\begin{enumerate}
    \item Introduce an auxiliary qubit at each vertex of the lattice initialized in the $\color{red}\ket{+}$ state, stabilized by {\color{red}{$X$}}.

    \item Apply a layer of ${\color{red}\mathrm{C}}\mathcal{C}$ gates that entangle the original $\ZZ_3$ qutrits with the new $\ZZ_2$ vertex qubits. Specifically, for each vertex, we apply two ${\color{red}\mathrm{C}}\mathcal{C}$ gates, where $\ZZ_2$ vertex qubit acts as the control and the right and top neighboring $\ZZ_3$ edge qutrits act as targets. 
    \begin{align}
        \label{eq:cc_guaging_1}
        \cbox{9}{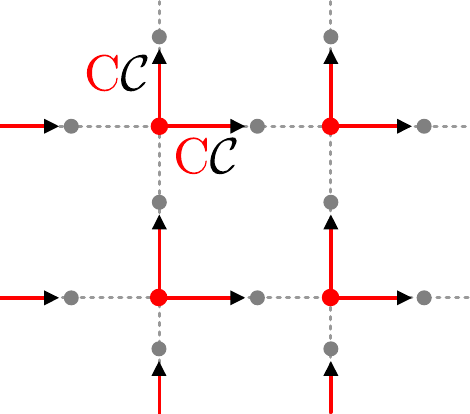}
    \end{align}
    Stabilizers of the joint state are now given by
    \begin{align}
        \label{eq:cc_gauging_2}
        \cbox{26}{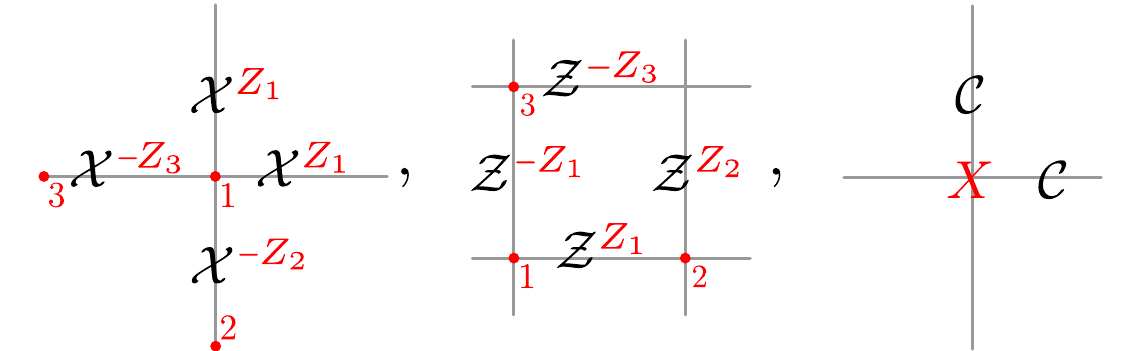}.
    \end{align}
    
    \item We further add a qubit on each edge initialized in the $\color{blue}\ket{0}$ state, stabilized by {\color{blue}$Z$}. We apply {\color{red}C}$\color{blue}X$ gates between vertex and neighboring qubits with vertex qubits as the control and edge qubits as the target.
    \begin{align}
        \label{eq:cc_gauging_3}
        \cbox{8.5}{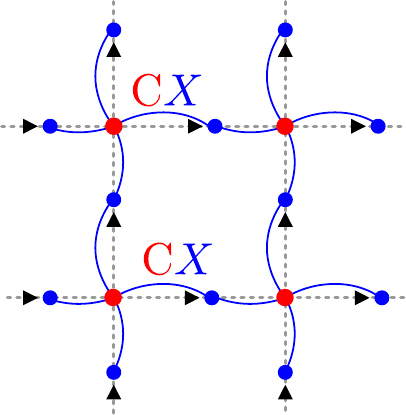}
    \end{align}
    The expanded set of stabilizers for the state augmented with edge qubits is given by
    \begin{align}
        \label{eq:cc_gauging_4}
        \cbox{40}{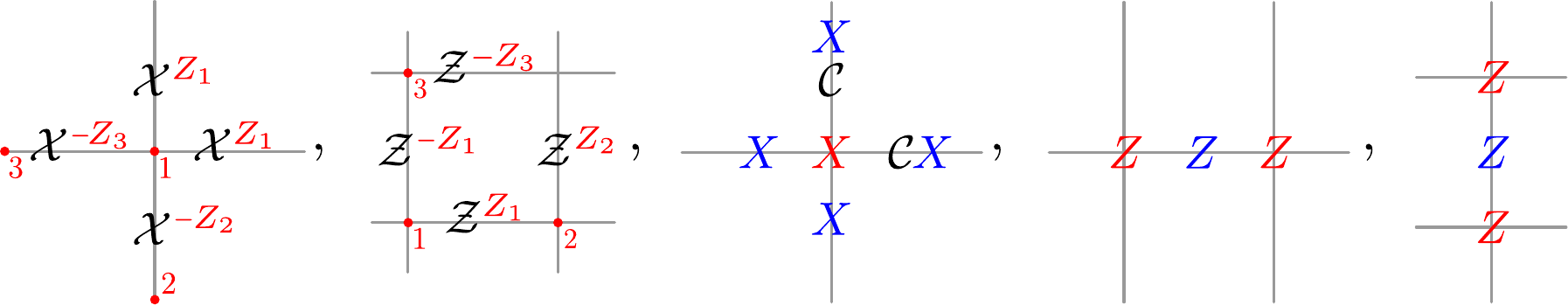}.
    \end{align}
    These stabilizers of the state do not all commute with each other. However, by writing these stabilizers as projectors over the trivial (+1) eigenspace, we obtain a set of commuting projectors with a \emph{simpler} form (see~\cite{verresen2021efficiently} for details)
    \begin{align}
        \label{eq:cc_gauging_5}
        \cbox{28}{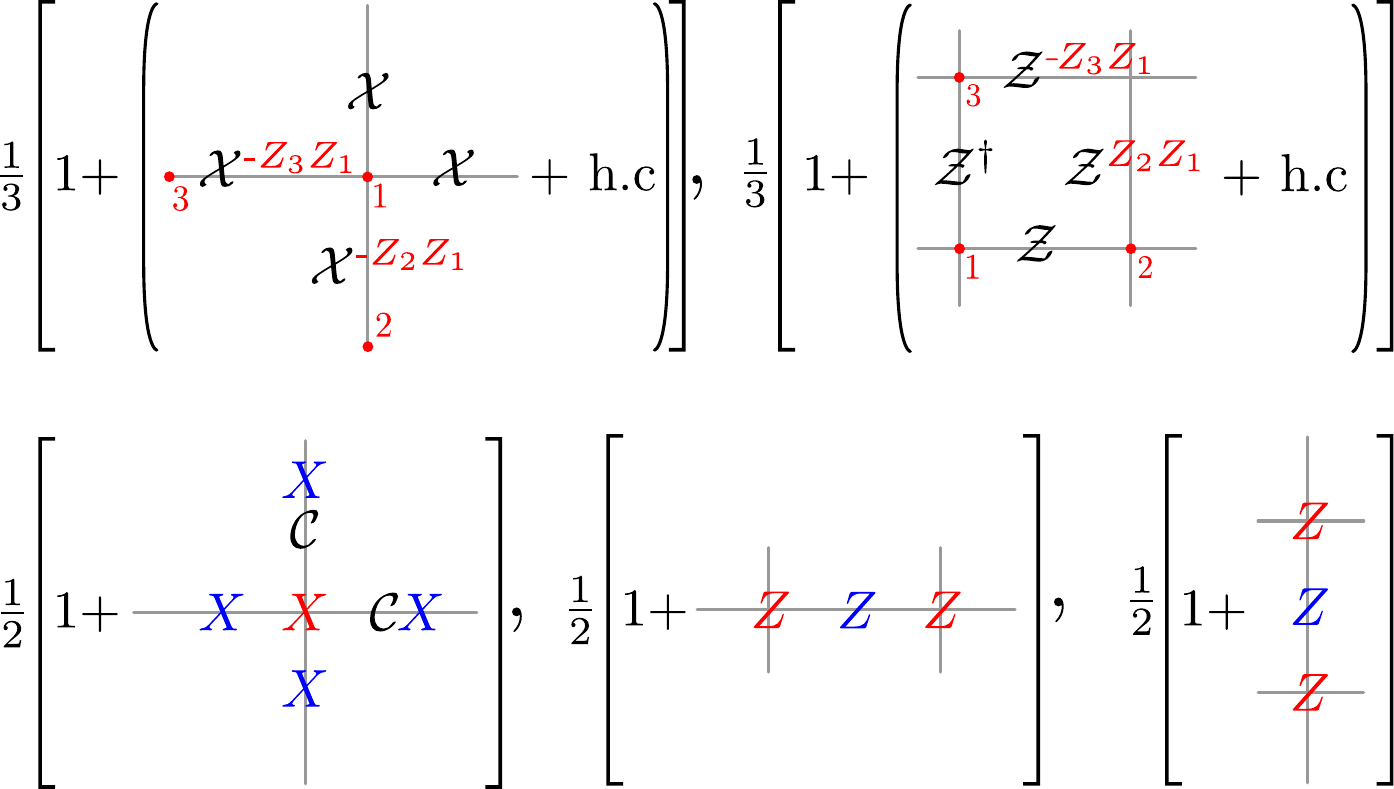}
    \end{align}	
    Moreover, the vertex qubit dependence of $\ZZ_3$ projectors can be eliminated by writing vertex qubit variables in terms of edge qubits as determined by {\color{red}C}$\color{blue}X$ gates between vertex and edge qubits.
    \begin{align}
        \label{eq:cc_gauging_6}
        \cbox{28}{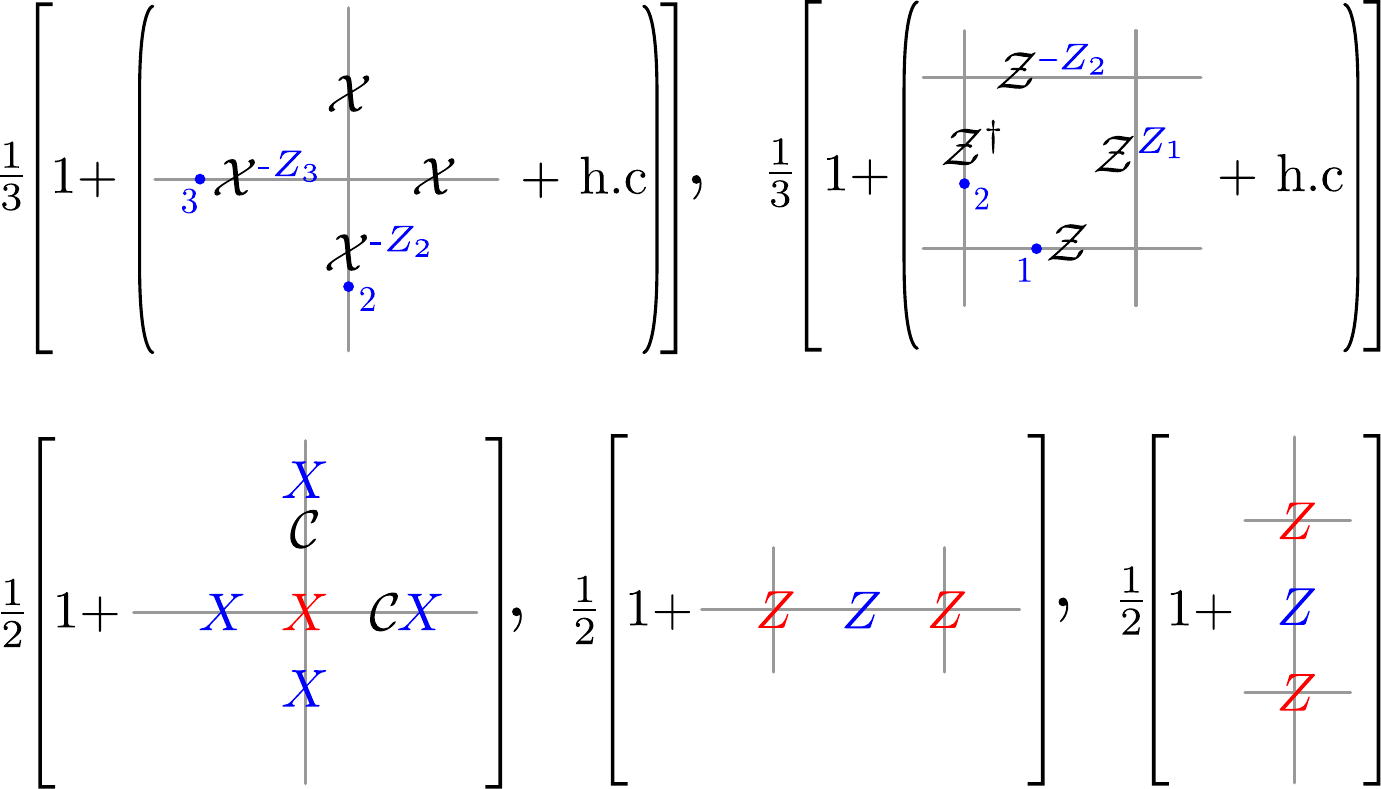}
    \end{align}	
    \item Finally, we measure vertex qubits in $X$-basis and enforce local $\ZZ_2$ gauge invariance. This projects the state into the subspace where the original $\ZZ_3$ vertex constraints are now tied to the state of the $\ZZ_2$ gauge qubits. These measurements yield $\pm 1$ outcomes. Based on the outcomes, we apply corrective $Z$'s to the $\ZZ_2$ edge qubits. This feed-forward ensures that the system is deterministically prepared in the $+1$ eigenspace of all the projectors, namely
    \begin{align}
        \label{eq:cc_gauging_7}
        \cbox{32}{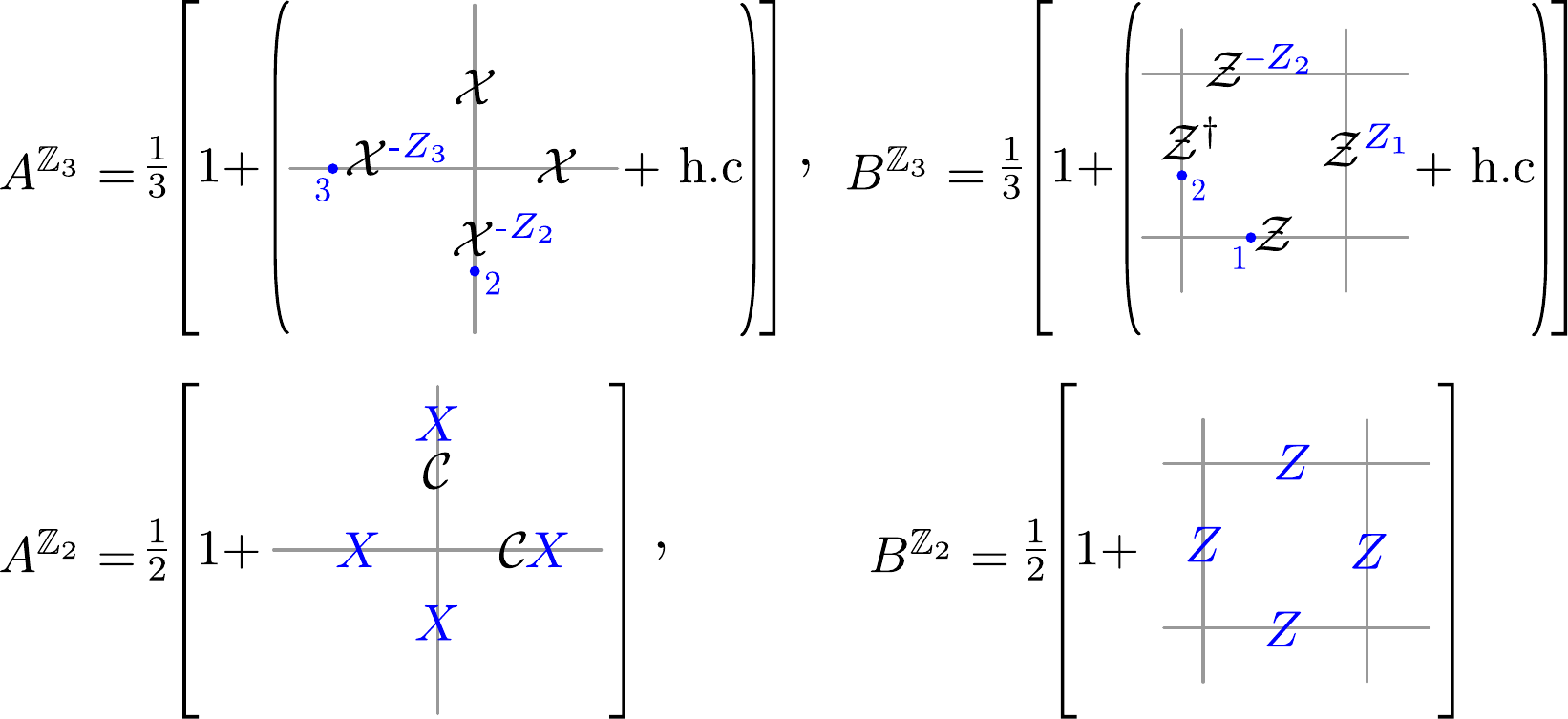}
    \end{align}
    
\end{enumerate}
The resulting projectors can be shown to be equivalent to the canonical quantum double projectors (see Section~\ref{app:z2z3-decomposition}) and the state is the ground state of the \ds3 model. While this method is optimal in terms of circuit depth, its reliance on mid-circuit measurements, feed-forward, and ancilla qubits affects the fidelity of the prepared state.

\subsubsection{Using unitary circuit}\label{app:unitary-prep}
We developed a purely unitary protocol that requires no measurements or feed-forward. The charge-conjugation gauging procedure, as described above, effectively amounts to measuring \emph{dressed} vertex stabilizers of the $\ZZ_2$ toric code using an ancilla, after preparing the $\ZZ_3$ toric code, which is given as	
\begin{align}
    \label{eq:unitary_gauging_dressed_stabilizer}
    \cbox{6}{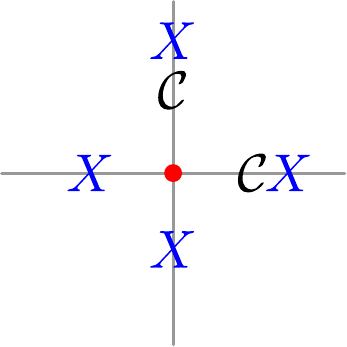}
\end{align}
We can omit the need for introducing the ancilla (red vertex) qubit in the stabilizer measurement by using the following circuit identity.
\begin{align}
    \label{eq:a_unitary_circuit_identity}
    \cbox{32}{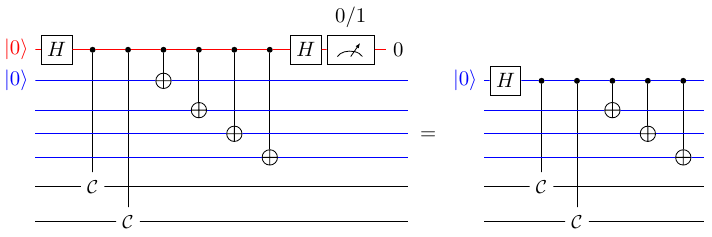}
\end{align}
where blue and red lines correspond to qubits, and black lines correspond to qutrits. This identity shows that a circuit involving an ancilla measurement (left) can be replaced by a purely unitary sequence of gates (right), provided that one of the target qubits is initialized in the $|0\rangle$ state. This constraint forces the preparation of the dressed $\ZZ_2$ vertex stabilizers to be performed sequentially, similar to the unitary preparation of the $\ZZ_3$ toric code ground state in Section~\ref{subsec:z3_unitary_prep}. We select a representative qubit, prepare it in the qubit $|+\rangle$ state, and use it as a control to entangle its neighbors. This process is repeated for all vertices in a specific order, shown in Fig.~\ref{fig:z3z2_prep_order}b, ensuring that the final state is the ground state of the \ds3 model.

\section{$C_2$ and $C_3$ Flux Insertion and Single Non-Abelian Anyon}
\label{sec:single-anyon}
This section details the quantum circuits required to prepare the ground states of the \ds3\ model threaded with non-trivial topological flux. Specifically, we focus on the $C_2$ and $C_3$ flux sectors. Furthermore, we outline the steps for preparing a single $C_3$ non-Abelian anyon, a capability that is fundamentally impossible in Abelian topological phases.

To access the non-trivial $C_2$ and $C_3$ sectors, we employ the symmetry un-gauging and re-gauging strategy described in Ref.~\onlinecite{lyons-2025}. Construction circuits are provided for a $3\times 3$ lattice on a torus.
\begin{figure}[!ht]
	\centering
	\includegraphics[width=0.25\linewidth]{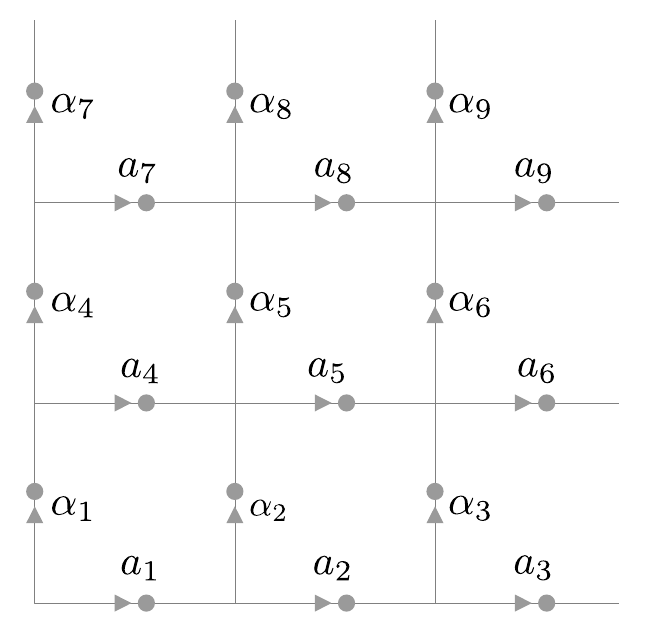}
	\caption{\textbf{3$\times$3 square lattice with labeled edges} }
	\label{fig:3x3_with_labels}
\end{figure}
\subsection{$C_3$ flux}	
The preparation of a ground state with $C_3$ flux, for instance $\ket{\text{gs}_{C_3,\textrm{hori}}}$, where the flux is threaded through bonds in the horizontal direction, involves:
\begin{itemize}
    \item Ungauging the $\ZZ_2$ charge-conjugation symmetry.
    \item Ungauging the $\ZZ_3$ symmetry to access the `paramagnet' qutrits.
    \item Applying a sequence of gates along a non-trivial horizontal loop to prepare the qutrits in the state $\frac{1}{\sqrt{2}}(\ket {11\ldots}_{\!\blacktriangle} + \ket {22\ldots}_{\!\blacktriangle})$.
    \item Re-gauging the $\ZZ_2$ and $\ZZ_3$ symmetries to return to the \ds3 model with the desired $C_3$ flux.
\end{itemize}
The following is the optimized circuit for preparing $\ket{\text{gs}_{C_3,\textrm{hori}}}$ on a $3\times3$ lattice (see Fig.~\ref{fig:3x3_with_labels} for qudit labels) by applying it to the trivial ground state, $\ket{\text{gs}}$.
\begin{align}
    \label{eq:single_anyon_C3}
    U_{C_3,\textrm{hori}} &= \cbox{38}{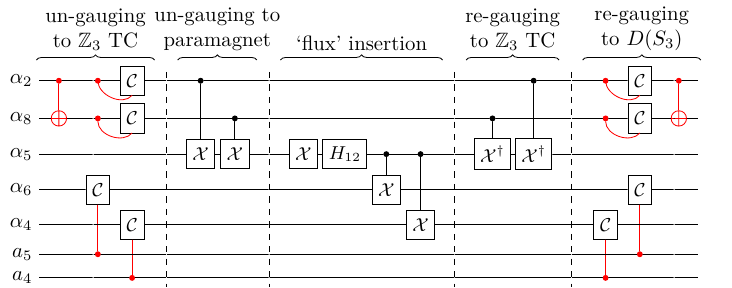},
\end{align}
In the circuits above, each black line represents a qubit–qutrit pair encoding an $S_3$ qudit. Gate actions on the qubit lines are shown in red, and the gates act as follows:
\begin{itemize}
    \item CNOT acts on the qubit lines of the control and target qudits.
    \item C$\mathcal{X}$ acts on the qutrit lines of the control and  target qudits.
    \item C$C$ acts on the qubit line of the control and qutrit line of the target qutrit. Hence, the "self-loop" C$C$ gates at the un-gauging and re-gauging step act on distinct qubit and qutrit lines of a qudit.
    \item $\mathcal{X}$ and $H_{12}$ act on the qutrit degrees of freedom of the qudits.
\end{itemize}
For detailed definitions and physical implementations, see Section~\ref{app:s3_encoding_and_circs}.

\subsection{$C_2$ flux}
The construction of the ground state with $C_2$ flux requires ungauging to the $\ZZ_3$ toric code and then applying the projector $\frac{1}{3}(II\ldots + \mathcal{XX}\ldots + \mathcal{X}^\dagger \mathcal{X}^\dagger\ldots)$ across a non-trivial loop on qutrits, , as well as $X$ gates on the qubits. This requires an ancilla qutrit and complex feed-forward operations for deterministic preparation, as shown below, which creates the ground state $\ket{\text{gs}_{C_2, \textrm {vert}}}$ by acting on horizontal bonds across a non-trivial loop in the vertical direction.
\begin{align}\label{eq:single_anyon_C2}
    U_{C_2,\textrm{vert}}\! &=\!\!\!\!\! 
    \cbox{44}{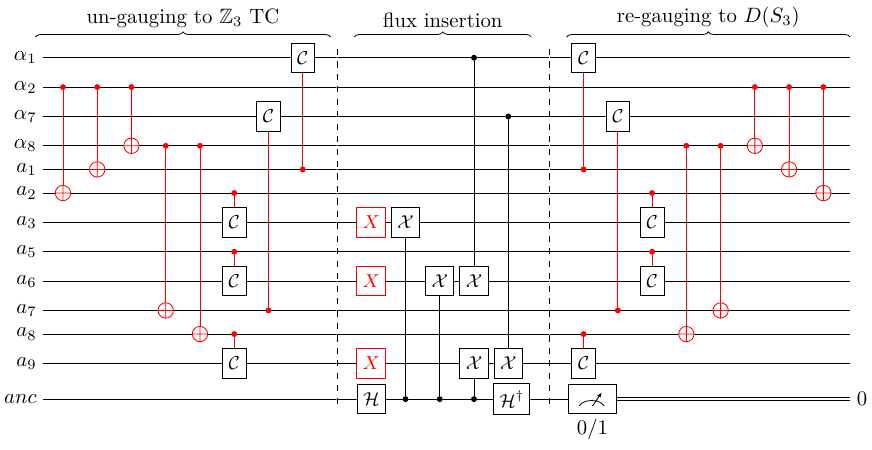}
\end{align}
To simplify and avoid mid-circuits measurement, we constructed an optimized initialization circuit. The key insight is that the flux insertion operation in the above circuit can be commuted through the $\ZZ_3$ gauging unitary used in the standard \ds3 state preparation. This shifts the flux-insertion step to an earlier stage, where it acts on a simple product state, thereby eliminating the need for an ancilla.
The resulting initialization circuit, 
\begin{align}\label{eq:single_anyon_C2_initialization}
    U_{C_2,\textrm{vert}}^{\textrm{init.}} &= 
    \cbox{22}{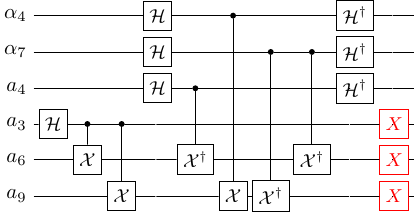}\quad ,
\end{align}
is applied to the qudits \textit{before} the main \ds3 preparation unitary (see Section~\ref{app:s3_encoding_and_circs}). This circuit effectively implements the $+1$ post-selected projector by using one of the qutrits as a control (i.e. $a_4$ in Eq.~\eqref{eq:single_anyon_C2_initialization}) for a series of C$\mathcal{X}$ gates. The final \ds3 ground state prepared after this initialization has $C_2$ flux. Note that the correctness of this optimized initialization circuit depends on the ordering of plaquettes used during the $\ZZ_3$ toric code preparation (see Fig.~\ref{fig:z3z2_prep_order}).

\subsection{Single $C_3$ flux anyon on the torus}\label{app-subsec:single-C3}

Distinct, non-commuting flux sectors enable the creation of a single non-Abelian anyon through a mechanism unique to non-Abelian topological order. When a non-Abelian flux pair is created in a ground state that already carries a different non-trivial flux, braiding between the two flux types changes the fusion channel of the pair. Consequently, fusing the pair does not return the system to the vacuum but instead leaves behind a single isolated anyon.

\begin{figure}[!ht]
	\centering
	\includegraphics[width=0.62\linewidth]{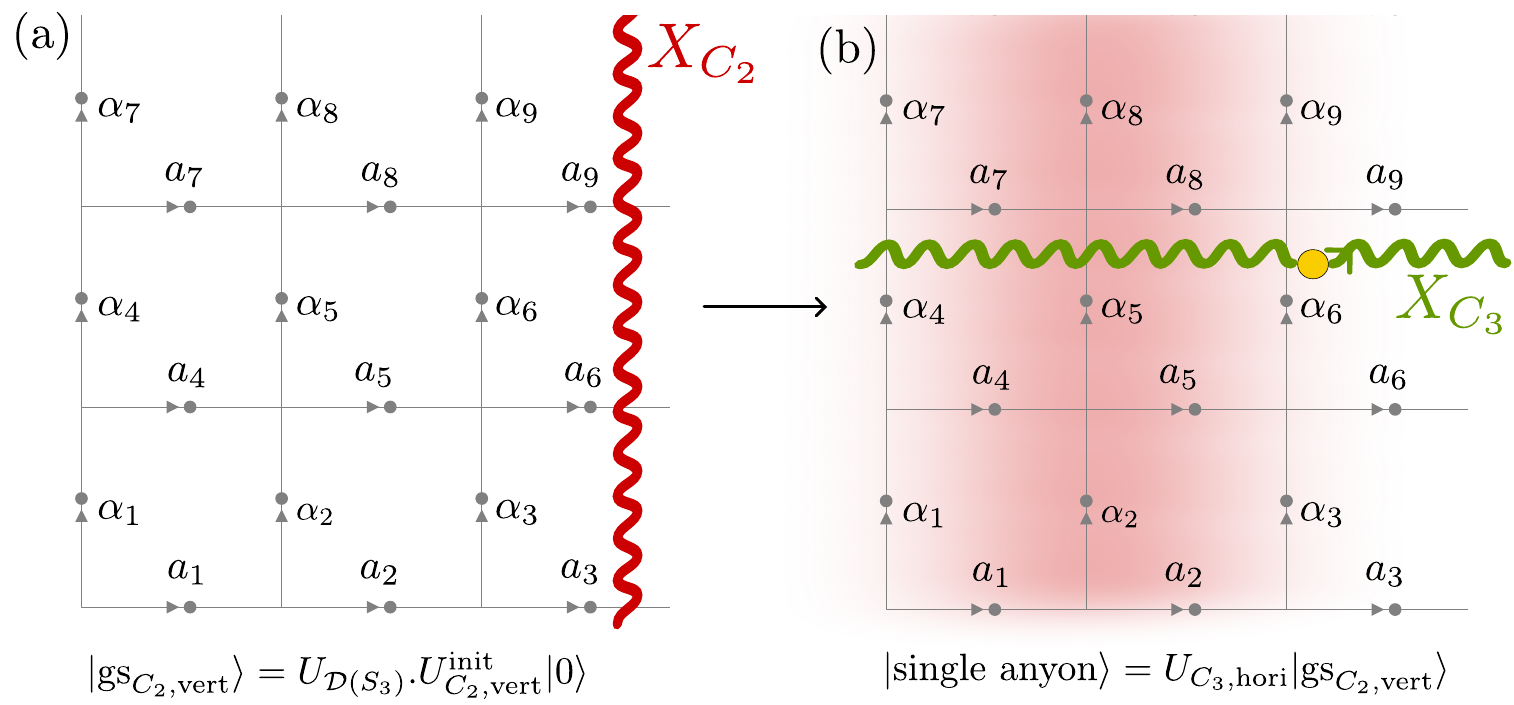}
	\caption{\textbf{Steps to create a single $C_3$ anyon} }
	\label{fig:single_anyon_steps}
\end{figure}

In our implementation, we begin from the ground state $\ket{\mathrm{gs}_{C_2,\mathrm{vert}}}$ (see Fig.~\ref{fig:single_anyon_steps}a) and create a pair of $C_3$ fluxes that encircle the torus non-trivially along the horizontal direction (see Fig.~\ref{fig:single_anyon_steps}b). Because the $C_2$ and $C_3$ fluxes braid non-trivially, the $C_3$ pair is forced out of the vacuum fusion channel. Upon fusion, a remnant $C_3$ non-Abelian anyon with quantum dimension $2$ remains localized at the fusion site, consistent with the fusion rule
\begin{equation}
	C_3 \times C_3 = [+] + [-] + C_3.
\end{equation}

This process can also be understood at the level of the $\ZZ_3$ toric code by ungauging the $\mathbb{Z}_2$ charge-conjugation symmetry. In this picture, the $C_3$ flux pair becomes the superposition of $m$ and $\bar{m}$ anyons in the state $\ket{m}_1\ket{\bar m}_2 + \ket{\bar m}_1\ket{m}_2$, while the $C_2$ flux becomes a charge-conjugation defect~\cite{iqbal_qutrit_2025}. The $C_2$ non-contractible loop becomes a charge-conjugation domain wall, which toggles $m \leftrightarrow \bar{m}$. So moving the second particle of the pair across the charge-conjugation domain wall maps the state $\ket{m}_1\ket{\bar m}_2 + \ket{\bar m}_1\ket{m}_2 \to \ket{m}_1\ket{m}_2 + \ket{\bar m}_1\ket{\bar m}_2$. Finally, fusing particle 1 and 2 together gives $\ket{m}+\ket{\bar{m}}$; this is why the remnant particle is a $C_3$ flux. We note that the location of the single anyon excitation is where the fusion happened, not where the $C_3$ flux crosses the support of the $\hat{X}_{C_2}$ operator.

We can also create a single non-Abelian $[2]$ charge on the torus. One way to see it is by replacing the $m,\bar{m}$ pair with $e,\bar{e}$ pair while keeping the charge-conjugation domain wall the same. Running through the same argument as above, we obtain the state $\ket{e}+\ket{\bar{e}}$. Upon regauging the charge-conjugation symmetry, $\ket{e},\ket{\bar{e}}$ becomes internal states of the $[2]$ charge, thus we obtain a single $[2]$ charge. Another more efficient perspective is that $S_3$ quantum double admits a $\mathbb{Z}_2$ anyon automorphism duality that interchange $C_3$ flux with $[2]$ charge (this is inherited from the EM duality at the level of the qutrit toric code). Therefore by applying this duality transformation to the single $C_3$ flux state, we obtain the single $[2]$ charge state.

\section{Fidelity Bounds for \ds3 Ground State}
\label{app:fidelity_s3}
This section establishes lower and upper bounds on the fidelity of the 
prepared state, $\rho$, with respect to the target ground state $|\text{gs}\rangle$.
The ground state is defined as the +1 eigenstate of the local vertex projectors ($A_v^{\mathbb{Z}_2}, A_v^{\mathbb{Z}_3}$) and local plaquette projectors ($B_p^{\mathbb{Z}_2}, B_p^{\mathbb{Z}_3}$). 
Furthermore, within the ground state manifold, $|\text{gs}\rangle$ is uniquely identified by the +1 eigenspace of the horizontal and vertical non-contractable $S_3$ flux projectors, $Z_{\textrm{hori}}$ and $Z_{\textrm{vert}}$ (see Section~\ref{app:z2z3-decomposition}). We can express the projector onto the ground state as:
\begin{equation}
    \label{eq:state_projector}
    \ket{\text{gs}}\!\bra{\text{gs}} = 
    \underbrace{
        \prod_{v} {A_v^{\mathbb{Z}_2}}
    }_{:= P_A^{\mathbb{Z}_2}} .\ 
    \underbrace{
        \prod_{v} {A_v^{\mathbb{Z}_3}}
    }_{:= P_A^{\mathbb{Z}_3}} .\ 
    \underbrace{
        \prod_{p} {B_p^{\mathbb{Z}_2}} .\ 
        \prod_{p} {B_p^{\mathbb{Z}_3}} .\
        Z_{\textrm{hori}} .\  Z_{\textrm{vert}}
    }_{:= P_B}
\end{equation}
While all projectors commute, allowing for simultaneous measurement, we group them into `product' projectors $P_A^{\mathbb{Z}_2}$, $P_A^{\mathbb{Z}_3}$, and $P_B$. These groupings are chosen to be as large as possible, subject to the constraint that their destructive measurement at the end of the circuit incurs minimal gate overhead.

For the experimentally prepared density matrix $\rho$, the fidelity with respect to the target state $\ket{\text{gs}}$ is:
\begin{align}\label{eq:fidelity_exact}
    \bra{\text{gs}}\rho\ket{\text{gs}} &= \text{Tr}\left(\rho  \ket{\text{gs}}\!\bra{\text{gs}}  \right) \nonumber \\
    &= \text{Tr}\left(\rho \ P_A^{\mathbb{Z}_2} \ P_A^{\mathbb{Z}_3} \ P_B \right)
\end{align}
We define the expectation values of the product projectors as:
\begin{align}\label{eq:projector_exp_vals}
    \begin{split}
        p_A^{\mathbb{Z}_2} &:= \text{Tr}(\rho \ P_A^{\mathbb{Z}_2} ) \\
        p_A^{\mathbb{Z}_3} &:= \text{Tr}(\rho \ P_A^{\mathbb{Z}_2} ) \\
        p_B &:= \text{Tr}(\rho \ P_B )
    \end{split}
\end{align}
To derive a lower bound for the fidelity, we use the following inequality which holds for a set of $n$ commuting projectors $\{P_i\}$:
\begin{equation}
    \prod_{i=1}^n P_i \geq \left(\sum_{i=1}^n P_i\right) - (n-1)\mathbb{I}.
    \label{eq:projector_inequality}
\end{equation}
Taking the trace of both sides with the density matrix $\rho$, and applying Eqs.~\eqref{eq:fidelity_exact} and \eqref{eq:projector_exp_vals}, we obtain:
\begin{align}\label{eq:lower_bound}
    \text{Tr}\left(\rho \ P_A^{\mathbb{Z}_2} \ P_A^{\mathbb{Z}_3} \ P_B \right)
    &\geq \ (p_A^{\mathbb{Z}_2} + p_A^{\mathbb{Z}_B} + p_B - 2 )
\end{align}
Substituting the experimentally measured expectation values for the $3\times 3$  lattice gives:
\begin{align}
    \bra{\text{gs}}\rho\ket{\text{gs}} &\geq  0.93(3) + 0.84(3) + 0.81(4) - 2 \\
    &= 0.58(6), 
\end{align}
and the per-qudit fidelity is
\begin{align}
    f:=\bra{\text{gs}}\rho\ket{\text{gs}}^{1/18} &\geq  0.970(5).
\end{align}

Similarly, to get the upper bound for fidelity, we use the fact that the product of projectors is bounded by any individual projector:
\begin{equation}
    \prod_{j=1}^n P_j \leq P_i \quad \forall i.
\end{equation}
Taking the trace with $\rho$ yields:
\begin{align}\label{eq:upper_bound}
    \text{Tr}\left(\rho \ P_A^{\mathbb{Z}_2} \ P_A^{\mathbb{Z}_3} \ P_B \right) &\leq \text{min} \left\{ p_A^{\mathbb{Z}_2},\ p_A^{\mathbb{Z}_B},\ p_B \right\}.
\end{align}
Substituting the expectation values, we obtain:
\begin{equation}
    \bra{\text{gs}}\rho\ket{\text{gs}} \leq 0.81(4).
\end{equation}
In summary, the fidelity per qudit, $f$, for the ground state of the \ds3 model on the $3\times3$ lattice is bounded by:
\begin{equation}
    0.970(5) \leq f \leq 0.988(3).
\end{equation}

\section{Bureau of standards: theory}\label{app:bureau-theory}

\begin{figure}
    \centering
    \includegraphics[width=0.75\linewidth]{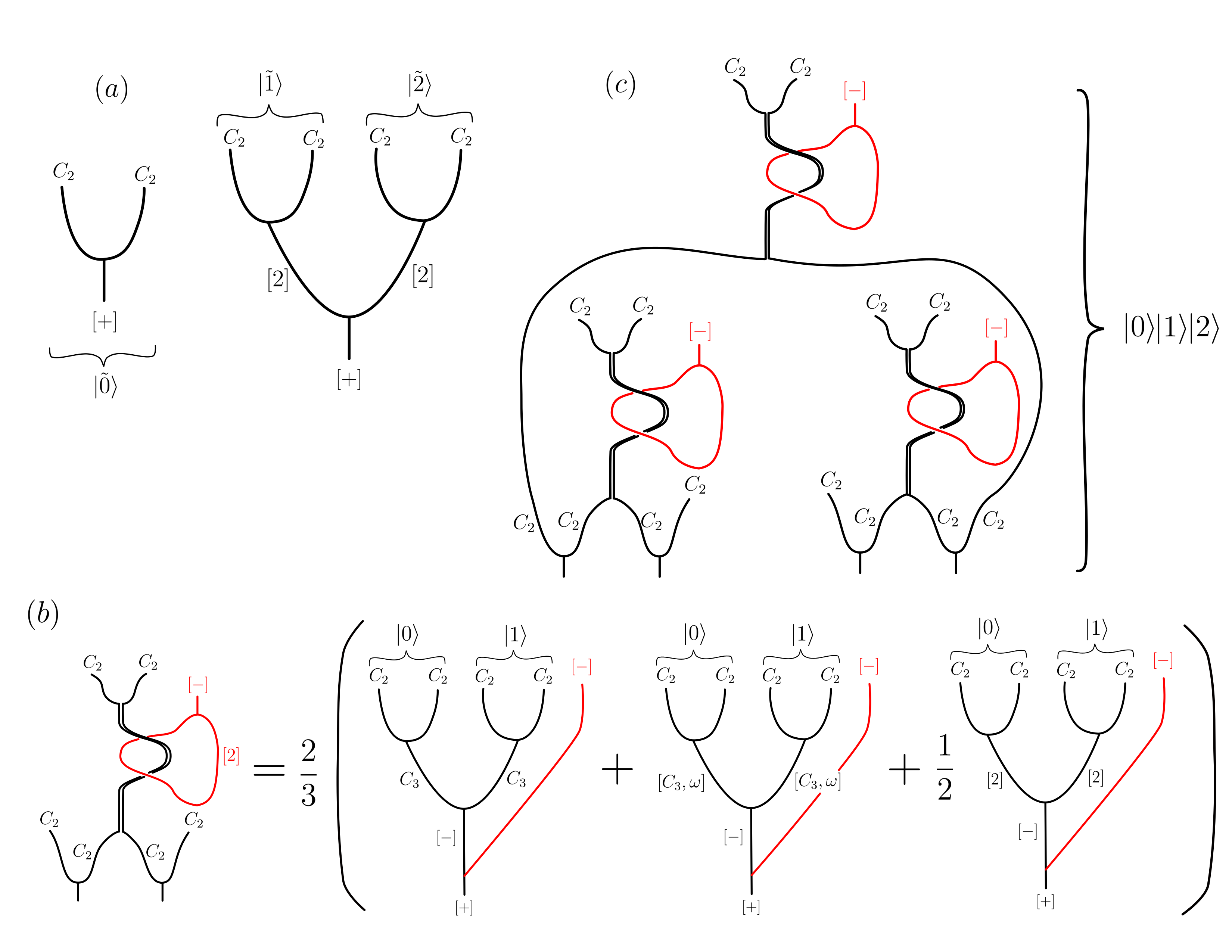}
    \caption{\textbf{Reformulating our approach as a fusion tree.} (a) Qutrit dual basis fusion trees; a $\ket{\tilde{0}}$ state is the vacuum state of two $C_2$ fluxes, while a reference pair of $\ket{\tilde{1}}$ and $\ket{\tilde{2}}$ state can be prepared by splitting a vacuum pair of $[2]$ charges into four $C_2$ fluxes. Any further logical qubits will have to be compared with the $\ket{\tilde{1}}$ and $\ket{\tilde{2}}$ states to set their values in the dual basis. (b) The first step in setting up the computation basis bureau of standards. We assume the left pair is set as our reference $\ket{0}$, and we want to compare the right pair of fluxes in the computational basis. If the $[2]$ charge braiding yields a remnant $[-]$, the right pair cannot be a $\ket{0}$, and so we can set it to be a $\ket{1}$. By re-writing the fusion and braiding diagram, we can express this $\ket{0} \ket{1}$ state in a more conventional fusion tree form, as a superposition over different intermediate fusion channels for the two logical flux pairs. (c) The fusion and braiding diagram corresponding to a full set of reference states, showing the three comparison measurements that ensure all three pairs of fluxes are in different computational basis states.}
    \label{fig:app-bureau-of-standards}
\end{figure}

As touched upon in the main text, we are encoding a logical qutrit inside the zero-flux fusion subspace of two $C_2$ fluxes, defined by the following fusion rule:
\begin{equation}
    C_2 \times C_2 = [+] + [2].
\end{equation}
One qutrit basis state is supplied by the $[+]$ channel, and the other two are furnished by the $[2]$ channel---see Fig.~\ref{fig:app-bureau-of-standards}(a). We note that this is a somewhat different logical encoding strategy that might be familiar to some readers, as we are resolving the $[2]$ charge outcome into its internal states. This type of encoding is sometimes known as a ``split'' fusion channel encoding \cite{cui_universal_2015}, to contrast with the usual fusion channel encoding where every anyon type has a single logical state associated with it. 

The split fusion channel encoding treats the internal states of the $[2]$ charge as distinct logical states: however, these internal states have no definite labeling, and are only physically meaningful as relative states. This subtlety necessitates the use of a ``bureau of standards'' which enables us to set a consistent convention of which state is which. 

It is important to emphasize that, although we are using the internal states of the $[2]$ charge as logical states, our encoding is \emph{still topologically protected}, and cannot be accessed (and hence corrupted) locally. This is because the $[2]$ charge is itself stored nonlocally between the $C_2$ fluxes that form the logical qutrit. Ultimately, this split fusion channel kind of encoding is equivalent to a more traditional fusion tree approach: we must simply take into account the reference states which form the bureau of standards, and the probe anyons that perform the comparison measurements, when writing down the fusion tree. 

We build the bureau of standards for the $Z$-basis by starting with a pair of $C_2$ fluxes created from the vacuum. At this point, these fluxes must be in the $\ket{\tilde{0}}$ state:
\begin{equation}
    \ket{\tilde{0}} = \frac{1}{\sqrt{3}} \left ( \ket{\sigma, \sigma} + \ket{\mu \sigma, \mu \sigma} + \ket{\bar{\mu} \sigma, \bar{\mu} \sigma} \right).
\end{equation}
This initial pair of fluxes will act as our reference $\ket{0}$ state: any subsequent pair of fluxes must be entangled with this pair so that each branch of the wavefunction has a consistent convention. For instance, if we create a second pair of fluxes to act as reference $\ket{1}$, we could entangle with the first ``$\ket{0}$'' in the following way: 
\begin{equation}
    \frac{1}{\sqrt{3}} ( \ket{\sigma, \sigma} \ket{\mu \sigma, \mu \sigma} + \ket{\mu \sigma, \mu \sigma} \ket{\bar{\mu} \sigma, \bar{\mu}\sigma} + \ket{\bar{\mu} \sigma \bar{\mu}\sigma} \ket{\sigma, \sigma} ).
\end{equation}
To create this entangled state, we first notice that if we take one flux from each pair, the combined flux of the two anyons will be $C_3$-valued. So we can start with our reference $\ket{0}$ state and a new vacuum pair of $C_2$ fluxes, performing a $[2]$ charge braiding experiment around one half of each pair. If the $[2]$ charge fuses to a remnant $[-]$ charge, then the total flux encircled must have been a $C_3$, and we have projected into the desired state. In other words, we have performed a $Z$-basis comparison between the reference $\ket{0}$ and the new pair, found that they must be in different states, and so labeled the new pair as a $\ket{1}$. 

This procedure can be expressed as a fusion and braiding diagram, see Fig.~\ref{fig:app-bureau-of-standards}(b). Using the $F$- and $R$-symbol data for $S_3$, this diagram can be converted into a pure fusion diagram, demonstrating that the split fusion channel encoding can be understood as a usual fusion tree encoding with an expanded set of anyons. Fig.~\ref{fig:app-bureau-of-standards}(c) shows the braiding and fusion diagram expressing the full bureau of standards protocol to create all three computational basis reference states, which can be simplified into a fusion diagram involving the three pairs of logical fluxes and the three remnant $[-]$ charges.

\section{Protocol implementation details}\label{app:protocol-imple}

\subsection{Pull-through gate}\label{app:pullthrough-imple}

\subsubsection{Theory prediction for local plaquette projector}\label{app:local-plaquette-theory-predict}

In Fig.~\ref{fig:pull-through} caption, we note that the theoretical prediction for the expectation value local plaquette projector $B^{\mathbb{Z}_3}$ is $\frac{1}{3}$. This is due to the fact that, for the local flux label, the $C_2$ flux we create has a equal and symmetric weight on all three labels: $\sigma,\mu\sigma,\bar\mu\sigma$. The $B^{\mathbb{Z}_3}$ projector projects to the sector where the qutrit value is $0$, i.e. to the label $\sigma$. Therefore, the expectation value of $B^{\mathbb{Z}_3}$ is $\frac{1}{3}$.

\subsubsection{Theory prediction for nonlocal correlator}\label{app:nonlocal-correlator-theory-predict}

In the caption of Fig.~\ref{fig:pull-through}, we also note that the nonlocal correlator $\left\langle \Pi^{\mathbb{Z}_3}_{W_2W_3}\right\rangle$ has a predicted value of $1/3$ at $t=0$. The logical state is $\ket{\tilde0}_L \ket{0}_L = \frac{1}{\sqrt{3}}(\ket{0}_L+\ket{1}_L+\ket{2}_L)\ket{0}_L$. We see that there is equal weight between the $\mathcal{Z}$-basis eigenstates. Therefore, the expectation value of the nonlocal correlator
\begin{equation}
\left\langle \Pi^{\mathbb{Z}_3}_{W_2W_3}\right\rangle = \bra{\tilde0}\bra{0}\Pi^{\mathbb{Z}_3}_{W_2W_3} \ket{\tilde0}\ket{0} 
= \left(\frac{1}{\sqrt{3}} \bra{0}\bra{0}\right)\left(\frac{1}{\sqrt{3}} \ket{0}\ket{0}\right)
= \frac{1}{3}.
\end{equation}

At $t=6$, the logical state is the qutrit Bell state $\frac{1}{\sqrt{3}}(\ket{0}_L\ket{0}_L+\ket{1}_L\ket{2}_L+\ket{2}_L\ket{1}_L)$. The $\Pi^{\mathbb{Z}_3}_{W_2W_3}$ correlator is designed to evaluate the $\mathbb{Z}_3$ part of the global flux and check whether it multiply to the trivial group element. For example, the state $\ket{1}_L\ket{2}_L$ has global flux $\mu\sigma$ and $\bar{\mu}\sigma$, such that the total qutrit flux is $\mu\cdot \bar\mu = e$; similarly for the $\ket{2}_L\ket{1}_L$ state. Therefore, the state is manifestly in the $+1$ eigenstate of the $\Pi^{\mathbb{Z}_3}_{W_2W_3}$ correlator.

\subsubsection{Circuits}

The system is on a 3$\times$3 square lattice, where the indexing of qutrits (denoted by $a_i,\alpha_i$) and qubits (denoted by $b_i,\beta_i$) on each edge is as follows: 

        
        



\begin{figure}[!ht]
	\centering
	\includegraphics[width=0.32\linewidth]{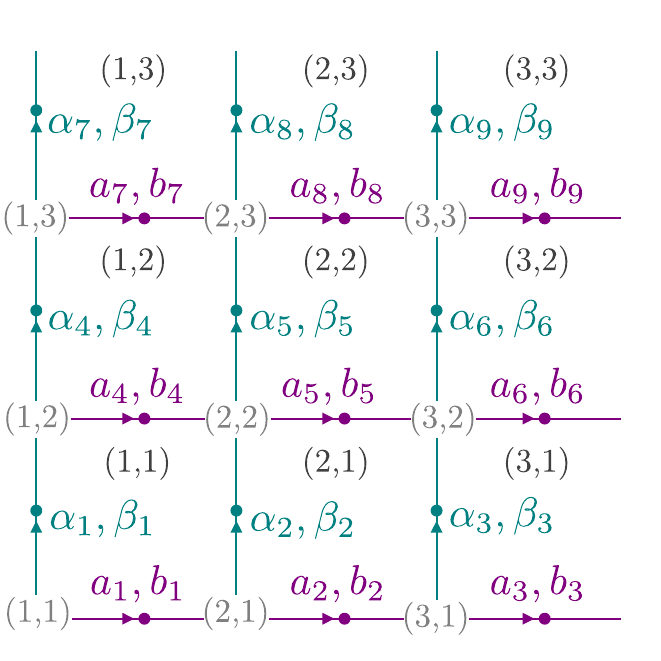}
	\caption{Labeling of vertices, edges, and plaquettes on a 3$\times$3 square lattice on torus. The coordinate of vertex and plaquette are of the form $(x,y)$ indexed from 1. The edges are indexed from left to right and bottom to top (alternating between horizontal and vertical edges).}
	\label{fig:3x3_with_double_labels}
\end{figure}

\subsubsection{Step 1: Initialize control qutrit}\label{app-subsub:pull-through-step1}

We initialize a pair of $C_2$ flux (at the (1,1) and (2,1) plaquettes) to be the control qutrit in the $\ket{\tilde0}_L$ state at $t=0$ depicted in the main text Fig.~\ref{fig:pull-through}. For the circuit below, the ancilla is initialized in the $\ket{0}_a$ state, such that the first qutrit $\mathcal{H}$ transforms the ancilla to be in the $\ket{\tilde0}_a$ state. Then, the flux's internal state is entangled with the ancilla via action of ribbon operator with internal state conditioned on the ancilla state; we refer to this as state injection (see circuit below). This creates the state $\ket{0}_L\ket{0}_a+\ket{1}_L\ket{1}_a+\ket{2}_L\ket{2}_a$. The action is derived from the unitary circuit representation of $C_2$ ribbon operators (an example is shown in Eq.~\ref{eq:appendix_C2_circuit}). Significant simplifications in the circuit (in particular, there are no ungauging and regauging layers) derives from the fact that there is no direct triangle in the ribbon. 

\begin{center}
\begin{tikzpicture}
\draw[decoration=brace, decorate]
    		($(1,.3)$) -- ($(2.5,.3)$)
    		node[midway, above=1pt, text width=3.0cm, align=center] {Conditioned \\ribbon operator};
\begin{yquant}[horizontal]
    [purple] qubit {$\ket{anc}$} anc[1];
    [teal] qubit {$\ket{\alpha_2}$} alp2[1];
    [teal] qubit {$\ket{\beta_2}$} beta2[1];

    box {$\mathcal{H}$} anc;

    barrier (anc, alp2, beta2);

    box {$\mathcal{C}$} alp2;
    box {$X$} beta2;
    box {$\mathcal{X}$} alp2 | anc;

\end{yquant}
\end{tikzpicture}
\end{center}

After creating the entangled state between the data qutrit and the ancilla, we need to disentangle the ancilla to initialize the data qutrit in the $\ket{\tilde1}_L$ state. To disentangle the ancilla, we need to apply a $C_L\mathcal{X}^\dagger_a$ gate so as to obtain the state $\ket{\tilde1}_L\ket{0}_a$. To control on the logical qutrit, we need to conditioned on the value of the flux loop (based at the origin) around the $(3,1)$ plaquette. The following circuit implements the $C_L\mathcal{X}^\dagger_a$ gate to disentangles ancilla from the logical state.

\begin{center}
    \begin{tikzpicture}
\draw[decoration=brace, decorate]
    		($(0,.3)$) -- ($(5.2,.3)$)
    		node[midway, above=1pt, text width=3.0cm, align=center] {Disentangle ancilla};
\begin{yquant}[horizontal]

[orange] qubit {$\ket{\alpha_2}$} a[1];
[orange] qubit {$\ket{\beta_2}$} b[1];
[orange] qubit {$\ket{a_4}$} a[+1];
[orange] qubit {$\ket{b_4}$} b[+1];
[orange] qubit {$\ket{\alpha_1}$} a[+1];
[orange] qubit {$\ket{\beta_1}$} b[+1];
[orange] qubit {$\ket{a_1}$} a[+1];
[orange] qubit {$\ket{b_1}$} b[+1];

[purple] qubit {$\ket{anc}$} anc[1];

box {$\mathcal{X}^{\dagger}$} anc[0] | a[0];
box {$\mathcal{C}$} anc[0] | b[0];
box {$\mathcal{C}$} anc[0] | b[1];
box {$\mathcal{X}$} anc[0] | a[1];
box {$\mathcal{C}$} anc[0] | b[2];
box {$\mathcal{X}$} anc[0] | a[2];
box {$\mathcal{X}^\dagger$} anc[0] | a[3];
box {$\mathcal{C}$} anc[0] | b[3];

\end{yquant}
\end{tikzpicture}
\end{center}

\subsubsection{Initialize target qutrit}

We initialize a second $C_2$ flux pair (at the (3,1) and (3,2) plaquettes) to serve as the target qutrit in the $\ket{0}$ state at $t=0$ depicted in the main text Fig.~\ref{fig:pull-through}. As this is a $\mathcal{Z}$-basis eigenstate, no ancilla is needed to initialize the state. The circuit is conjugated by gates that change the orientation of the edge for the qudits $\ket{a_6,b_6}$ such that the orientation of the triangle operator is aligned. 

\begin{center}
\begin{tikzpicture}
\begin{yquant}[horizontal]
    [teal] qubit {$\ket{a_6}$} a6[1];
    [teal] qubit {$\ket{b_6}$} b6[1];

    [violet] qubit {$\ket{\alpha_1}$} alp1[1];
    [violet] qubit {$\ket{\beta_1}$} beta1[1];

    box {$\mathcal{C}$} a6;
    box {$\mathcal{C}$} a6 | b6;

    barrier (a6, b6, alp1, beta1);

    box {$\mathcal{C}$} a6;
    box {$X$} b6;
    box {$\mathcal{C}$} alp1 | beta1;
    box {$\mathcal{X}$} a6 | alp1;
    box {$\mathcal{C}$} alp1 | beta1;

    barrier (a6, b6, alp1, beta1);

    box {$\mathcal{C}$} a6;
    box {$\mathcal{C}$} a6 | b6;

\end{yquant}

\draw[decoration=brace, decorate]
		($(0,.3)$) -- ($(1.5,.3)$)
		node[midway, above=1pt, text width=3.0cm, align=center] {Flip \\orientation};
        
\draw[decoration=brace, decorate]
		($(1.7,.3)$) -- ($(3.9,.3)$)
		node[midway, above=1pt, text width=3.0cm, align=center] {ribbon\\ operator};

\draw[decoration=brace, decorate]
        ($(4.1,.3)$) -- ($(5.5,.3)$)
		node[midway, above=1pt, text width=3.0cm, align=center] {Revert \\orientation};

\end{tikzpicture}
\end{center}

\subsubsection{Coherent moving of flux example}\label{app-subsub:moving-example-circ}

\begin{figure}[!h]
	\centering
    \includegraphics[width=0.2\linewidth]{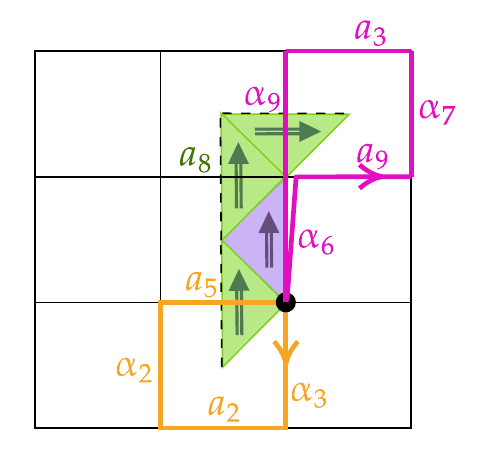}
	\caption{\textbf{Schematic for the pull-through step at $t=1$, moving a flux of the control flux pair.} }\label{fig:appendix_coherent-moving}
\end{figure}

Here, we provide an example circuit for moving flux coherently, when we move control qutrit from the (2,1) plaquette to the (3,3) plaquette as illustrated in Fig.~\ref{fig:appendix_coherent-moving}. This is the step at $t=1$ depicted in the main text Fig.~\ref{fig:pull-through}. For all the remaining steps of the protocol (from $t=2$ to $t=6$), similar circuits are used to coherently move the fluxes to complete the braiding. 

First, the local flux internal state of the flux (measured in a clockwise loop depicted in orange based at the black dot) is entangled with the ancilla. Then, conditioned on the ancilla, a corresponding ribbon operator is applied to cancel out the flux at the initial plaquette. Finally, the ancilla is disentangled by applying a $C_L\mathcal{X}^\dagger_a$ gate (flux value measured in a counterclockwise loop depicted in magenta) \textit{based at the same base point} (black dot). We emphasize that the base point has to be the same as the start of the ribbon to ensure that the ancilla is properly disentangled (see Section~\ref{app:coherent-moving} for explanation).

The ancilla is initialized in the $\ket{0}$ state. The circuit on the left panel entangles the ancilla state with the internal local flux state of one of the control qutrit by implementing a $C_L\mathcal{X}_a$ gate. For the circuit in the middle panel, conditioned on the ancilla state, it applies the corresponding ribbon operator to cancel out the flux at the initial plaquette. Since the ribbon starts with a dual triangle, whose action is decoupled from the rest of the ribbon, its action is directly conditional on the ancilla state, just as the first direct triangle. Finally, for the circuit on the right panel, the ancilla is disentangled by implementing a $C_L\mathcal{X}^\dagger_a$ gate.

\begin{center}
\begin{tikzpicture}[scale=0.8]
\draw[decoration=brace, decorate]
    		($(0,.3)$) -- ($(5.3,.3)$)
    		node[midway, above=1pt, text width=3.0cm, align=center] {Entangle internal state with ancilla};            
\begin{yquant}[horizontal]
[orange] qubit {$\ket{\alpha_3}$} a[1];
[orange] qubit {$\ket{\beta_3}$} b[1];
[orange] qubit {$\ket{a_2}$} a[+1];
[orange] qubit {$\ket{b_2}$} b[+1];
[orange] qubit {$\ket{\alpha_2}$} a[+1];
[orange] qubit {$\ket{\beta_2}$} b[+1];
[orange] qubit {$\ket{a_5}$} a[+1];
[orange] qubit {$\ket{b_5}$} b[+1];

[purple] qubit {$\ket{anc}$} anc[1];

box {$\mathcal{C}$} anc[0] | b[0];
box {$\mathcal{X}$} anc[0] | a[0];
box {$\mathcal{C}$} anc[0] | b[1];
box {$\mathcal{X}$} anc[0] | a[1];
box {$\mathcal{X}^\dagger$} anc[0] | a[2];
box {$\mathcal{C}$} anc[0] | b[2];
box {$\mathcal{X}^\dagger$} anc[0] | a[3];
box {$\mathcal{C}$} anc[0] | b[3];

\end{yquant}
\end{tikzpicture}
\begin{tikzpicture}[scale=0.8]
\draw[decoration=brace, decorate]
    		($(0,.3)$) -- ($(5,.3)$)
    		node[midway, above=1pt, text width=3.0cm, align=center] {Conditioned ribbon operator};
\begin{yquant}[horizontal]

[teal] qubit {$\ket{a_5}$} a5[1];
[teal] qubit {$\ket{b_5}$} b5[1];
[teal] qubit {$\ket{a_8}$} a8[1];
[teal] qubit {$\ket{b_8}$} b8[1];
[violet] qubit {$\ket{\alpha_6}$} alp6[1];
[violet] qubit {$\ket{\beta_6}$} beta6[1];
[violet] qubit {$\ket{\alpha_9}$} alp9[1];
[violet] qubit {$\ket{\beta_9}$} beta9[1];

[purple] qubit {$\ket{anc}$} anc[1];

box {$\mathcal{C}$} a5 | b5;
box {$\mathcal{C}$} a8 | b8;

box {$\mathcal{X}$} alp6 | anc;

box {$\mathcal{C}$} alp9;
box {$X$} b5;
box {$X$} b8;
box {$X$} beta9;

box {$\mathcal{X}$} a5 | anc;

box {$\mathcal{C}$} alp6 | beta6;
box {$\mathcal{X}$} a8 | alp6;
box {$\mathcal{X}$} alp9 | alp6;
box {$\mathcal{C}$} alp6 | beta6;

box {$\mathcal{X}^\dagger$} alp6 | anc;

box {$\mathcal{C}$} a5;
box {$\mathcal{C}$} a5 | b5;
box {$\mathcal{C}$} a8;
box {$\mathcal{C}$} a8 | b8;

\end{yquant}
\end{tikzpicture}
\begin{tikzpicture}[scale=0.8]
\draw[decoration=brace, decorate]
    		($(0,.3)$) -- ($(7.8,.3)$)
    		node[midway, above=1pt, text width=3.0cm, align=center] {Disentangle ancilla};
\begin{yquant}[horizontal]
    [magenta] qubit {$\ket{\alpha_6}$} alp6[1];
    [magenta] qubit {$\ket{\beta_6}$} beta6[1];
    [magenta] qubit {$\ket{a_9}$} a9[1];
    [magenta] qubit {$\ket{b_9}$} b9[1];
    [magenta] qubit {$\ket{\alpha_7}$} alp7[1];
    [magenta] qubit {$\ket{\beta_7}$} beta7[1];
    [magenta] qubit {$\ket{a_3}$} a3[1];
    [magenta] qubit {$\ket{b_3}$} b3[1];
    [magenta] qubit {$\ket{\alpha_9}$} alp9[1];
    [magenta] qubit {$\ket{\beta_9}$} beta9[1];

    [purple] qubit {$\ket{anc}$} anc[1];

    box {$\mathcal{X}^\dagger$} anc[0] | alp6;
    box {$\mathcal{C}$} anc[0] | beta6;

    box {$\mathcal{X}^\dagger$} anc[0] | a9;
    box {$\mathcal{C}$} anc[0] | b9;

    box {$\mathcal{X}^\dagger$} anc[0] | alp7;
    box {$\mathcal{C}$} anc[0] | beta7;

    box {$\mathcal{C}$} anc[0] | b3;
    box {$\mathcal{X}$} anc[0] | a3;

    box {$\mathcal{C}$} anc[0] | beta9;
    box {$\mathcal{X}$} anc[0] | alp9;

    box {$\mathcal{C}$} anc[0] | beta6;
    box {$\mathcal{X}$} anc[0] | alp6;

\end{yquant}
\end{tikzpicture}
\end{center}

\subsection{$\mathcal{X}$-basis measurement}\label{app:xmeasure-imple}

Below, we detail the circuit for implementing the $\mathcal{X}$-basis measurement protocol for the case where the data logical qutrit is in the $\ket{\tilde 1}$ state. For the case where the data logical qutrit is in the $\ket{\tilde 0}$ state, the only difference is in step 1, where we omit the $\mathcal{Z}$ gate in the initial state injection stage; the rest of the circuit remains the same. 

\subsubsection{Step 1: Create data logical $\ket{\tilde{1}}$}\label{app:example-tilde1-logical}

First, we create a $C_2$ flux pair at the $(3,1)$ and $(3,2)$ plaquettes (see Fig.~\ref{fig:3x3_with_double_labels} for labels); this flux pair serves as the data qutrit and is initialized in the $\ket{\tilde{1}}_L$ logical state by the end of step 1 using state injection similar to the procedure in the step 1 of pull-through gate (see Section~\ref{app-subsub:pull-through-step1}). The circuit implementing the entangling is shown on the left panel, where the qudits in violet (teal) font are acted on by direct (dual) triangle of the ribbon operator.

The circuit on the right panel disentangle the ancilla by applying a $C_L\mathcal{X}^\dagger_a$ gate to obtain the (periodic boundary condition is used to shorten the flux loop).

\begin{center}
\begin{tikzpicture}
\begin{yquant}[horizontal]
[teal] qubit {$\ket{a_6}$} a6[1];
[teal] qubit {$\ket{b_6}$} b6[1];

[violet] qubit {$\ket{\alpha_1}$} alp1[1];
[violet] qubit {$\ket{\beta_1}$} beta1[1];

[purple] qubit {$\ket{anc}$} anc[1];

box {$\mathcal{H}$} anc;
box {$\mathcal{Z}$} anc;
box {$\mathcal{X}$} alp1 | anc;

barrier (a6,b6,alp1,beta1,anc);

box {$\mathcal{C}$} a6;
box {$\mathcal{C}$} a6 | b6;

barrier (a6,b6,alp1,beta1,anc);

box {$\mathcal{C}$} a6;
box {$X$} b6;
box {$\mathcal{C}$} alp1 | beta1;
box {$\mathcal{X}$} a6 | alp1;
box {$\mathcal{C}$} alp1 | beta1;

barrier (a6,b6,alp1,beta1,anc);

box {$\mathcal{C}$} a6;
box {$\mathcal{C}$} a6 | b6;

barrier (a6,b6,alp1,beta1,anc);

box {$\mathcal{X}^\dagger$} alp1 | anc;

\end{yquant}		
		\draw[decoration=brace, decorate]
		($(0,.3)$) -- ($(2.1,.3)$)
		node[midway, above=1pt, text width=3.0cm, align=center] {State\\ injection};
        
        \draw[decoration=brace, decorate]
		($(2.2,.3)$) -- ($(3.9,.3)$)
		node[midway, above=1pt, text width=3.0cm, align=center] {Flip \\orientation};

        \draw[decoration=brace, decorate]
		($(4,.3)$) -- ($(6.3,.3)$)
		node[midway, above=1pt, text width=3.0cm, align=center] {Conditioned\\ ribbon \\operator};

        \draw[decoration=brace, decorate]
		($(6.4,.3)$) -- ($(8.2,.3)$)
		node[midway, above=1pt, text width=3.0cm, align=center] {revert \\ orientation};

        \draw[decoration=brace, decorate]
		($(8.3,.3)$) -- ($(9.3,.3)$)
		node[midway, above=1pt, text width=3.0cm, align=center] {State\\ injection};

\end{tikzpicture}
\begin{tikzpicture}
\begin{yquant}[horizontal]

[orange] qubit {$\ket{\alpha_1}$} a[1];
[orange] qubit {$\ket{\beta_1}$} b[1];
[orange] qubit {$\ket{a_6}$} a[+1];
[orange] qubit {$\ket{b_6}$} b[+1];
[orange] qubit {$\ket{\alpha_3}$} a[+1];
[orange] qubit {$\ket{\beta_3}$} b[+1];
[orange] qubit {$\ket{a_3}$} a[+1];
[orange] qubit {$\ket{b_3}$} b[+1];

[purple] qubit {$\ket{anc}$} anc[1];

box {$\mathcal{X}^{\dagger}$} anc[0] | a[0];
box {$\mathcal{C}$} anc[0] | b[0];

box {$\mathcal{C}$} anc[0] | b[1];
box {$\mathcal{X}$} anc[0] | a[1];

box {$\mathcal{C}$} anc[0] | b[2];
box {$\mathcal{X}$} anc[0] | a[2];

box {$\mathcal{X}^\dagger$} anc[0] | a[3];
box {$\mathcal{C}$} anc[0] | b[3];

\end{yquant}
\end{tikzpicture}
\end{center}

\subsubsection{Step 2: Create measurement pure flux}\label{app:example-0tilde-logical}

We then create the second flux pair in the $\ket{\tilde{0}}_L$ logical state at the $(1,2)$ and $(3,3)$ plaquette, using similar techniques as in step 1. 

\begin{center}
\begin{tikzpicture}
\begin{yquant}[horizontal]
    [purple] qubit {$\ket{anc}$} anc[1];
    
    [teal]   qubit {$\ket{\alpha_5}$} alp5[1];
    [teal]   qubit {$\ket{\beta_5}$}  beta5[1];
    
    [violet] qubit {$\ket{a_5}$} a5[1];
    [violet] qubit {$\ket{b_5}$} b5[1];

    [violet]   qubit {$\ket{\alpha_6}$} alp6[1];
    [violet]   qubit {$\ket{\beta_6}$}  beta6[1];

    [teal] qubit {$\ket{a_8}$} a8[1];
    [teal] qubit {$\ket{b_8}$} b8[1];

    [teal]   qubit {$\ket{\alpha_9}$} alp9[1];
    [teal]   qubit {$\ket{\beta_9}$}  beta9[1];

    box {$\mathcal{H}$} anc;

    box {$\mathcal{X}$} a5 | anc;

    barrier (a5,b5,a8,b8,alp5,beta5,alp6,beta6,alp9,beta9,anc);

    box {$\mathcal{C}$} a8;
    box {$\mathcal{C}$} a8 | b8;

    barrier (a5,b5,a8,b8,alp5,beta5,alp6,beta6,alp9,beta9,anc);

    cnot beta6 | b5;
    box {$\mathcal{C}$} alp6 | b5;
    
    box {$\mathcal{X}$} alp6 | a5;

    barrier (a5,b5,a8,b8,alp5,beta5,alp6,beta6,alp9,beta9,anc);

    box {$\mathcal{C}$} a8;
    box {$\mathcal{C}$} alp5;
    box {$\mathcal{C}$} alp9;
    box {$X$} beta5;
    box {$X$} b8;
    box {$X$} beta9;
    box {$\mathcal{X}$} alp5 | anc;
    box {$\mathcal{C}$} alp6 | beta6;
    box {$\mathcal{X}$} a8 | alp6;
    box {$\mathcal{X}$} alp9 | alp6;
    box {$\mathcal{C}$} alp6 | beta6;

    barrier (a5,b5,a8,b8,alp5,beta5,alp6,beta6,alp9,beta9,anc);

    box {$\mathcal{X}^\dagger$} alp6 | a5;
    box {$\mathcal{C}$} alp6 | b5;
    cnot beta6 | b5;

    barrier (a5,b5,a8,b8,alp5,beta5,alp6,beta6,alp9,beta9,anc);

    box {$\mathcal{C}$} a8;
    box {$\mathcal{C}$} a8 | b8;

    barrier (a5,b5,a8,b8,alp5,beta5,alp6,beta6,alp9,beta9,anc);

    box {$\mathcal{X}^\dagger$} a5 | anc;

\end{yquant}
    \draw[decoration=brace, decorate]
		($(0,.3)$) -- ($(1.5,.3)$)
		node[midway, above=1pt, text width=3.0cm, align=center] {State\\ injection};
        
        \draw[decoration=brace, decorate]
		($(1.6,.3)$) -- ($(3.3,.3)$)
		node[midway, above=1pt, text width=3.0cm, align=center] {Flip \\orientation};

        \draw[decoration=brace, decorate]
		($(3.4,.3)$) -- ($(11.3,.3)$)
		node[midway, above=1pt, text width=3.0cm, align=center] {Conditioned ribbon operator};

        \draw[decoration=brace, decorate]
		($(11.4,.3)$) -- ($(13.1,.3)$)
		node[midway, above=1pt, text width=3.0cm, align=center] {revert \\ orientation};

        \draw[decoration=brace, decorate]
		($(13.2,.3)$) -- ($(14.3,.3)$)
		node[midway, above=1pt, text width=3.0cm, align=center] {State\\ injection};
\end{tikzpicture}
\end{center}

Similar to the second part of step 1, we disentangle the ancilla from the logical state by applying a $C_L\mathcal{X}^\dagger_a$ gate, conditioned on the value of the flux loop (based at the origin) around the $(1,2)$ plaquette using periodic boundary condition.

\begin{center}
\begin{tikzpicture}
\begin{yquant}[horizontal]

[orange] qubit {$\ket{\alpha_5}$} a[1];
[orange] qubit {$\ket{\beta_5}$} b[1];
[orange] qubit {$\ket{a_7}$} a[+1];
[orange] qubit {$\ket{b_7}$} b[+1];
[orange] qubit {$\ket{\alpha_4}$} a[+1];
[orange] qubit {$\ket{\beta_4}$} b[+1];
[orange] qubit {$\ket{a_4}$} a[+1];
[orange] qubit {$\ket{b_4}$} b[+1];

[purple] qubit {$\ket{anc}$} anc[1];

box {$\mathcal{X}^{\dagger}$} anc[0] | a[0];
box {$\mathcal{C}$} anc[0] | b[0];

box {$\mathcal{C}$} anc[0] | b[1];
box {$\mathcal{X}$} anc[0] | a[1];

box {$\mathcal{C}$} anc[0] | b[2];
box {$\mathcal{X}$} anc[0] | a[2];

box {$\mathcal{X}^\dagger$} anc[0] | a[3];
box {$\mathcal{C}$} anc[0] | b[3];

\end{yquant}
\end{tikzpicture}
\end{center}

\subsubsection{Step 3-5: Coherent moving fluxes}

We schematically illustrate the coherent moving circuits for step 3-5, similar to those implemented in Section~\ref{app-subsub:moving-example-circ}.

\begin{figure}[!h]
	\centering
	\includegraphics[width=0.7\linewidth]{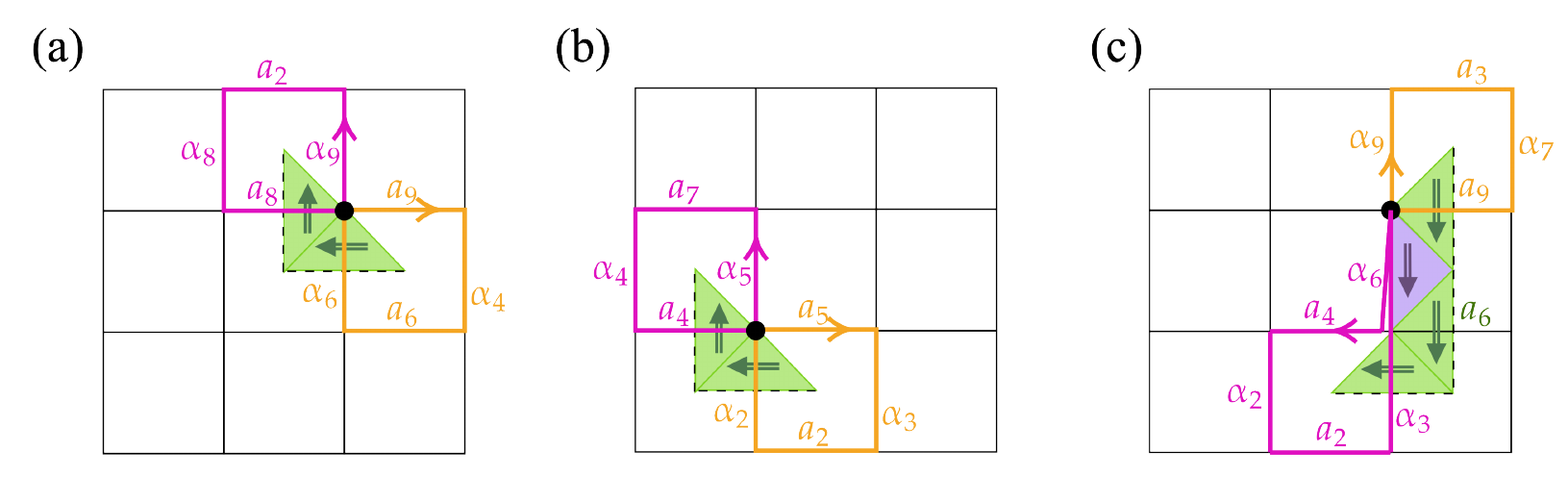}
	\caption{\textbf{Schematics for the coherent moving circuits for $\mathcal{X}$-basis measurement.} (a) Step 3, (b) step 4, and (c) step 5.}
\end{figure}

\subsubsection{Step 6: Fuse $C_2$ flux}

For the last step of the $\mathcal{X}$-basis measurement protocol, it involves fusing two $C_2$ flux, which warrants more explanation as the circuit towards the end is different from that of the coherent moving circuits in previous steps. 

The entangling (left panel) and conditioned ribbon operator (middle panel) of the circuit proceeds in the same way as the coherent moving circuits, as shown below: 

\begin{center}
\begin{tikzpicture}
\draw[decoration=brace, decorate]
    		($(0,.3)$) -- ($(5.3,.3)$)
    		node[midway, above=1pt, text width=3.0cm, align=center] {Entangle internal state with ancilla};
\begin{yquant}[horizontal]
[orange] qubit {$\ket{\alpha_2}$} a[1];
[orange] qubit {$\ket{\beta_2}$} b[1];

[orange] qubit {$\ket{a_2}$} a[+1];
[orange] qubit {$\ket{b_2}$} b[+1];

[orange] qubit {$\ket{\alpha_3}$} a[+1];
[orange] qubit {$\ket{\beta_3}$} b[+1];

[orange] qubit {$\ket{a_5}$} a[+1];
[orange] qubit {$\ket{b_5}$} b[+1];

[purple] qubit {$\ket{anc}$} anc[1];


box {$\mathcal{X}^\dagger$} anc[0] | a[3];
box {$\mathcal{C}$} anc[0] | b[3];

box {$\mathcal{C}$} anc[0] | b[2];
box {$\mathcal{X}$} anc[0] | a[2];

box {$\mathcal{C}$} anc[0] | b[1];
box {$\mathcal{X}$} anc[0] | a[1];

box {$\mathcal{X}^\dagger$} anc[0] | a[0];
box {$\mathcal{C}$} anc[0] | b[0];

\end{yquant}
\end{tikzpicture}
\begin{tikzpicture}
\draw[decoration=brace, decorate]
    		($(0,.3)$) -- ($(3.2,.3)$)
    		node[midway, above=1pt, text width=3.0cm, align=center] {Conditioned ribbon operator};
\begin{yquant}[horizontal]

[purple] qubit {$\ket{anc}$} anc[1];

[teal] qubit {$\ket{\alpha_2}$} alp[1];
[teal] qubit {$\ket{\beta_2}$} beta[1];
[teal] qubit {$\ket{a_4}$} alp[+1];
[teal] qubit {$\ket{b_4}$} beta[+1];


box {$\mathcal{C}$} alp[0] | beta[0];
box {$\mathcal{C}$} alp[1] | beta[1];

box {$\mathcal{X}$} alp[0] | anc[0];
box {$\mathcal{X}$} alp[1] | anc[0];

align alp,beta;

box {$\mathcal{C}$} alp[0] | beta[0];
box {$\mathcal{C}$} alp[1] | beta[1];

box {$X$} beta;

\end{yquant}
\end{tikzpicture}
\end{center}

At this point, the flux pair have fused together and leaves behind a remnant charge at the fusion site. The ancilla is entangled with the system in a non-obvious basis. First, a qutrit Hadamard $\mathcal{H}^\dagger$ is applied, because this would reset the ancilla from the $\ket{\tilde0}$ state back to the $\ket{0}$ state if the ancilla is not entangled; the state is then in the basis where the internal state of the remnant charge is entangled with the ancilla as follows: 
\begin{equation}
    \ket{0}_a\ket{+} + \ket{1}_a \ket{2-} + \ket{2}_a\ket{2+}.
\end{equation}

At this point, we could in principle end the protocol; projective measurement on the ancilla is equivalent to measuring the local charge at the fusion site. However, the ancilla is entangled with the system and cannot be reused until projective measurement. 

We embrace a quixotic spirit by asking the question: can the ancilla be disentangled? Besides freeing the ancilla for other parts of the protocol, this also has the additional benefit of allowing the usual local vertex projectors ($A^{\mathbb{Z}_2}_v$ and $A^{\mathbb{Z}_3}_v$) to directly determine the charge violations.

Here, we make use of the fact that for the $\ket{\tilde1}_L$ input state of the $\mathcal{X}$-basis measurement protocol, there is no weight in the vacuum fusion channel, implying our state simplifies to 
\begin{equation}
    \ket{1}_a \ket{2-} + \ket{2}_a\ket{2+}.
\end{equation}

To disentangle the ancilla from the above state, we need to implement a special $C_{a}A^{\sigma}$ gate on the vertex of the fusion site: apply identity when the control is in $\ket{1}$ state, and apply $A^{\sigma}$ when the control is in $\ket{2}$ state (since $A^\sigma\ket{2-}=\ket{2+}$). To decompose this further, the action of $A^{\sigma}$ involves $X$ gates on all edges of a vertex, and additional $\mathcal{C}$ gates on the oppositely aligned edges. The circuit implementation is as follows:

\begin{center}
    \begin{tikzpicture}
        \draw[decoration=brace, decorate]
            		($(0,.3)$) -- ($(1,.3)$)
            		node[midway, above=1pt, text width=3.0cm, align=center] {Change \\basis};
        \draw[decoration=brace, decorate]
            		($(1.1,.3)$) -- ($(6.4,.3)$)
            		node[midway, above=1pt, text width=6.0cm, align=center] {$C_{a}A^\sigma$};
        \draw[decoration=brace, decorate]
            		($(6.5,.3)$) -- ($(8.5,.3)$)
            		node[midway, above=1pt, text width=6.0cm, align=center] {Reset ancilla};
        \begin{yquant}[horizontal]
        
        [purple] qubit {$\ket{anc}$} anc[1];

        [orange] qubit {$\ket{\alpha_5}$} ap[1];
        [orange] qubit {$\ket{\beta_5}$} bp[1];
        
        [orange] qubit {$\ket{a_5}$} ap[+1];
        [orange] qubit {$\ket{b_5}$} bp[+1];
        
        [orange] qubit {$\ket{\beta_2}$} bp[+1];
        [orange] qubit {$\ket{b_4}$} bp[+1];
        
        
        box {$\mathcal{H}^\dagger$} anc[0];

        barrier (anc,ap,bp);
        
        box {$X_{12}$} ap[0] | anc[0];
        cnot bp[0] | anc[0];
        
        box {$X_{12}$} ap[1] | anc[0];
        cnot bp[1] | anc[0];
        
        box {$X_{12}$} bp[2] | anc[0];
        box {$X_{12}$} bp[3] | anc[0];

        barrier (anc,ap,bp);
        
        box {$H_{12}$} anc[0];
        box {$\mathcal{X}^\dagger$} anc[0];
        
        \end{yquant}
        \end{tikzpicture}
\end{center}

The definitions and physical implementations of the the qutrit--control qubit--target $\textrm{C}_2\textrm{NOT}$ gate and the qutrit--control qutrit--target C$X_{12}$ gate are given in Eqs.~\eqref{eq:qubit-CNOT} and ~\eqref{eq:CX12}, respectively.

Finally, after applying the $C_{a_{12}}A^{\sigma}$ gate, the ancilla is disentangled and is in the $\ket{1}+\ket{2}$ state. Therefore, a $H_{12}$ gate (defined as qubit Hadamard on the $\set{\ket1,\ket2}$ subspace in Eq.~\eqref{eq:qutrit-hc}) followed by a $\mathcal{X}^\dagger$ are applied to revert the ancilla back to the $\ket0$ state.

\subsection{$\mathcal{Z}$-basis measurement}\label{app:zmeasure-imple}
\begin{figure}[!ht]
	\centering
	\includegraphics[width=0.23\linewidth]{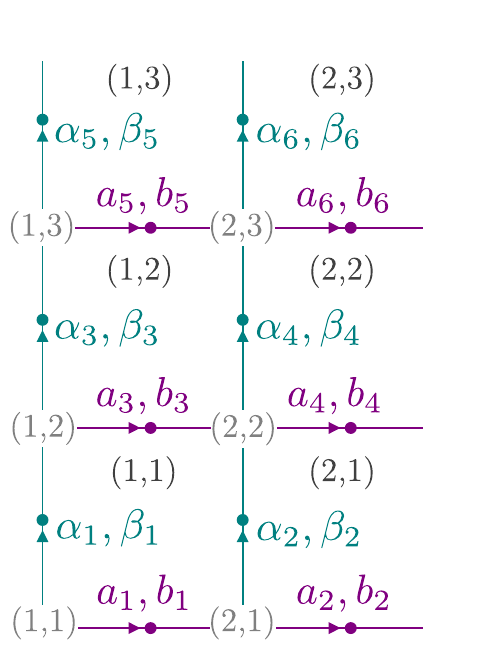}
	\caption{Labeling of vertices, edges, and plaquettes on a $3 \times 2$ periodic square lattice. Vertex and plaquette coordinates are $(x,y)$, indexed from $1$. Edges are ordered left to right and bottom to top, alternating between horizontal and vertical.}
	\label{fig:3x2_with_double_labels}
\end{figure}

\subsubsection{Example of generalized ribbon operator}\label{app:example-generalized-ribbon}

Here, we provide an example of how the generalized ribbon operator (see Section~\ref{app:generalized-ribbon}) is used to specify a logical qutrit in the absolute encoding~\ref{subsec:logical-encoding}. It involves prepending the usual ribbon with a ``direct string" that connects from the starting vertex of the ribbon to the origin. Operationally, this means that the ungauging (regauging) step of the ribbon operator starts (ends) at the origin. 

As an example, we show a circuit implementation of step 1 of the $\mathcal{Z}$-basis measurement protocol~\ref{subsec:z-measurement}, where we create the data qutrit in the $\ket{0}_L$ state at the (1,2) and (2,2) plaquettes: 

\begin{center}
    \begin{tikzpicture}
        \draw[decoration=brace, decorate]
            		($(0,.3)$) -- ($(2.2,.3)$)
            		node[midway, above=1pt, text width=3.0cm, align=center] {Ungauging};
        \draw[decoration=brace, decorate]
            		($(2.3,.3)$) -- ($(4.5,.3)$)
            		node[midway, above=1pt, text width=6.0cm, align=center] {Flux insertion};
        \draw[decoration=brace, decorate]
            		($(4.6,.3)$) -- ($(6.9,.3)$)
            		node[midway, above=1pt, text width=6.0cm, align=center] {Regauging};
        \begin{yquant}[horizontal]

        [violet] qubit {$\ket{\alpha_1}$} a[1];
        [violet] qubit {$\ket{\beta_1}$} b[1];
        [violet] qubit {$\ket{a_3}$} a[+1];
        [violet] qubit {$\ket{b_3}$} b[+1];
        
        [teal] qubit {$\ket{\alpha_4}$} alp[1];
        [teal] qubit {$\ket{\beta_4}$} beta[1];

        cnot b[1] | b[0];
        box {$\mathcal{C}$} a[1] | b[0];
        box {$\mathcal{X}$} a[1] | a[0];
        
        barrier (a,b, alp, beta);

        box {$X$} beta[0];
        box {$\mathcal{C}$} alp[0];
        box {$\mathcal{C}$} a[1] | b[1];
        box {$\mathcal{X}$} alp[0] | a[1];
        box {$\mathcal{C}$} a[1] | b[1];

        barrier (a,b, alp, beta);

        box {$\mathcal{X}^\dagger$} a[1] | a[0];
        box {$\mathcal{C}$} a[1] | b[0];
        cnot b[1] | b[0];
        
        \end{yquant}
        \end{tikzpicture}
\end{center}

We can decompose the above circuit into direct string and the following triangle operators in two ways: 1) the direct string has support on the $(\alpha_1,\beta_1)$ and $(a_3,b_3)$ edges, and the first triangle operator is the dual triangle acting on the $(\alpha_4,\beta_4)$ edge 2) the direct string has support on just the $(\alpha_1,\beta_1)$ edge, and the following is a ribbon consisting of two triangle operators (a direct triangle acting on the $(a_3,b_3)$ edge and a dual triangle acting on the $(\alpha_4,\beta_4)$ edge). In either decomposition, the direct string is necessary, since there is no way to express the above circuit using only consecutive triangle operators. 

Similar circuits are used for creating logical states in Section~\ref{sec:bureau} and \ref{sec:magic}.

\subsubsection{$[2]$ charge braiding}\label{app:example-2-charge-braiding}

When we braid $[2]$ charges around the two fluxes for the $\mathcal{Z}$-basis comparison measurement, we can perform it in one shot, i.e. implementing a single closed ribbon operator without the need for coherent moving. An example circuit is given below, at the step where we compare the data qutrit with the reference $\ket{0}_L$ qutrit:

\begin{center}
    \begin{tikzpicture}
        \draw[decoration=brace, decorate]
            		($(0,.3)$) -- ($(1.5,.3)$)
            		node[midway, above=1pt, text width=3.0cm, align=center] {Flip \\orientation};
        \draw[decoration=brace, decorate]
            		($(1.6,.3)$) -- ($(5.7,.3)$)
            		node[midway, above=1pt, text width=3.0cm, align=center] {Ungauging};
        \draw[decoration=brace, decorate]
            		($(5.8,.3)$) -- ($(6.9,.3)$)
            		node[midway, above=1pt, text width=6.0cm, align=center] {FDLU};
        \draw[decoration=brace, decorate]
            		($(7,.3)$) -- ($(11.1,.3)$)
            		node[midway, above=1pt, text width=3.0cm, align=center] {Regauging};
        \draw[decoration=brace, decorate]
            		($(11.2,.3)$) -- ($(12.7,.3)$)
            		node[midway, above=1pt, text width=3.0cm, align=center] {Revert \\orientation};
        \begin{yquant}[horizontal]

        [violet] qubit {$\ket{a_2}$} a[1];
        [violet] qubit {$\ket{b_2}$} b[1];
        [violet] qubit {$\ket{\alpha_1}$} a[+1];
        [violet] qubit {$\ket{\beta_1}$} b[+1];
        [violet] qubit {$\ket{\alpha_3}$} a[+1];
        [violet] qubit {$\ket{\beta_3}$} b[+1];
        [violet] qubit {$\ket{a_6}$} a[+1];
        [violet] qubit {$\ket{b_6}$} b[+1];
        [violet] qubit {$\ket{\alpha_4}$} a[+1];
        [violet] qubit {$\ket{\beta_4}$} b[+1];
        [violet] qubit {$\ket{\alpha_2}$} a[+1];
        [violet] qubit {$\ket{\beta_2}$} b[+1];

        box {$\mathcal{C}$} a[3],a[4],a[5];
        box {$\mathcal{C}$} a[3] | b[3];
        box {$\mathcal{C}$} a[4] | b[4];
        box {$\mathcal{C}$} a[5] | b[5];

        barrier (a,b);

        cnot b[1] | b[0];
        cnot b[2] | b[1];
        cnot b[3] | b[2];
        cnot b[4] | b[3];
        cnot b[5] | b[4];

        align a;
        box {$\mathcal{C}$} a[1] | b[0];
        box {$\mathcal{C}$} a[2] | b[1];
        box {$\mathcal{C}$} a[3] | b[2];
        box {$\mathcal{C}$} a[4] | b[3];
        box {$\mathcal{C}$} a[5] | b[4];

        barrier (a,b);
        box {$\mathcal{C}$} a;
        barrier (a,b);

        box {$\mathcal{C}$} a[1] | b[0];
        box {$\mathcal{C}$} a[2] | b[1];
        box {$\mathcal{C}$} a[3] | b[2];
        box {$\mathcal{C}$} a[4] | b[3];
        box {$\mathcal{C}$} a[5] | b[4];

        cnot b[5] | b[4];
        cnot b[4] | b[3];
        cnot b[3] | b[2];
        cnot b[2] | b[1];
        cnot b[1] | b[0];

        barrier (a,b);
        box {$\mathcal{C}$} a[3],a[4],a[5];
        box {$\mathcal{C}$} a[3] | b[3];
        box {$\mathcal{C}$} a[4] | b[4];
        box {$\mathcal{C}$} a[5] | b[5];
        
        \end{yquant}
        \end{tikzpicture}
\end{center}

It is slightly modified from the $[2]$ charge unitary ribbon given in Eq.~\ref{eq:appendix_2charge_circuit}. The orientation of some direct edges are flipped at first, such that all direct edges' orientation is aligned with respect to the direction of the ribbon operator. The orientation is reverted after the application of the unitary ribbon. 

Because of the nontrivial braiding of the $[2]$ charge with a $C_3$ flux (e.g. when performing comparison measurement between the data qutrit and the reference $\ket{1}_L$ state), a closed loop ribbon operator is not ``rotationally-invariant" along the loop. The start and end point of the ribbon operator determines the location of any remnant charge (in this case, a possible remnant $[-]$ charge) that breaks this rotational invariance. 

Similar circuits are used for $\mathcal{Z}$ comparison measurements implemented in Section~\ref{sec:bureau} and \ref{sec:magic}.

\subsection{$\mathcal{Z}$ bureau of standards}\label{app:bureau-imple}

The circuits used in $\mathcal{Z}$ bureau of standards are very similar to that used in $\mathcal{Z}$-basis measurement (see Section~\ref{app:zmeasure-imple}). The main difference is that the initial states are all in the $\ket{0}_L$ state (an example circuit creating such state is given in Section~\ref{app:example-0tilde-logical}. Subsequent $[2]$ charge braiding circuits are adapted from the example given in Section~\ref{app:example-2-charge-braiding}.

\subsubsection{Multi-verse interpretation of relative encoding}

First, we outline an argument by Mochon~\cite{mochon_smaller_groups_2004} that illustrates how we can interpret the relative encoding as the multi-verse of possible flux labels, such that the absolute encoding is one ``branch" of this multi-verse. Consider the state $\ket{\tilde0}_L = \frac{1}{\sqrt3}(\ket0+\ket1+\ket2)$. If we forget about the coherence between the $\mathcal{Z}$ eigenstates, it is a completely mixed state $\rho = \frac{1}{3}(\ketbra{0}{0}+\ketbra{1}{1}+\ketbra{2}{2})$. Equivalently, we can obtain this mixed state $\rho$ by starting with a qutrit Bell state $\ket{0}\ket{0} + \ket{1}\ket{1} + \ket{2}\ket{2}$ and trace out the second qutrit. Then, we can use it as a $\mathcal{Z}$-basis reference state. Without loss of generality, we can assert it to be the $\ket{0}_L$ in the relative encoding. Then, other $\mathcal{Z}$ states can be instantiated by $\mathcal{Z}$-basis comparison measurement with the reference state.

For example, given a reference $\ket{0}_L$ in the relative encoding, if we instantiate another $\ket{0}_L$ state, this is actually the state $\psi \sim \ket{0}\ket{0}+\ket{1}\ket{1}+\ket{2}\ket{2}$ in the absolute encoding. What matters is that the logical qutrit values are the same for every ``branch" of this state. 

\subsubsection{$[2]$ charge braiding once suffices for instantiating $\mathcal{Z}$ reference states}

Next, we explain why we only perform $[2]$ charge braiding once rather than twice to instantiate the $\mathcal{Z}$ bureau of standards reference states.

For a full $\mathcal{Z}$-basis measurement, we need to braid a $[2]$ charge pair twice to ensure that we obtain the state with the correct coherence in the orthogonal subspace after the measurement. For example, in Section~\ref{sec:magic}, two $[2]$ charge braiding are implemented to ensure that after projecting out of the $\ket{2}_L$ subspace, the resulting state $\ket{0}_L+\omega\ket{1}_L$ has the correct relative phase factor. 

However, for the $\mathcal{Z}$ bureau of standards protocol (Section~\ref{sec:bureau}), we do not require performing such $[2]$ charge braiding twice. This is because the coherence between the different ``branches" of the relative encoding is not important; all that matters is that the absolute flux label of the states are distinct pair-wise for the three reference states. Hence, we do not need to repeat braiding with the $[2]$ charge twice to obtain the correct logical reference states for the $\mathcal{Z}$ bureau of standards. 

\subsection{Magic state creation}\label{app:magic-imple}

The circuits used in creating the magic state ($\ket{0}+\omega\ket{1}$) are very similar to that used in $\mathcal{Z}$-basis measurement (see Section~\ref{app:zmeasure-imple}). The main difference is that the initial states are in the logical $\ket{2}_L$ and $\ket{1}_L$ state. For the logical $\ket{2}_L$, it is a slight modification of the circuit given in \ref{app:example-generalized-ribbon}: the circuit is conjugated with $\mathcal{X}^\dagger$ (see remark on $C_2$ ribbon in Section~\ref{app:unitary-ribbon} for explanation). For the logical $\ket{\tilde1}_L$ state, an example circuit is given in \ref{app:example-tilde1-logical}. Subsequent $[2]$ charge braiding circuits are adapted from the example given in Section~\ref{app:example-2-charge-braiding}.

\clearpage

\section{Extended Data Figures and Tables}\label{app:extended-data}
\begin{figure*}[!ht]
		\centering\includegraphics[width=0.55\linewidth]{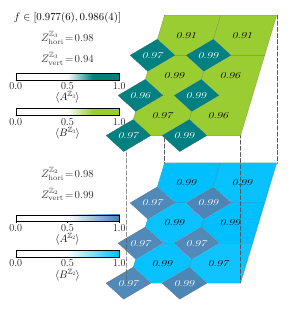}
		\caption{\textbf{Ground state on the $3\times 2$ lattice.} Experimentally measured values of local and $Z$-logical projectors for a state prepared via a fully unitary protocol (see Section~\ref{app:unitary-prep}). The $\ZZ_3$ toric code is first prepared unitarily, followed by the unitary gauging of the $\ZZ_2$ charge-conjugation symmetry. The per-qudit fidelity of the resulting state is denoted by $f$. The average (maximum) standard error is 0.018 (0.034) for local projectors and 0.018 (0.031) for $Z$-logical projectors.
		}\label{fig:appendix_gs_3x2}
\end{figure*}

\begin{figure*}[t]
		\centering    \includegraphics[width=0.87\linewidth]{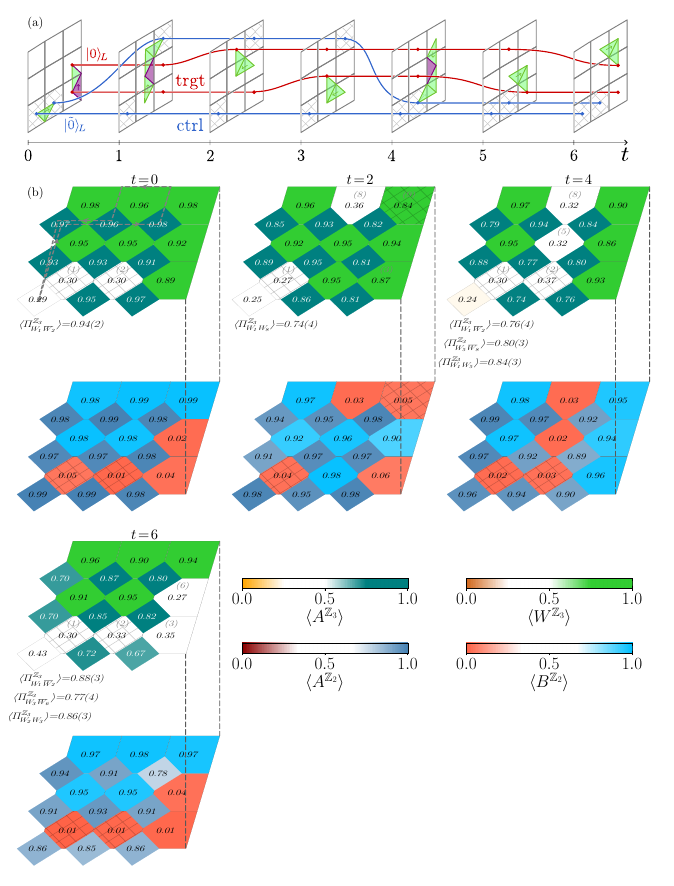}
		\caption{\textbf{Extended data for pull-through gate.} (a) A control qutrit, initialized in the $\ket{\tilde{0}}_L$ state, is shown with blue worldlines; plaquettes hatched by $\times$ indicate the locations of its endpoints at each time slice. The control qutrit is moved around a target qutrit initialized in the $\ket{0}_L$ state and shown with red worldlines. The ribbon operator implementing the motion is constructed from elementary triangle operators (see Section~\ref{app:ribbon-operators}). 
		(b) Measured expectation values of the local projectors $A^{\ZZ_2}$, $A^{\ZZ_3}$, and $B^{\ZZ_2}$, together with the nonlocal projector $W^{\ZZ_3}$, shown at different stages of the protocol. For all local and $W^{\ZZ_3}$ measurements, the average standard error is $0.023$, with a maximum of $0.052$. Also shown are the $W$-flux correlators $\Pi^{\ZZ_3}_{W_{p_1}W_{p_2}}$, which certify inter- and intra-pair delocalized $\mathcal{Z}$ correlations of the control and target qutrits throughout the protocol. In the noise-free limit these correlators equal $1$; the measured values are consistent with this expectation, indicating high-fidelity state preparation.
		}
        \label{fig:appendix_pull_through}
\end{figure*}

\begin{figure*}[t]
	\centering    \includegraphics[width=0.85\linewidth]{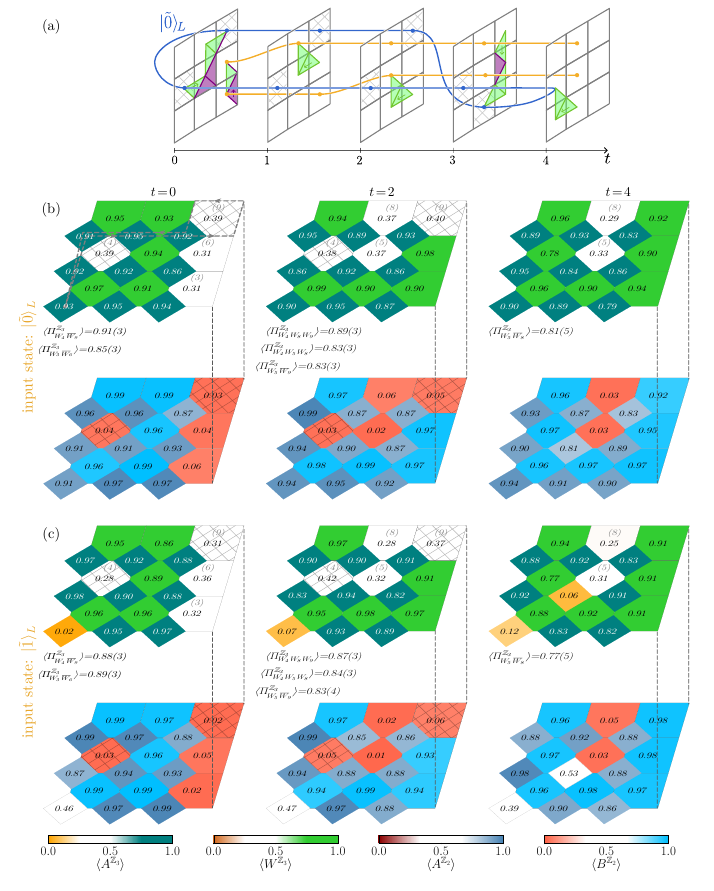}
	\caption{\textbf{Extended data for $\mathcal{X}$-basis measurement.} (a) Spacetime diagram of a ``measurement'' qutrit braided around the flux endpoint of an input qutrit. Worldlines are shown in blue (measurement) and yellow (input). Plaquettes marked by $\times$ denote the qutrit flux endpoints on the $3\times3$ lattice. (b,c) Measurement results for the $\ZZ_2$ and $\ZZ_3$ projectors. The input qutrit is prepared in the $\ket{\tilde{0}}_L$ state in panel (b) and in the $\ket{\tilde{1}}_L$ state in panel (c). In each panel, the lower grid shows the local $A^{\ZZ_2}$ and $B^{\ZZ_2}$ projectors, while the upper grid shows the local $A^{\ZZ_3}$ and nonlocal $\Pi^{\ZZ_3}_W$ projectors (cf Section~\ref{app:z2z3-decomposition} for precise $W$-flux paths). The vertex projectors $A^{\ZZ_2}$ and $A^{\ZZ_3}$ remain close to $+1$ at all sites except the origin and the remnant charge site, while the local $B^{\ZZ_2}$ plaquette stabilizers remain near $0$ at qutrit endpoints and reveal no logical information. The nonlocal $W$-flux projectors, which distinguish the logical states $\ket{0}_L$, $\ket{1}_L$, and $\ket{2}_L$, remain near $1/3$ due to the superposition of these states. Two-body $W$-flux correlators $\Pi^{\ZZ_3}_{W_{p_1}W_{p_2}}$ are close to $1$, indicating that the prepared state closely approximates the noiseless ideal. The local ($A^{\ZZ_2}$, $A^{\ZZ_3}$, $B^{\ZZ_2}$) and nonlocal ($W^{\ZZ_3}$) projectors have an average (maximum) standard error of $0.028$ ($0.072$).
	}
    \label{fig:appendix_x_basis_measurement}
\end{figure*}

\begin{figure*}[t]
\centering \includegraphics[width=0.85\linewidth]{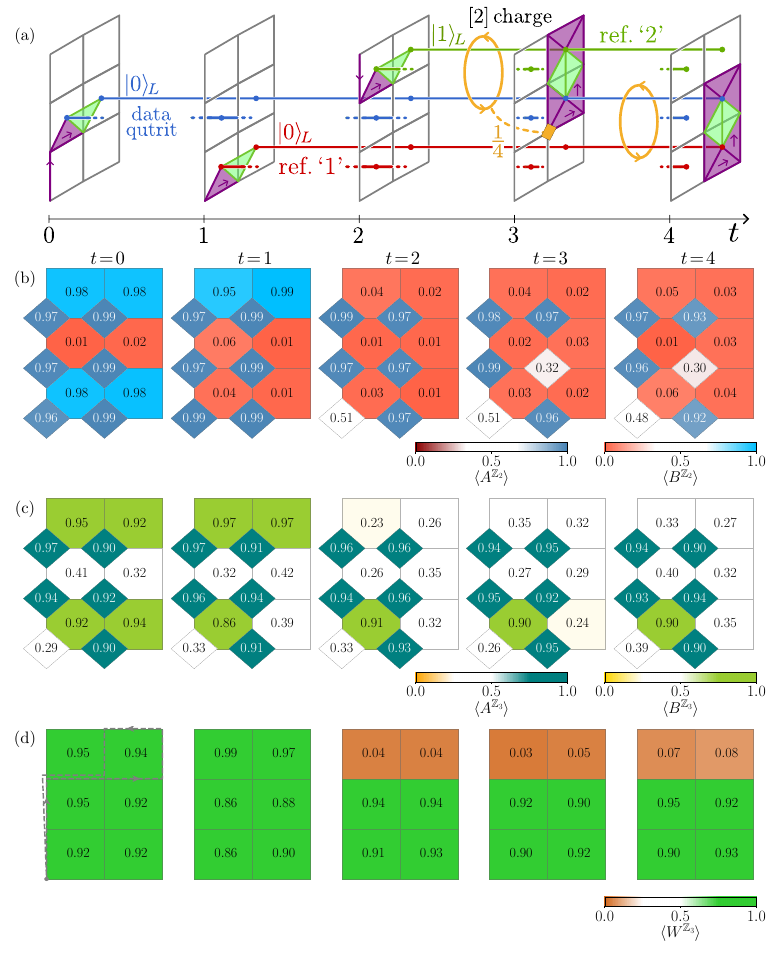}
\caption{\textbf{Extended data for $\mathcal{Z}$-basis measurement protocol.} (a) A data qutrit (blue) and two reference qutrits, labeled `1' (red) and `2' (green), are initialized in $\ket{0}_L$, $\ket{0}_L$, and $\ket{1}_L$ states at $t=0$, $t=1$, and $t=2$, respectively. $[2]$ charge pairs loop around the endpoints of data-reference qutrit pairs ($t=2$-$t=3$ for data and reference '2' qutrits, $t=3$-$t=4$ for data and reference '1' qutrits). (b) Expectation values of local $\ZZ_2$ vertex ($A^{\ZZ_2}$) and plaquette ($B^{\ZZ_2}$) projectors. $B^{\ZZ_2} \rightarrow 0$ upon logical qutrits initialization indicate presence of logical qutrit endpoints. The vertex projector at the first $[2]$ charge pair fusion site yields a value $\approx 1/4$, confirming remnant charge. Subsequent $[2]$ charge pair fusion between $t=3$ and $t=4$ leaves $A^{\ZZ_2}$ values unchanged, indicating no additional charge accumulation. (c) $\ZZ_3$ vertex ($A^{\ZZ_3}$) and plaquette ($B^{\ZZ_3}$) projector expectation values. $A^{\ZZ_3}$ remains constant ($\approx 1$, except $\approx 1/3$ at origin). $B^{\ZZ_3}$ projectors are measured $\approx 1/3$ at all qutrit initialization sites provide no information about the internal state of the qutrits. (d) Nonlocal $W^{\ZZ_3}$ projector values distinguish internal states: $W^{\ZZ_3} \approx 1$ for $\ket{0}_L$ state of data and reference `1' qutrits and $\approx 0$ for $\ket{1}_L$ state of reference `2' qutrit. Across all local and nonlocal projectors, the average standard error is $0.030$, with a maximum standard error of $0.064$.
}\label{fig:appendix_z_basis_measurement}
\end{figure*}

\begin{figure*}[t]
		\centering    \includegraphics[width=0.8\linewidth]{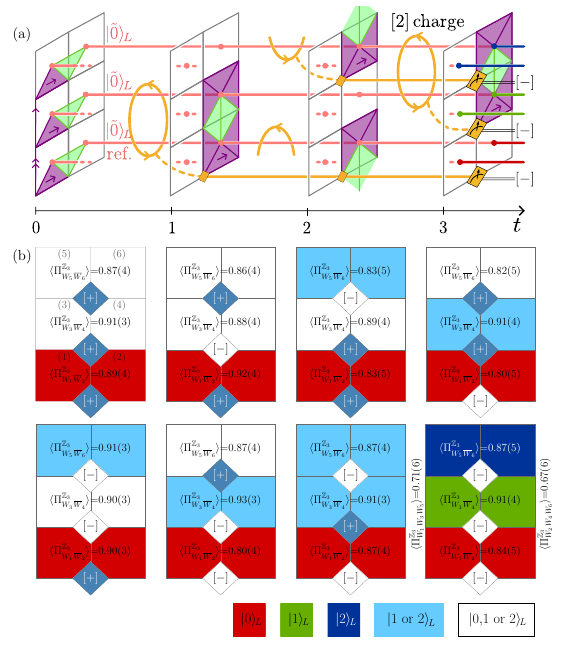}
		\caption{\textbf{Extended data for $\mathcal{Z}$ bureau of standards protocol.}(a) Spacetime diagram illustrating the action of ribbon operators at different steps of the protocol (see Section~\ref{app:bureau-imple} for details). (b) Expectation values of the $\Pi^{\ZZ_3}_{W_{p1} W_{p2}}$ projector, which quantify the fidelity of intra-pair $\mathcal{Z}$-correlations. The data is shown for different sets of `partially' prepared basis states, where each set is defined by post-selection on specific outcome of the remnant charge ($A^{\ZZ_2}$) measurement. 
        }
    \label{fig:appendix_z_bureau_of_standards}
\end{figure*}

\begin{figure*}[t]
		\centering\includegraphics[width=0.6\linewidth]{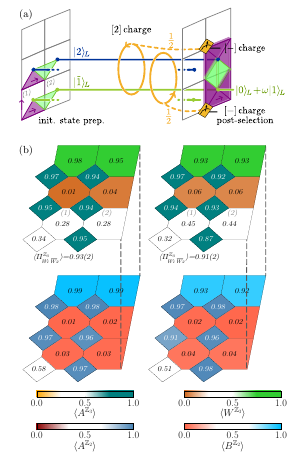}
		\caption{\textbf{Extended data for magic state preparation protocol.} (a) Schematic illustrating the creation of a magic state by inducing projection via measurement (see Section~\ref{app:magic-imple} for details). (b) Expectation values of local projectors $A^{\ZZ_2}$, $A^{\ZZ_3}$, and $B^{\ZZ_2}$, and the nonlocal projector $W^{\ZZ_3}$ after initial state preparation and magic state preparation. For plaquettes (1) and (2), the measured $\Pi^{\ZZ_3}_W$ values are 0.28 and 0.28 after initial state preparation, in close agreement with the theoretical value of 1/3. Following the complete protocol, these values change to 0.45 and 0.44, consistent with the theoretical value of 1/2. The good quality of the prepared states is further supported by measurements of $\Pi^{\ZZ_3}_{W_{1} W_{2}}$, which are close to unity. Across all $W^{\ZZ_3}$ and local projector measurements, the average standard error is $0.017$, and the maximum standard error is $0.057$.
		}\label{fig:appendix_magic_state}
\end{figure*}

\clearpage 

\begin{table}[hbt!]
    \label{table_gate_counts}
    \bgroup
    \def\arraystretch{1.2}
    \begin{tabular}{ |c|c|c|c|c|c| } 
        \hline
        Experiment 
        &\begin{tabular}{@{}c@{}} Lattice \\ $\quad$size$\quad$ \end{tabular} 
        &\begin{tabular}{@{}c@{}} No. of native \\ $\quad$one-qubit  gates$\quad$ \end{tabular}  &\begin{tabular}{@{}c@{}}No. of native \\ $\quad$two-qubit gates$\quad$ \end{tabular}    
        &\begin{tabular}{@{}c@{}}Depth  \\ $\quad$(two-qubit gates)$\quad$\end{tabular}    \\ 
        \hline
        \hline
        \begin{tabular}{@{}c@{}} Ground state unitary prep.  \end{tabular} & $3\times3$ & 429 & 238 & 43 \\
        
        \hline
        \begin{tabular}{@{}c@{}} Single anyon  state prep. \end{tabular} & $3\times3$ & 453 & 308 & 76\\
        
        \hline
        \begin{tabular}{@{}c@{}} $\mathcal{Z}$-basis measurement  at $t=4$  \end{tabular} & $3\times2$ & 423 & 269 & 74\\ 
        
        \hline
        \begin{tabular}{@{}c@{}} $\mathcal{Z}$-basis bureau of standards at $t=3$ \end{tabular} & $3\times2$ & 646 & 437 & 107\\
        
        \hline
        \begin{tabular}{@{}c@{}} Magic state preparation\end{tabular} & $3\times2$ & 507 & 346 & 134 \\
                
        \hline
        \begin{tabular}{@{}c@{}} $\mathcal{X}$-basis measurement at $t=5$\end{tabular} & $3\times3$ & 938 & 744 & 232 \\ 
        
        \hline
        \begin{tabular}{@{}c@{}} Pull-through gate at $t=6$\end{tabular} & $3\times3$  & 1008 & 845 & 307\\ 
        
        \hline
        
    \end{tabular}
    \egroup
    \caption{\textbf{Circuit specifications}. Gate counts and depths for the circuits used in each experiment. Destructive measurement at the end of the circuit was performed to compute the plaquette projectors. Furthermore, each circuit was optimized using local qutrit basis transformations (see. Section~\ref{sec:app-local-basis}). For reference, the execution time for a single shot of the $\mathcal{X}$-basis measurement at $t=5$ was approximately $5.0$ s, and for the pull-through gate experiment at $t=6$ it was approximately $5.9$ s on Quantinuum's H2-1 quantum computer.
    }
\end{table}

\end{document}